# Contents

# SQL and NoSQL Databases Software architectures performance analysis and assessments - A Systematic Literature review

Wisal Khan[1*], Teerath Kumar[2], Zhang Cheng[1*], Kislay Raj[2], Arunabha M Roy[3], Bin Luo[1]

[1] School of Computer and Technology, Anhui University, Hefei 230039, Peoples Republic of China,
[2]School of Computing, Dublin City University, Dublin, Ireland.
[3] Aerospace Engineering Department, University of Michigan, Ann Arbor, MI 48109, USA

## Abstract

**Context:** The efficient processing of Big Data is a challenging task for SQL and NoSQL Databases, where competent software architecture plays a vital role. The SQL Databases are designed for structuring data and supporting vertical scalability. In contrast, horizontal scalability is backed by NoSQL Databases and can process sizeable unstructured Data efficiently. One can choose the right paradigm according to the organisation's needs; however, making the correct choice can often be challenging. The SQL and NoSQL Databases follow different architectures. Also, the mixed model is followed by each category of NoSQL Databases. Hence, data movement becomes difficult for cloud consumers across multiple cloud service providers (CSPs). In addition, each cloud platform IaaS, PaaS, SaaS, and DBaaS also monitors various paradigms.

**Objective:** This systematic literature review (SLR) aims to study the related articles associated with SQL and NoSQL Database software architectures and tackle data portability and Interoperability among various cloud platforms. State of the art presented many performance comparison studies of SQL and NoSQL Databases by observing scaling, performance, availability, consistency and sharding characteristics. According to the research studies, NoSQL Database-designed structures can be the right choice for big data analytics, while SQL Databases are suitable for OLTP Databases.

The researcher proposes numerous approaches associated with data movement in the cloud. Platform-based APIs are developed, which makes users' data movement difficult. Therefore, data portability and Interoperability issues are noticed during data movement across multiple CSPs. To minimize developer efforts and Interoperability, Unified APIs are demanded to make data movement relatively more accessible among various cloud platforms.

**Method:** The systematic literature review technique and approach are used in this paper to select the appropriate and related documents. Most of the articles investigated the technical reasons for both Databases and identified the scenarios when to use which Database. Data analysis, data collection process, and the required results are detailed in this paper.

**Results:** A total of 142 papers have been selected associated with the topic in this systematic literature review. 35% are journal documents, 52% are conferences, and 11% are technical reports and thesis. We also performed a performance analysis between the SQL and NoSQL document Databases. Besides, DBaaS and unified APIs approaches are investigated in terms of data portability and Interoperability to extract the desired results. We evaluated and analyzed the research papers accordingly and identified the state-of-the-art gaps.

**Conclusion:** According to our findings and analysis in this SLR, NoSQL Databases are not the alternative to SQL Databases. Each Database has its advantages in a particular scenario. The SQL and NoSQL Databases follow various data models and software architectures. In contrast, data movement is strenuous across multiple cloud platforms. DBaaS cloud architecture is used to transfer traditional Database architecture into cloud architecture. Different, unified APIs frameworks have been investigated to minimize data portability and Interoperability issues across various cloud platforms during data movement.

## 2. Introduction

The architecture of a particular software deals with the non-functional attributes such as reliability, usability, scaling, performance, Interoperability, portability, adaptability and data sharding. The tradeoffs among the set of quality attributes always exist and the challenging task for a software architect. The big data[1] systems are inherently distributed. Data sharding and replication within extensive data systems caused data availability and consistency issues. Database technologies are resulted in significant fluctuations due to enormous data applications growth. For more than one decade, the scale of NoSQL Databases arose vastly while full-fledged, traditional Database automation is spread persistently. The precise schema structure is being forced by traditional mock-up, which leads to scaling obscurity and prevents data modification over clusters. In contrast, easy prototypes are supported by NoSQL Databases. The main characteristics of NoSQL Databases architectures are:

- Schema-less structure
- Permitting data representations to grow effectively and dynamically and
- Scaling horizontally, by data replication collections and sharding over massive clusters.

In recent years, various organizations have been building a large amount of data, and relational Databases cannot process such amounts of data efficiently. Relational Databases have been performed tremendously in the last four decades. They are designed for structured data and follow the ACID (Availability, Consistency, Isolation, and Durability) property. While "Big Data" consists of tools and technologies that handle a large amount of data at any scale. Big data consists of 5Vs (Volume, Velocity, Variety, Veracity, and Value) and consists of a vast amount of varied nature of unstructured data. Many frameworks are used for big data processing, such as Hadoop/MapReduce, Spark, Flink, and Samza.

The issue of SQL queries performance and optimization for the enterprise, production, and parallel Databases and big data[2] is gained more attention in the last few years. The inefficient and not optimized queries may consume system and server resources and lead to table locking and data loss problems. Information mining, instead of information itself, means extracting the facts and logical correlation layout from the original data set. The term query optimization means to select the appropriate query execution plan with minimum cost and low usage of system resources. The data mining algorithms query the Databases deeply and profoundly to extract patterns and knowledge from the comprehensive data [3]. Alternative approaches such as XML and object Databases had never gained the same reputation as RDBMS technology.

- Over the last decade, science and web vendors have raised the question of the data store technology "one size fits all". This thinking led to a new alternative Database approach termed NoSQL means 'Not only SQL.' NoSQL describes the use of non-relational Databases among web developers [4]. In 1998 the term NoSQL [5] was used for the first time and gained more attention at the San Francisco conference of non-relational Databases. The essential characteristics of NoSQL Databases are described in figure 2.

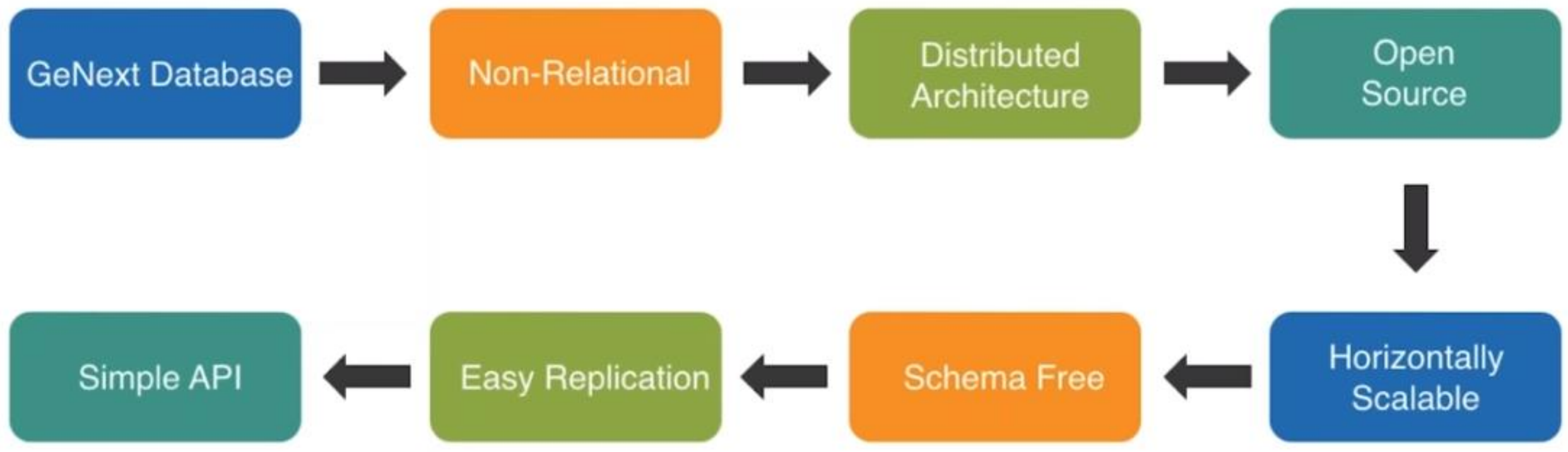

Figure 2: NoSQL – The essential features.

Eric Brewer presents the CAP (Consistency, Availability, and Partition Tolerance) theorem at ACM [6], [7]. The main characteristics of CAP theory are given in table 1.

| **Consistency** | **Availability** | **Partition Tolerance** |
|---|---|---|
| • Consistency means that the data in the Database remains consistent after the | • Availability means that the system will not have downtime (100% service uptime guaranteed) | • Partition tolerance means that the system continues to function even when the communication |
| execution of an operation.<br>• For example, after an update operation, all clients see the same data | • Every node (if not failed) always executes the query | between servers is unreliable.<br>• The servers may be partitioned into multiple groups that cannot communicate with one another |

Table 1: CAP Theory

Theoretically, it is impossible to fulfill all three requirements. Therefore, CAP supplies the essential need for a distributed system to follow two of the three elements. Hence, all the current NoSQL Databases support the different combinations of C, A and P of the CAP theorem, as describes in figure 2.

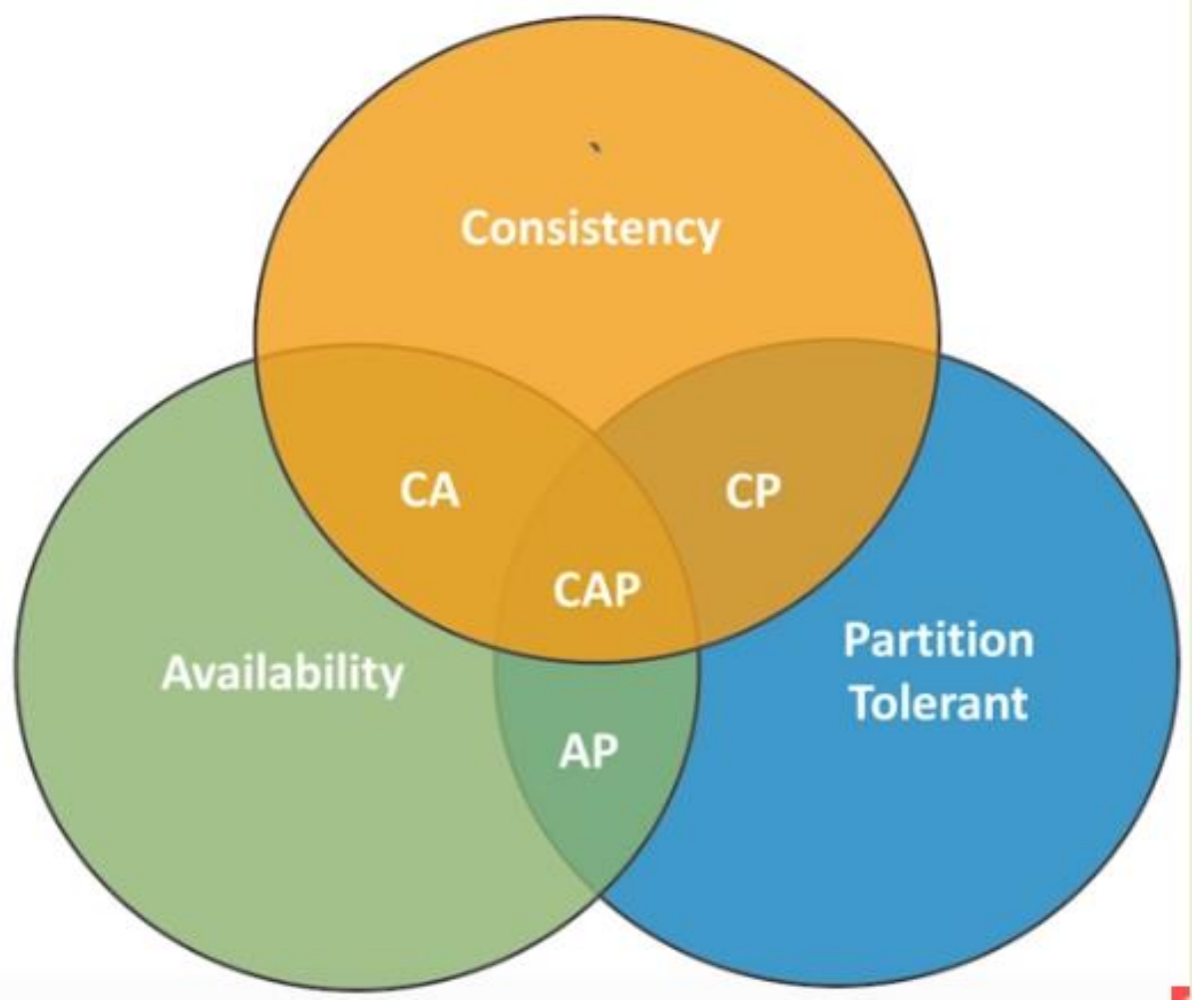

Figure 2: CAP Combinations.

The following are the advantages of NoSQL Databases:

- Volume: Data at rest - Terabytes to exabytes of existing data to process
- Velocity: Data in motion - Streaming data, milliseconds to seconds to respond
- Variability: Data in many forms – structured, unstructured, text, Etc.
- Veracity: Data in doubt – uncertainty due to latency, deception, ambiguities, Etc.
- Not built on tables and does not employ SQL to manipulate data.
- The schema comprises of key-value, document, columnar, graph, Etc.
- Alternative to traditional relational Databases.
- Database to handle unstructured, messy, and unpredictable data.
- Helpful for working with large sets of distributed data.

NoSQL Databases are of different kinds and follow unique models and architectures. Unlike relational Databases, it follows the relational data model. NoSQL Databases, Interoperability, portability, and security[8] approaches require adequate attention due to various models. Particularly on the cloud whenever storing NoSQL Database on it. Recently cloud service providers (CSP) [9][10] offer good services at low cost and can access the data at high speed. CSP provides greater scalability, availability, and confidentiality[11]. However, the third-party providers have full access to the client's data and are not considered the trusted source. Therefore, it is necessary to encrypt[12] the users' sensitive data before giving access to the data to the CSP. It is a challenging task because of the heterogeneous nature of NoSQL Databases. Moreover, Database as a Service (DBaaS)[13] is a cloud service platform that moves traditional architectures to cloud architectures.

The main contribution of this SLR is:

- This SLR is related to the SQL and NoSQL Databases architecture assessments, scaling capabilities, and performance analysis. In addition, data movement among various Databases across multiple cloud platforms has been explored.
- A total of 142 studies have been analyzed, ensuing the research as mentioned earlier questions.
- This article identifies the research gaps in the associated architectures and their causes.

## 2.1. State of the Problem

The system scale the experts have been developing has significantly increased over the past ten years. The most prominent example is the leading internet organisations. The scaling density is being felt at all levels within the business, Government, and defence departments. Whenever critics build scalable systems, they need to think differently about the solution and adopt the software architecture, method and mechanism that can ensure the scalability as the system demands. However, data (Big data) is a driving force behind scalability.

Big data[1] is everywhere, and scientists and researchers are trying to find the appropriate techniques to process such a massive amount of varied information efficiently. Big data analytics has proposed many approaches and methods [14]; traditional Databases have been used in the last three decades and have shown tremendous performance. They are designed specifically for structuring data and following the ACID property. Relational Databases are vertical scalable and can process a certain amount of data.

The prominent datum consists of 5Vs (Volume, Velocity, Variety, Veracity, and Value), and the relational Databases cannot manage such a massive amount of data. Hadoop framework has been proposed for processing big data using the MapReduce programming module. Hadoop follows the HDFS (Hadoop Distributed File System). Hadoop is using the Hive NoSQL Database.

NoSQL means Not Only SQL and is used for processing massive data. Unlike relational Databases, referenced as SQL Databases, NoSQL Databases follow the BASE property. NoSQL Databases are horizontally scalable Databases. Unlike SQL Databases, NoSQL Databases are of four different types, and each type is used according to the organisational needs and requirements. Four different NoSQL Databases are Key-Value, Document, Column, and Graph.

Each NoSQL Databases have its benefits and uses. For instance, a NoSQL graph Database is used to store data in nodes rather than tables due to the interconnected nature of the data. Relevant natural means when data is connected more, and more and a bunch of joins are required to retrieve the detail of a particular record. Unlike SQL Databases, relationships (joins) are saved among the nodes in NoSQL graph Databases and require constant execution time. In SQL Databases, join are calculated at run time and need more time when executing. The literature showed that the NoSQL graph Database performed better than relational Databases because of a graph Database's schema-less and dynamic architecture.

MongoDB has handled big data efficiently because of its high scalability characteristics. The literature has shown many comparisons of the NoSQL document Database and relational Database. According to the contributed research study in the mentioned field, the NoSQL document Database performed better for big data analytics. Considering OLTP Databases and complex schema structures, conversions to MongoDB may cause many problems, such as unique indices, composite keys, data inconsistency, and data duplications. Data aggregation in massive unstructured data is also a fundamental Database issue, particularly for BI users. The processing of extensive geospatial data efficiently is another hot problem for Databases.

Another critical issue is user-sensitive data security whenever data is transferred to the cloud storage. In this scenario, information is being managed by the third party (cloud service provider), not by the data owner. In this SLR, we will investigate the existing NoSQL Database security approaches and policies in-depth and identify the state-of-the-art gap. Due to NoSQL models' heterogeneity and varied natures, it is not easy to propose and design a unified cloud solution. Most CSPs are not following common standards and interfaces, and therefore, they are incompatible. The current cloud service provider (CSPs) architectures are built without considering Interoperability and portability problems.

## 2.2. Method

Our systematic literature review (SLR) is according to [15] and is commonly used. The evidence base systematic literature review is suggested by [16] for software engineering researchers. Almost all the software engineering researchers used the guidelines for the SLRs proposed and indicated by [15]. The main aims of EBSE are to synthesize high-quality primary studies from various sources based on evidence associated with specific research questions. Originally SLR process was used in medical sciences. Later on, SLR was used in other areas and fields too. Besides, SLR studies [17][18][19] are an advantageous way for future research, and directions as SLR captures maximum associated papers and articles related to a particular area. In practice, the research is conducted with a formal review.

Our proposed SLR highlighted SQL and NoSQL Database architecture assessments and performance evaluations. Also, data portability and Interoperability were explored across multiple cloud platforms in terms of data movement among various Databases. Numerous performance evaluation studies and surveys [19]–[25] have been investigated among SQL and NoSQL Databases like survey papers and empirical and constructive studies.

The main characteristics of our SLR, which differentiates from existing studies, are data collection via systemic process, associated covered studies list, the study scope focused on (performance analysis, Interoperability, and portability), and selected studies classifications and analysis.

The rest of the paper is organized as follows: section 3 presents relevant research questions, search strings, inclusion/exclusion criteria, data extraction, and classification. The results of selected papers are described in section 4, followed by discussion and research gaps in section 5. The conclusion and future are summarized in section 5.

## 3. Research Question

The purpose and the main focus and objective of the current SLR:

1. Addressed the existing SQL and NoSQL document approaches and techniques by considering big data processing.
2. The systematic literature review is associated with the SQL and NoSQL Databases.
3. Review selected study subsets in-depth.
4. The strength and weaknesses of SQL and NoSQL Databases are assessed based on the evidence collected and analysed from these studies.
5. Highlights the research gap in the area.
6. Identify future research directions.

The following research questions are formulated during our SLR to achieve the main objective of our study:

1. Considering Big Data (Structured and Unstructured Data): what is the need for NoSQL?
2. Why does the NoSQL Database follows the BASE property instead of the SQL Database ACID property?
3. Does DBaaS tackle data Interoperability and portability efficiently of various NoSQL Databases?

### 3.1. Search Criteria

This portion describes the search criteria of primary studies. All the associated literature is being identified and collected via the search criteria, which satisfy the inclusion and exclusion criteria. Various search techniques are used to search the related research, like electronic Database search, manual search, and snowballing. Besides these, associated journals and conference proceedings are also searched. While performing the systematic search, the protocol proposed by [15] was followed for the current SLR. Besides, we followed the seven phases for our SLR investigated in [17]. Figure 3 describes the selection of associated studies in various aspects.

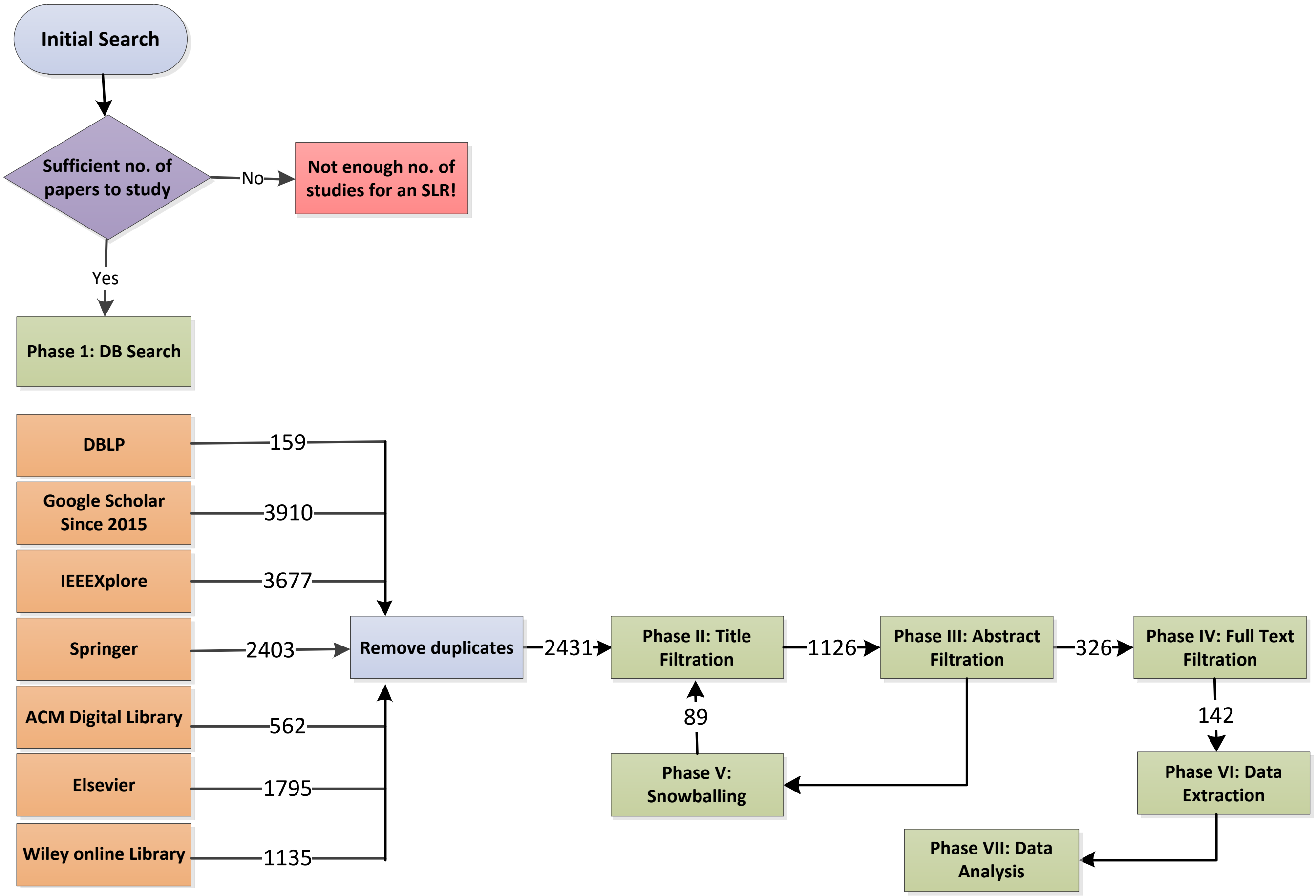

Figure 3. Selected studies process flow

First, we perform the Database search suggested by [26]. The search strings mentioned in the search strategy section were used for categorising and classifying, i.e., tools, methods, and framework.

### 3.1.1. Search Resources

In phase I, we derived a set of associated search strings. We used the derived set of strings to find the related papers. The largest Databases have been selected for finding the associated articles like:

- DBLP,
- IEEExplore Digital Library (ieeexplore.ieee.org)
- Google Scholar,
- ACM Digital Library (dl.acm.org)
- Springer (Springerlink.com)

- Elsevier (sciencedirect.com) and
- Wiley online library (onlinelibrary.wiley.com)

Article titles, abstracts, and keywords are considered only during the search process. Initially, we found around 13 thousand papers from the mentioned Databases by using the search strings listed in the search strategy section. In phase II, 2431 papers proceeded for inclusion and exclusion criteria after removing the repeated articles of phase I. Besides, we also performed backwards snowballing in Phase V. The selected study's reference list was analyzed during backward snowballing for searching associated items [27]. A total of 89 articles were chosen via snowballing after validating the inclusion/exclusion criteria repeatedly (Phase II to IV).

### 3.1.2. Search Strategy

Various combinations of search strings have been contrived. The devised search strings will run on the mentioned search resources to identify the associated literature.

- "SQL and NoSQL"
- "SQL or NoSQL"
- "Relational Database and Document Database"
- "Relational Database or Document Database"
- "Relational Database and NoSQL document Database."
- "Relational Databases and MongoDB."
- "Relational Databases or MongoDB."
- "Oracle and MongoDB comparison."
- "SQL and NoSQL Database comparisons."
- "Advantages of MongoDB over RDBMS"
- "relational and non-relational Databases."
- "cloud data portability and Interoperability."

## 3.2. Selection process and criteria

With the co-authors' help, we included all the associated papers by satisfying the inclusion/exclusion criteria after accumulating the studies. In phase 1, we evaluated the quality and characteristics according to our research questions. After the critical observation, we finalized the final list of selected and associated papers. The organizations of our research papers as per source are as follow:

- Step1: Total number of documents based on:
    1. Papers titles.
        i. Papers Abstract.
            a. Associated papers full reading
                1) Check the quality and impact of related papers

- ✓ Check the article in the catalog to avoid repetition
- ✓ Add item to the Finalized Papers catalog

2) Manual Search and Snowballing
3) Repeat the whole process, go to step1

Table 2 depicts the number of associated papers conferring the points as mentioned earlier after each phase filtrations.

| Search Source | Based On title (Phase II) | Based on Abstract (Phase III) | Papers full reading (Phase IV) |
|---|---|---|---|
| Total # of Papers | 1126 | 326 | 142 |

Table 2: Paper selection process and criteria

As shown in table 2, initially, based on the titles of the papers, 1126 papers were selected primarily. After reading the abstracts of 1126 candidate papers, 326 articles have been chosen by the first author. A total of 142 papers have been selected for full reading documents. In comparison, 89 of them are being chosen via snowballing. Most of the papers are based on empirical studies. The empirical research consists of the following:

- Research purpose
- Associated literature and supported theories
- Hypothesis measurement
- Proposed Method, design, approach, dimension and data collection
- Data Result analysis
- Conclusion

### 3.2.1. Inclusion Criteria

The following types of papers are included following our research questions:

- IC1: related SLRs and survey papers are added to conduct this SLR.
- IC2: New proposed techniques and approaches relevant to our proposed SLR.
- IC3: effective research methods presented in the proposed study are included.

To increase the reliability and efficiency of the SLR, the co-authors (2$^{nd}$, 3$^{rd}$) reviewed and investigated the impact and methodology of the included papers.

### 3.2.2. Exclusion Criteria

The research papers are excluded based on the following criteria's:

- EC1: papers not related to the mentioned domain are not included.
- EC2: irrelevant papers are excluded from our SLR.
- EC3: based on the title and abstract, some papers are also excluded.

- EC4: non-peer reviewed materials and papers are excluded.
- EC4: articles not written in English and duplicated articles are excluded.

To reduce the threat to the reliability of the SLR, the co-authors ($2^{nd}$, $3^{rd}$) rechecks the excluded papers according to the checklist of exclusion criteria.

The latest version of the proposed SLR will be used when it is published in the journal or conference. The document quality is checked by the co-authors ($2^{nd}$, $3^{rd}$)

## 3.3. Data Collection and Extraction

After collecting the 142 associated studies, review by the two reviewers. The reviewers extracted the data from the related articles [28], which satisfied the research questions. The following valuable data was obtained from each selected paper by the reviewers:

- Title of paper
- Abstract of paper
- Paper source (Journal or conference)
- Publication year
- Paper classification (type, scope)
- Related to the proposed SLR.
- Proposed SLR objectives and research question issues
- Each paper summary and method

We selected most of the empirical articles. The empirical studies consist of the following categories of the papers evaluations, discussion assessments, experiments, and reviews of existing techniques (SQL & NoSQL).

## 3.4. Data Analysis and classification

Furthermore, information about each of the extracted data is arranged and tabulated in accordance to answer the research questions as follows:

1. Considering Big Data (Structured and Unstructured Data): what is the need for NoSQL?
2. Why does the NoSQL Database follows the BASE property instead of the SQL Database ACID property?
3. Does DBaaS tackle data Interoperability and portability efficiently of various NoSQL Databases?

We authors, during analysis, grouped all extracted strategies into various categories. Additionally, the finding of our papers and summarized into three groups: research methods, research process phases, and evaluation.

### 3.5. Validity Threats and Evaluations

Threats, as suggested by [11], should be considered during the SLR process and evaluated their role consistently. The risks are categorized into different groups, as investigated by [29], [30]: descriptive validity, theoretical validity, interpretive validity, generalizability validity, and repeatability.

The SLR validity checks are analyzed by professor zhang cheng of Anhui University. His suggested changes are incorporated accordingly into the protocol.

## 4. Results

In this section, we summarized our selected studies in terms of publication year in Table 4 (Appendix A), paper category, and the number of selected studies from a specific digital library. Based on the selection process and criteria, we picked most of the empirical studies. As we have seen in the literature, the researcher used both types of Databases for their proposed methods and experiments. Besides empirical studies, we also found some survey papers associated with SQL and NoSQL Databases. After the process of associated review selection, we have identified three main categories of the selected studies and papers. Figure 4 depicts the studies category pie chart.

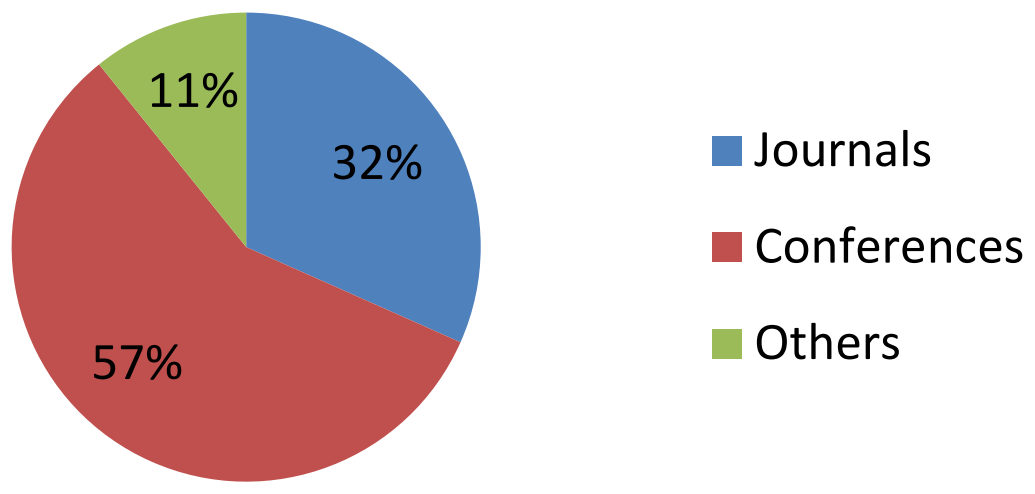

Figure 4: Categorization of selected studies pie chart

The publication years of each selected research associated with SQL and NoSQL Databases are between 2000 and 2019. As the NoSQL Databases got more attention after 2008. Figure 5 describes the number of papers per year.

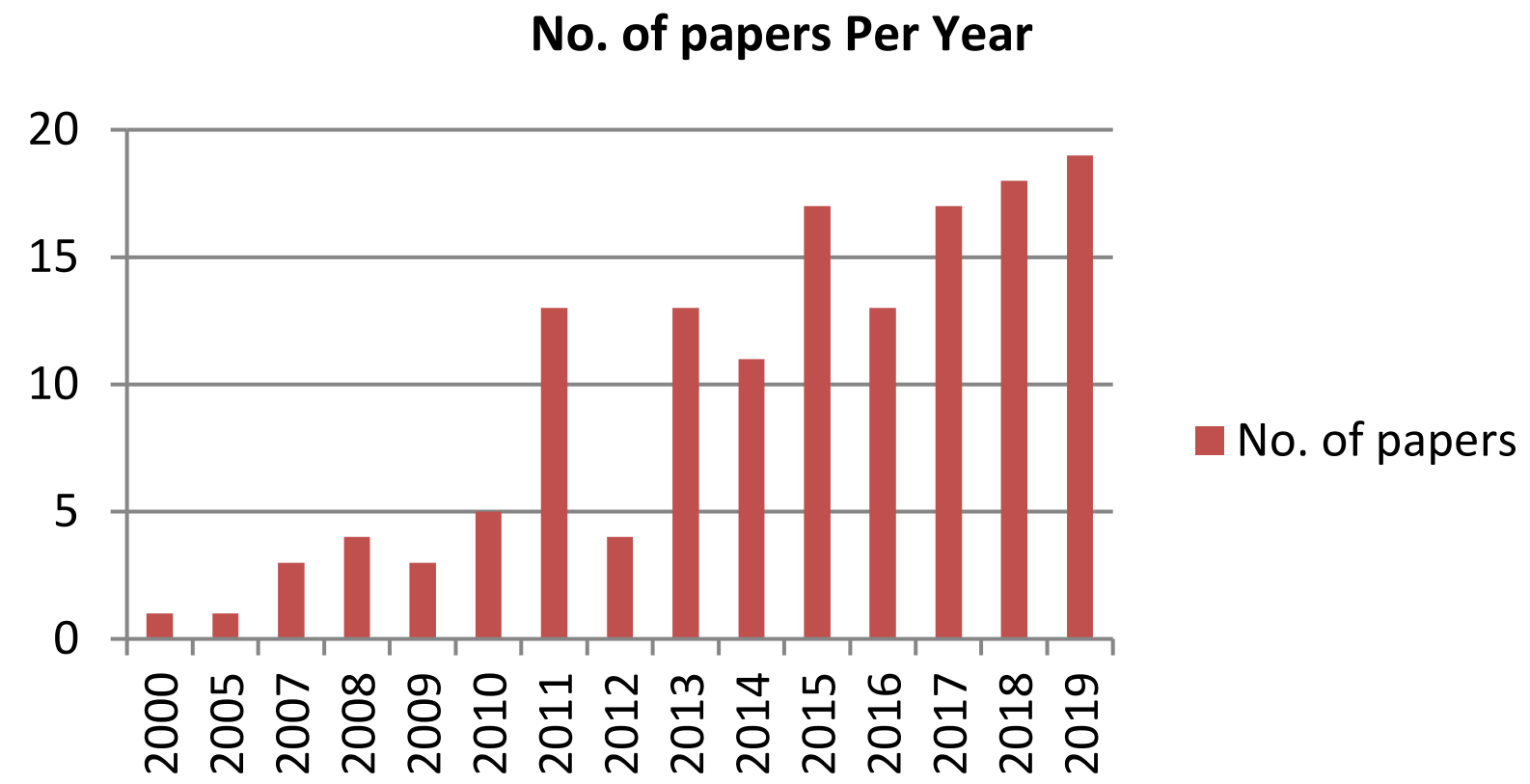

Figure 5. No. of papers per year

The number of papers related to relational and NoSQL Databases per digital library is presented in figure 6. Most of the related papers are found in IEEE and Springer digital libraries. Other options in figure 6 consist of articles such as white papers, book chapters, and technical reports from multiple publishers such as Oracle, MIT, Academic Journal, SciTePress, IJACSA, and IOPScience. Figure 7 illustrates the category of papers (Journal/Conference proceeding) of a specific digital library in figure 6.

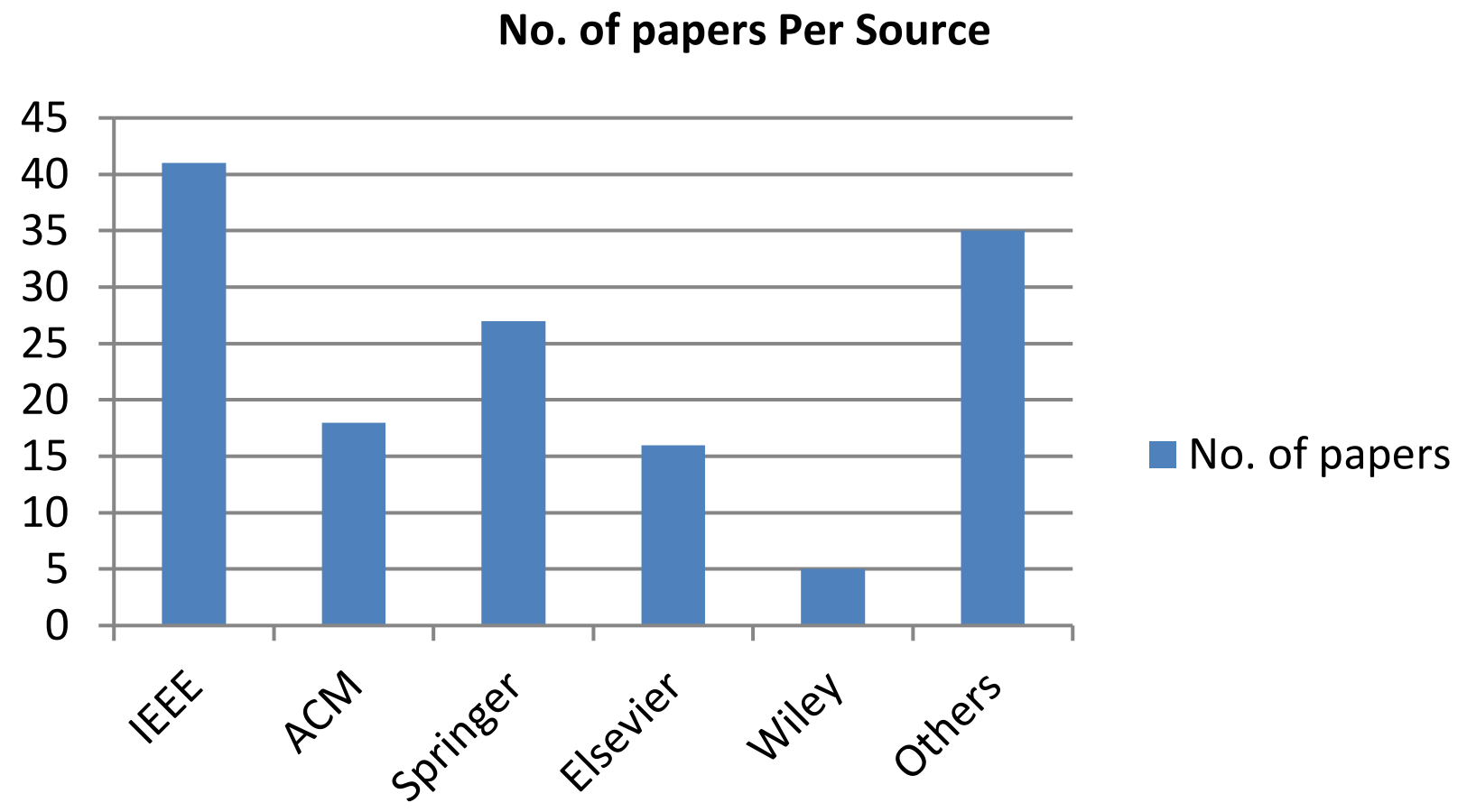

Figure 6. No. of papers per digital library.

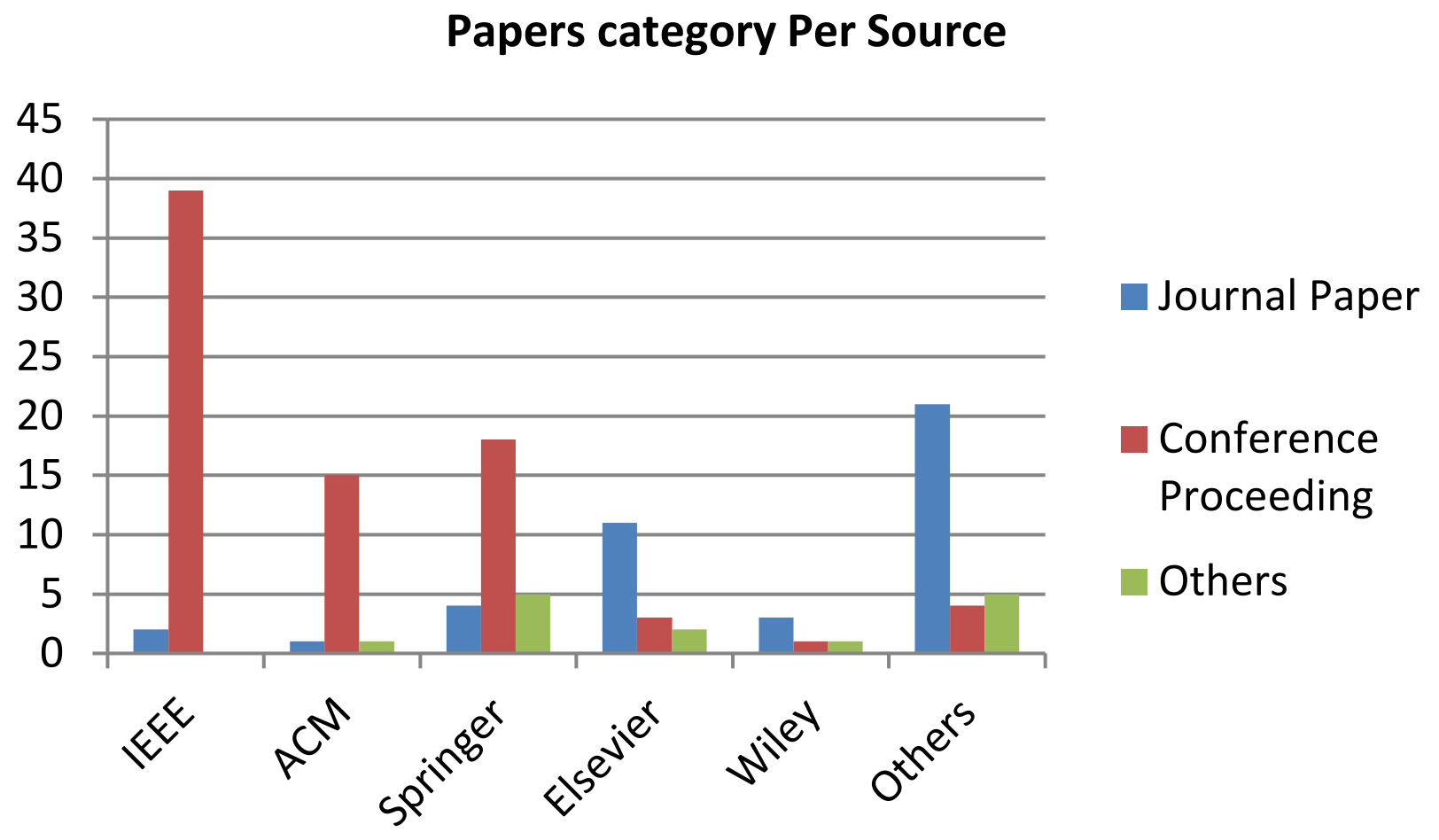

Figure 7. No. of papers category per source

During the analysis of the selected studies, many researchers used various flavours of Databases for performance comparisons and characteristics, such as MongoDB, MySQL, Cassandra, Couchbase, Oracle, SQL Server, PostgreSQL, Neo4j and Hbase. Figure 8 enumerates a different flavour of Databases used in the number of selected studies.

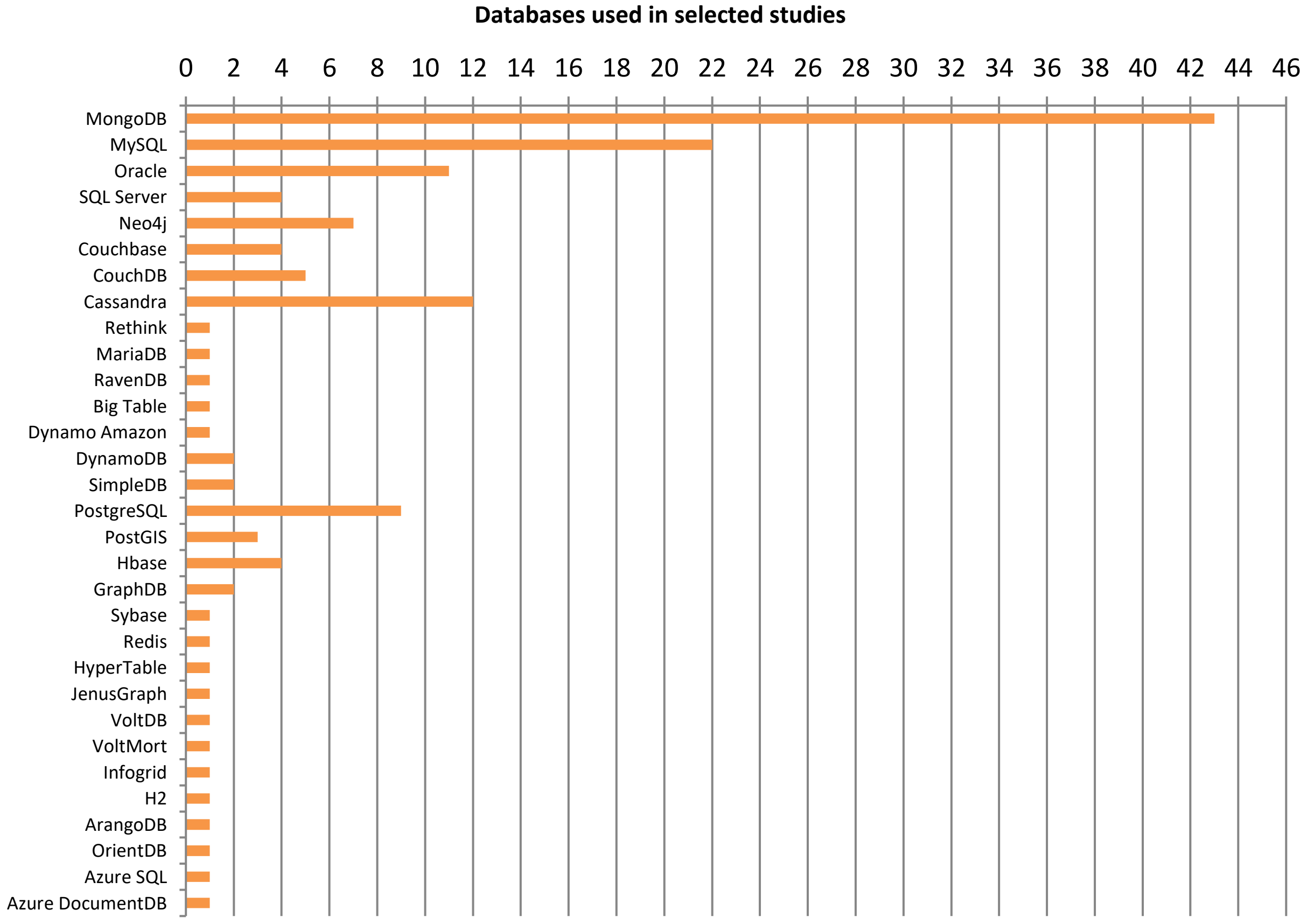

Most of the researchers used MongoDB, MySQL, Oracle, Cassandra, PostgreSQL, Neo4j, SQL server, and CouchDB for performance comparisons and characteristics analysis, as shown in figure 8.

## 4.1. Empirical Studies Analysis

The similarities, evaluations, experiments, classification, discussions, and surveys lie in the category of empirical studies and articles. The selected studies are analyzed accordingly, and analyzed how these articles are associated with our proposed research questions.

***RQ1: Considering Big Data (Structured and Unstructured Data): what is the need for NoSQL Databases? & RQ2: Why NoSQL Databases follows BASE property instead of SQL Database ACID property?***

Database Modeling is a way to understand how the Database will handle data and, more importantly, what type of data we can expect in a Database.

NoSQL means "Not Only SQL" [24] and is truly designed for managing large unstructured data and big data [31] analytics[14]. These Databases do not follow a fixed, rigid schema structure and have many flavours of query languages. At the same time, relational Databases have followed the standard SQL language for the last few decades. The document-oriented Database is one of the NoSQL Database types. Examples of documents oriented Databases are MongoDB and CouchDB. These Databases are used to store and manage document-oriented information. Document-oriented Databases store the data in JSON, BSON, XML, and PDF complex formats. MongoDB and CouchDB are open sources, and MongoDB [32] is specifically designed for the distributed environment and optimal for JSON. Various characteristics of NoSQL Databases have been investigated in [24]. For example MongoDB[20], [33] is developed in C++ and highly optimized for JSON. MongoDB follows schema-free or dynamic schema[17] structures[1] and does not follow predefined and fixed documents. Examining and accessing data is efficient and reliable in query processing, support indexing, and real-time aggregation. It also provides recovery and backup tools as well as end-to-end security. Figure 9 describes the architecture of SQL and NoSQL Databases.

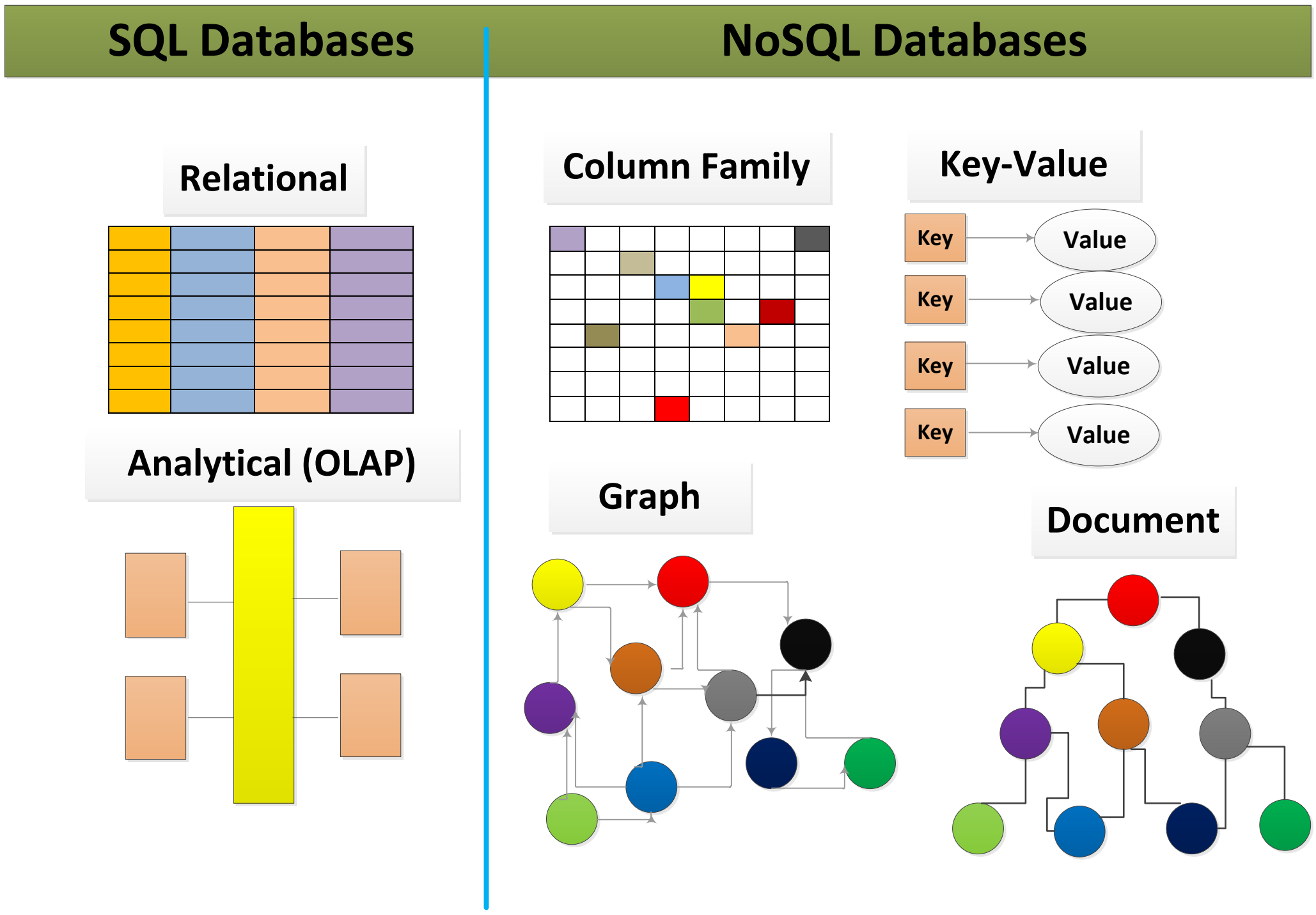

Figure 9: SQL and NoSQL Database architectures

Many researchers [21][33][26]–[33][42]–[50] did structure, design and performance comparison analysis of SQL (Oracle, MySQL and SQL-Server) and NoSQL (MongoDB, Neo4j) Databases. They used SQL and NoSQL Databases in their particular scenarios and evaluated the proposed methods, considering their results. Each NoSQL Database has its characteristics and uses. The NoSQL Databases are horizontally scalable, and due to these characteristics, the ACID property is not provided by the NoSQL Databases. For example, the Neo4j graph Database[51][52] is the right choice for connecting the nature of data. The literature has shown that the graph Database [53][54][55] handles and manages connected data very efficiently. MongoDB is mainly intended for documents. MongoDB exercises BSON format derived from JSON.

MongoDB deals with big data very efficiently and has shown tremendous performance. The study of [56] compared the NoSQL MongoDB and Postgre SQL Databases for SNS and stream sensor data. The MongoDB outperformed Postgre in their proposed use cases[56]. Unlike RDBMS, the NoSQL [57] Database can easily handle varied and complex structures and massive data in a cluster environment. In [35], the authors compared MongoDB and MySQL against the textbook management system. MongoDB outperformed MySQL in data retrieval and data insertion. Their proposed comparisons proved that NoSQL MongoDB is the appropriate choice for big data rather than MySQL. The article [43][58] investigated the need and requirements for handling sizeable unstructured data efficiently. They preferred to use the NoSQL MongoDB Database instead of using MySQL RDBMS. The relational and non-relational Database model has been considered in detail [44]. Depending on their analysis, the NoSQL MongoDB Database outperformed the MySQL Database. They suggested that for a small dataset with simple queries, MySQL is efficient.

While for large datasets with complex queries, MongoDB is a more appropriate choice. The authors of [40] compared the performance of Oracle RDBMS and the NoSQL MongoDB Database. Their analysis has shown that NoSQL Databases are not a suitable replacement for SQL Databases. According to them [40], one can choose a Database that considers their business requirements.

We have also found numerous migration papers [59]–[62][63]–[68] from a relational Database to NoSQL Databases and other papers regarding data scalability, high performance, and data availability [59], [69]–[74][75]–[78]. MongoDB can process big data efficiently due to its scalability, integrity, distribution, security [59], [69]–[74][19] characteristics and flexible architectures. Mentioned papers suggested it is better to migrate from a relational Database to NoSQL MongoDB. MongoDB store data in BSON (Binary JSON (JavaScript Object Notation)) format. MongoDB can store many documents in one collection and can access the data quickly without using joints as required in relational Databases. The varied nature of data, such as structured, semi-structured, and unstructured data, can easily manage and process the MongoDB. Relational Databases are particularly designed for structured data and can process a certain amount of data of vertical nature. The NoSQL Databases will not replace the traditional systems, but they are performed better and become more valuable in certain scenarios. The study [60][79] claimed the relational Database system is not good while scaling large data. In the article [61], the authors proposed an algorithm for automatic mapping from MySQL RDBMS to MongoDB NoSQL Database using metadata stored in MySQL RDBMS. The main characteristics of relational Database systems[80] are atomicity, consistency, isolation, and durability do not guarantee to provide these characteristics of the NoSQL Databases. NoSQL Databases follow BASE [81] property. Data availability, efficient data scaling, effective storage management, and better performance are the milestones of the NoSQL[60] umbrella. [23] investigates the transition reasons from RDBMS to NoSQL Databases. While in [82][83], the ACE (availability, consistency, and efficiency) properties have been discussed in both Databases for the big data system. The automatic schema transformation from SQL to NoSQL has been investigated in [84].

In article [70],[85],[86] various NoSQL Databases (Couchbase, MongoDB, RethinkDB, in-memory) performance is evaluated on real-data application. Various parameters have been considered during their proposed experiments, for example, the response time of diverse Databases. The study [71] made the Oracle NoSQL and MongoDB performance comparisons. They investigated Databases in diverse dimensions such as storage model, scalability, concurrency, and replication. Finally, they conclude that MongoDB is more widespread than the Oracle NoSQL Database. The rank of MongoDB is 5, and Oracle NoSQL is 78 for DBEngine.com.[1] According to paper [73], the Oracle RDBMS aggregation outperformed the MongoDB MapReduce. While MongoDB outperformed Oracle RDBMS in query response time of insertion, updating, and retrieving of data. The study of [40] has shown that the Oracle aggregation performed better than MongoDB aggregation in SUM, COUNT and AVG use cases. The dataset used by the [40] is contained around 500 thousand records and is not large enough for big data analytics. The SQL select statement's process flow diagram in Oracle 11g RDBMS is shown in Figure 10.

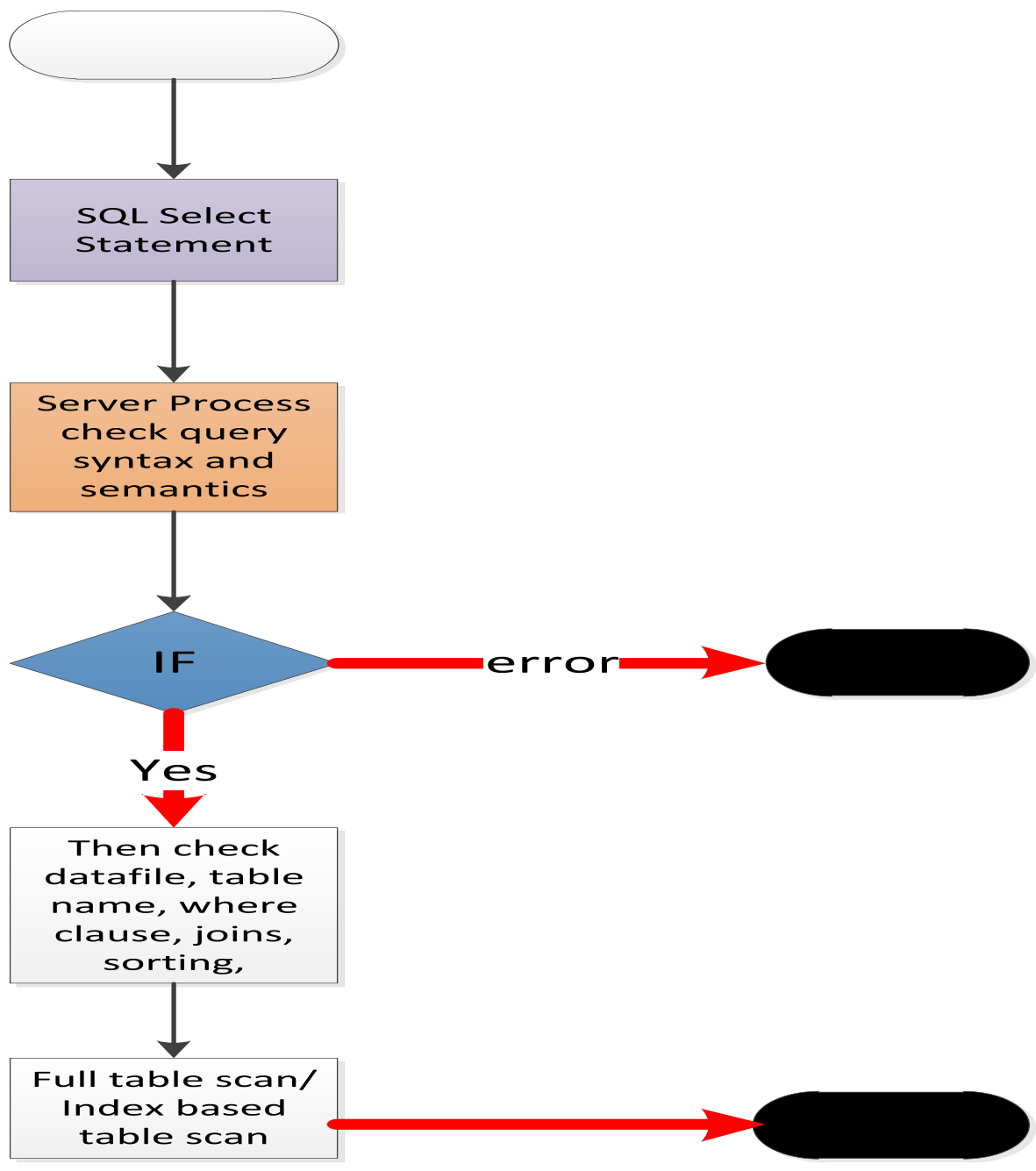

Figure 10: SQL Select statement process flow diagram

MongoDB data retrieval is simple and easy because of the sub-document structure and does not require to checking constraints or any clause, unlike the SQL select statement of Oracle 11g RDBMS. SQL insert statement processes flow given in figure 11.

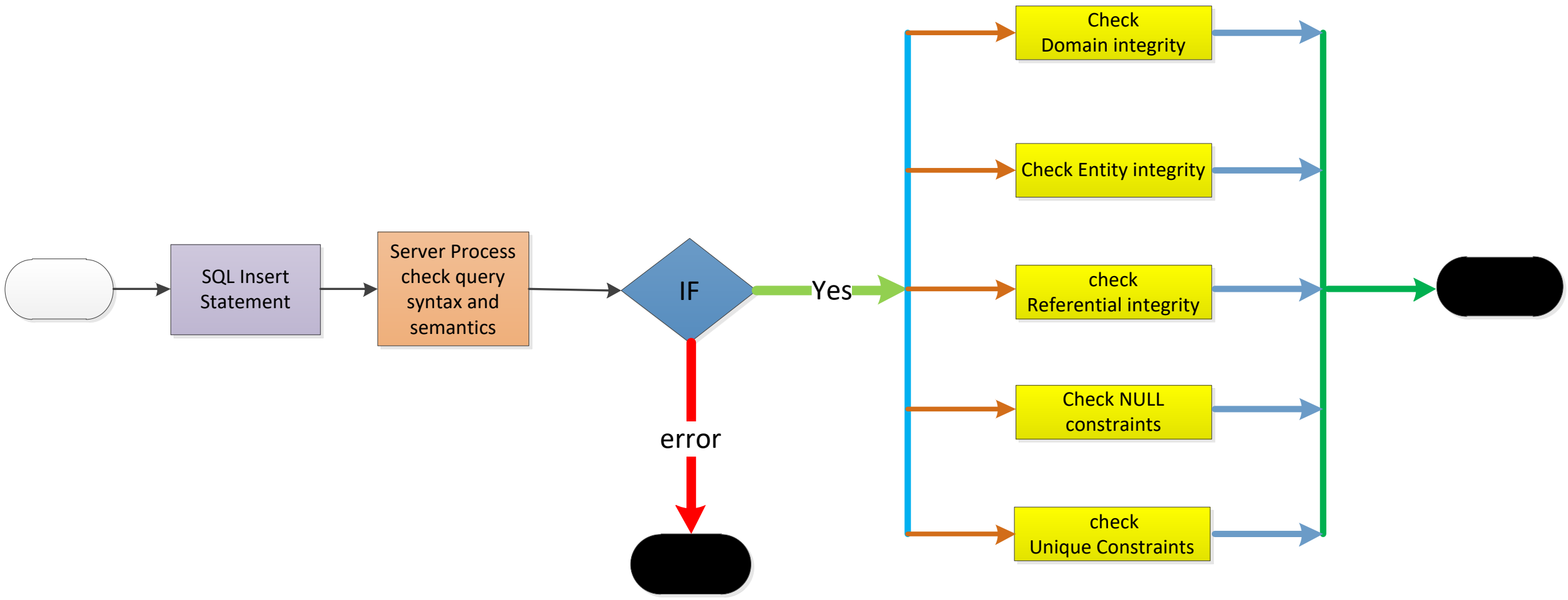

Figure 11: SQL insert statement process flow diagram

The MongoDB insertion does not need to check the mentioned steps in figure 2 of the SQL insert statement; therefore, MongoDB outperformed Oracle RDBMS in the insertion scenario.

In Contrast, MapReduce is the programming module [87] and performs well in distributing scenarios and is more suitable for big data[88] analysis against simple aggregation [73] in cluster (Multiple Server) environments. Before starting the map stage, map-reduce operations can perform any arbitrary sorting and limit on the documents in a single collection they use as their input. MapReduce needs two functions Map and Reduce; MongoDB applies the map phase of each input document in MapReduce. The map function returns pairs of critical values. The reduction phase, which MongoDB uses, gathers and condenses the aggregated data. The outcomes are then kept in a collection by MongoDB. For example, in the Chicago crime dataset, the map function calculates all the crimes against each day, and then the reduce function uses the day as the key and extracts the required values (Key: values). The authors of [89] added the merge operation into MapReduce architecture and derived the MapReduce-Merge framework. The performance of MapReduce was improved, and able to calculate relational algebra due to the merge operation and can process the data in the cluster. Others proposed [90] the MRShare framework to optimize MapReduce efficiency and improve query performance in the cluster. Therefore Oracle RDBMS aggregation outperformed MongoDB in MapReduce aggregation.

Oracle RDBMS has been using most industries for the last decade. Instead of having a tremendous performance in processing structured data, the SQL RDBMSs cannot process the big data efficiently and effectively because of their single image abstract system architecture.

Various NoSQL Database characteristics over/on Hadoop [91][92][93] and classifications have been investigated in the study [72]. According to the study [72] [94], the NoSQL Databases can be categorized by considering their various characteristics[95], such as scalability, nature of data (connected nature, document type of data), Etc. The main characteristics of MongoDB and Oracle RDBMS are shown in table 3.

| **SQL** | **MongoDB** |
|---|---|
| Rigid schema | Flexible schema |
| Table | Collection |
| row | Document |
| column | Field |
| Multiple joins are required to get the complete detail of a single person, student, or customer, which causes data accessibility slow. | Data accessibility is so fast as all data is stored in one single document |
| Vertical scalability | Horizontal scalability |
| Complex Data Sharding | Data Sharding become simples |
| Considering data storage efficiency | Considering speed, performance, developer time |
| Performed well in aggregation | Performed better in a retrieval |
| Not performed Read/write very fast in Big data analytics | Performed Read/write the very fast cause of memory mapping function in Big data analytics. |
| Support relational algebra [60] | Support relational algebra [60] |

Table 5 (Appendix A) describes the comparisons of NoSQL Databases (MongoDB, Neo4j) with the Relational Databases (Oracle, MySQL, and SQL-Server), while Table 6 (Appendix A) describes the main strength of both Oracle relational Database and MongoDB NoSQL Database.

References [96],[97] claimed that the amount of geospatial and geolocated data is increased significantly. In various domains (emergency management, archaeology, IoT, smart cities) also, many applications have been designed to process and exploit geospatial data [98]–[103]. Strictly speaking, efficient DBMS is required to process the extensive geospatial data potentially well [22].

NoSQL Database is the appropriate choice for web applications[57], [104]–[107] for large datasets instead of using SQL Databases. NoSQL Databases also handled and managed geolocation and spatial data [108][109]. Many studies and researchers [110]–[112][113] claimed that large and unstructured data are being processed by the NoSQL Database potentially well. According to the study [114], spatial data was noticed in the literature at the beginning of a Geographic Information System. The studies [115] and [116] investigated two issues regarding efficient geospatial query data processing. The first one is that the geospatial queries require a very different optimization strategy from traditional optimization. Later considering the big analytics, all the tested approach and techniques of geospatial queries have been various data sizes significantly not large enough. Geospatial data has various characteristics, unlike alphanumeric data. Many research studies [117]–[124][125][113][126] claimed extensive geospatial data processing requires established approaches to process the massive amount of geospatial data potentially well. The article [127] has shown that famous RDBMS systems face many issues and problems while processing geospatial data. The authors of [108] did a performance comparison analysis of the NoSQL MongoDB document Database and PostGIS RDBMS. Their experiments have shown that MongoDB outperformed PostGIS in processing geospatial data.

The oracle spatial storage data model [128] comprises two main components: location and shape. The SDO_GEOMETRY data type is used in the mentioned storage model: index Engine and Geometry Engine used by the query analysis. Geocoder is used to convert an address into SDO_GEOMETRY data to enable location. Oracle Maps and Map Viewer are used for visualization.

Besides SQL Oracle spatial Database, SQL Server 2016[1] or Azure SQL Database[2] and its cloud version have many geo-functions for geospatial data analytics. While NoSQL Databases, e.g., Azure DocumentDB[3] and MongoDB [4], support geospatial features. Database-as-a-Service (DBaaS) model used by the Azure SQL Database also has Microsoft SQL Server's same functionalities and provides cloud services. PostgreSQL is the open-source relational Database[5].

While PostGIS is a spatial Database[6] to exploit geospatial data and is the extension of PostgreSQL. The NoSQL Database Azure DocumentDB is designed by Microsoft and has the same geospatial operations and functions as MongoDB. Standard data types GeoJSON is supported by MongoDB. The authors of [129] did a performance analysis of geospatial data. According to their results, the Azure DocumentDB is faster than the Azure SQL Database but less scalable than the Azure SQL Database. Table 7 describes the main geospatial characteristics of well-known SQL and NoSQL Databases.

[1] "Microsoft SQL Server 2016." [Online]. Available: https://www.microsoft.com/en-us/cloud-platform/sql-server
[2] Microsoft Azure SQL Database." [Online]. Available: https://azure.microsoft.com/en-us/services/sql-Database
[3] Microsoft Azure DocumentDB." [Online]. Available: https://azure.microsoft.com/en-us/services/documentdb
[4] MongoDB." [Online]. Available: https://www.mongodb.com
[5] PostgreSQL." [Online]. Available: https://www.postgresql.org
[6] "PostGIS." [Online]. Available: http://postgis.net

| **Database** | **Oracle** | **PostGIS** | **Azure SQL** | **MongoDB** | **DocumentDB** |
|---|---|---|---|---|---|
| **Geometry Objects Supported** | The point, LineString, Polygon, MultiPoint, MultiLinePoint, MultiPolygon, GeometryCollection | The point, LineString, Polygon, MultiPoint, MultiLinePoint, MultiPolygon, GeometryCollection | The point, LineString, Polygon, MultiPoint, MultiLinePoint, MultiPolygon, GeometryCollection | The point, LineString, Polygon, MultiPoint, MultiLinePoint, MultiPolygon, GeometryCollection | The point, LineString, Polygon, MultiPoint, MultiLinePoint, MultiPolygon, GeometryCollection |
| **Geometry Functionalities Supported** | For geometry instances, Oracle has the support of Open Geospatial Consortium (OGC) | For geometry instances, PostGIS has the support of Open Geospatial Consortium (OGC) | For geometry instances, Azure SQL has the support of Open Geospatial Consortium (OGC) | Inclusion, Intersection, Distance / Proximity | Inclusion, Distance / Proximity |
| **Spatial Indexes Support** | B-Trees, Parallel index builds for spatial R-tree indexes | GiST index, R-Tree index, B-Tree index | B-Trees, 2d plane index | 2d index, 2dsphere index | 2d plane index, quadtree |
| **GeoServer Compatibility** | Yes | Yes | Yes | Yes | Yes |
| **DaaS** | Yes | No | Yes (Cloud Computing Platform) | Yes | Yes |
| **Horizontal Scalability** | No | No | No | Yes | Yes |

Table 8: SQL and NoSQL Databases geospatial characteristics.

### 4.1.1. NoSQL MongoDB Data Modeling

According to the study of [42][130][131] the MongoDB, architecture consists of the shard nodes, configuration servers, and routing servers or mongos components described by figure 12.

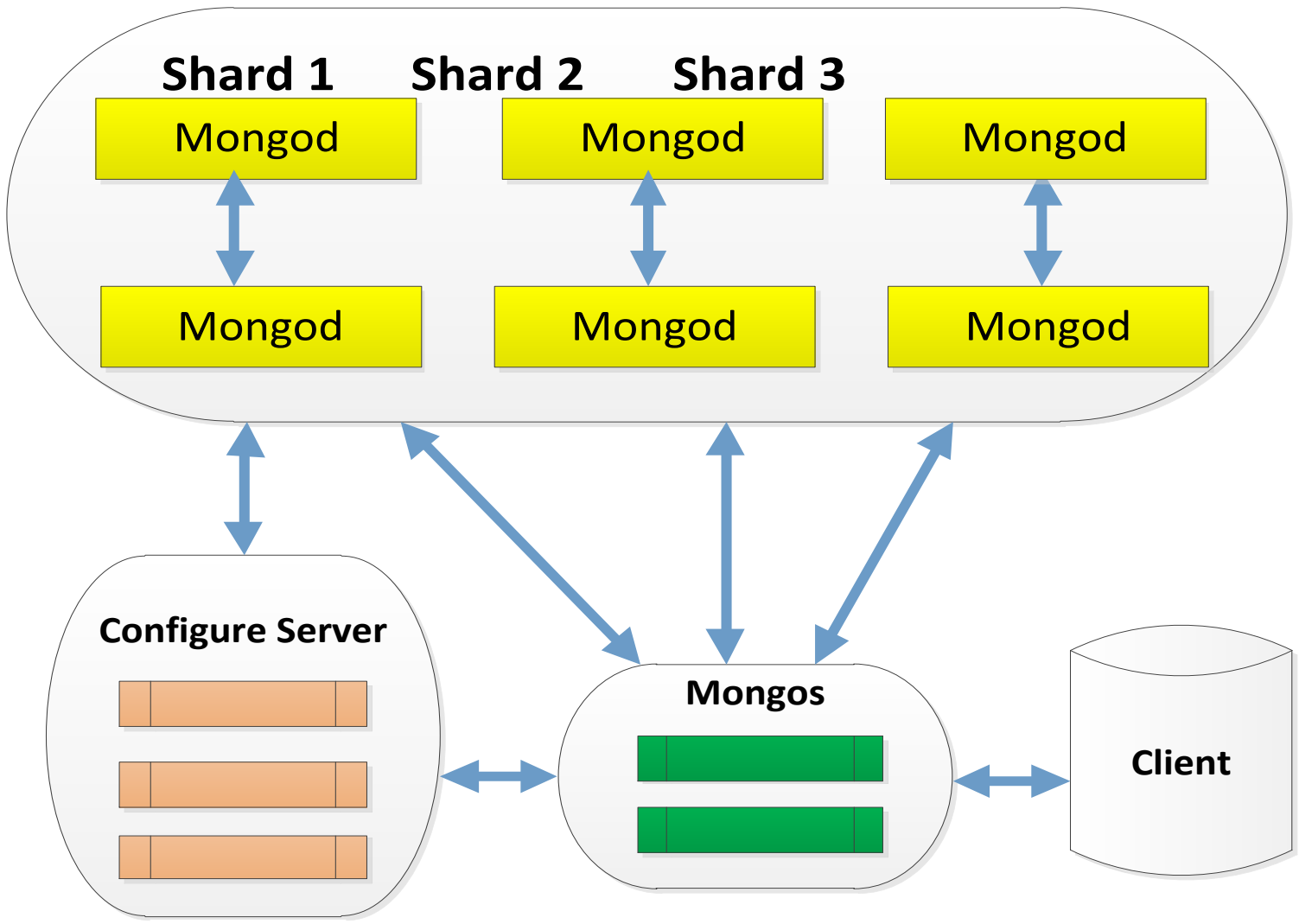

Figure 12: MongoDB Architecture

A shard is used to store the actual data, and one or more shards are required to build the underline MongoDB cluster. In failure, replicated nodes hold data for that particular shard. The appropriate shard is used according to the data transaction (read/write). The replicated node uses one or more servers. The replicated node follows the structure of the primary-secondary server. One of the secondary servers becomes the primary server in the case of primary server failure. All the transactions (read/write) are processed via the central server. All the distributed read transactions are eventually consistent in the cluster. Metadata is stored by the group of servers called configuration servers in the cluster. These servers indicate which data portion is associated with which shard. The client requested that the routing servers or MongoDb perform the task. Before the client acknowledges, the MongoDb send the particular user task to the specific shard according to the task type and merges the results. Mongos[21] can be run in parallel because they are stateless by themselves.

MongoDB uses memory-mapped files to maximize the use of the memory that is available and generally improves performance. MongoDB[132] Database adopts B-trees for indexing. A particular partition collection can be owned by the user in a MongoDB cluster[133] via a user-specified shard key

***RQ3: Does DBaaS tackle data Interoperability and portability efficiently of various NoSQL Databases?***

The article investigated many papers on DBaaS architecture. Based on the analysis and extraction, the cloud DBaaS technique for relational Databases is not equally suitable for NoSQL Databases because of varied natures. The scientists and researchers keep trying to find a unified[134]–[138] DBaaS[139] solution, both SQL and NoSQL Databases, to minimise the expert's efforts and enhance security. Also, no need to re-engineer the applications across various CSPs via unified Application Program Interfaces (APIs). The main challenging tasks are data Interoperability and portability among various cloud service providers. The three models [136], IaaS, PaaS, and SaaS define Interoperability differently. The Interoperability problem is minimised via open standards [137], but it is mainly designed for the IaaS layer. Unified APIs are generally required to move data among various cloud vendors to reduce data portability [138]. Each cloud vendor follows a different data storage model. Figure 12 describes the abstract level architecture of data movement whenever the developer switches from one CSP to another.

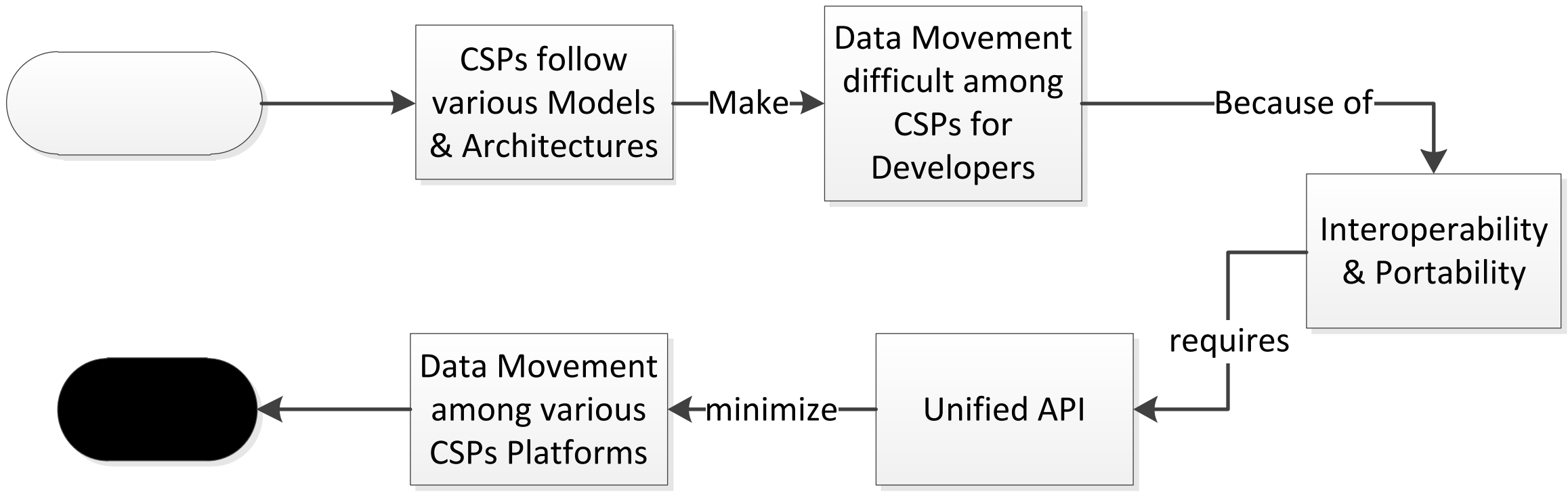

Figure 13: Data Movement inside CSPs

In the modern IT age, integrated and consistent Databases [140] and efficient management [141] are central and critical issues. The persistent data store guarantees physical data independence, and declarative query processing, Etc., are the main characteristics included by a Database system (DBS). Various methods for the different domains have been seen in the Database field. ACID, Object-Oriented model, XML, and Data warehousing are the various data management systems that follow the relational data model approach. While NoSQL Databases, mainly designed for big data processing, support the BASE [81] property.

On the one hand, cloud service providers are providing new cloud services [142] and functionalities at low cost with high efficiency to their consumers. On the other hand, various cloud service providers are offering the same functions with different models and interfaces [143], which finally leads to Interoperability [144], incompatibility, and portability issues[145]. These are the challenging problems for cloud service providers while adopting and facilitating cloud technology [146]. The three cloud service models [136], IaaS, PaaS, and SaaS Interoperability have various meanings and use for multiple purposes. Cloud consumers should be switched from one CSP to another for the following reasons [137]: downtime or failure higher rate, contract termination, business strategy changes, better alternatives with low cost, and legal issues. Cloud customers cannot move their data quickly among various cloud services. The cloud service models try to control the customer's ability because their core architectures lack Interoperability. Vendor lock-in is used for this problem and is a high-security risk for cloud paradigms [147]. The open standard should be used among various CSPs to address Interoperability issues, and ultimately data movement will become easy. Therefore, it is not easy to maintain the same data and application [148] on various cloud platforms for a developer.

Many open standards[1] and projects[149]–[151] have been investigated to handle the Interoperability among the various CSPs such as Open Virtualization Format (OVF)[2], Cloud

Cloud Infrastructure Management Interface (CIMI)[1], Open Cloud Computing Interface (OCCI)[2], and Cloud Data Management Interface (CDMI)[3]. These standards only supported the IaaS layer. Similarly, several projects such as mOSAIC[4][149], MODACLOUDS[5][150], and the Clous4SOA[151] have been developed to address the semantic Interoperability problem at the PaaS layer. The mentioned projects are also insufficient to address the Interoperability issue while having different APIs for each PaaS.
CIMI standard worked correctly with IaaS API and minimized the Interoperability between the infrastructure CSPs and the cloud customer. The Interoperability is decreased via OCCI among various CSPs with reasonable outcomes. Due to the lack of Interoperability functionalities inside the architectures of these standards, they are not much active across different CSPs platforms. Another issue is the lack of standard interfaces and various APIs.

Other researchers [137][152][153] proposed design patterns, cloud middle infrastructures, Service Delivery Cloud Platform (SDCP), and migration tools to transfer data across multiple cloud platforms. Their proposed techniques reduce the consumers' time but do not address the portability challenge across various CSPs due to following multiple APIs. The cloud service vendors, such as Amazon Web Service (AWS), Microsoft Azure, Google App Engine (GAE), Etc., facilitate the cloud consumers to develop and deploy the applications. Also, provide the Database as a Service (DBaaS) cloud platform to encourage the Database developers. The Interoperability issue also arises during data porting in DBaaS, between SQL [154] and NoSQL [155] Databases and within each category of NoSQL Databases. Each class of NoSQL Databases follows different data models and incompatible storage structures. Unified APIs are demanded to control data portability across multiple data storage services.

DBaaS[139] is a highly scalable and available backend cloud service for software developers to host their data and applications. Therefore, DBaaS is currently the most attractive solution for cloud users [154]. Cloud Service Providers (CSP) offer Database as a Service (DBaaS) [156] cloud service to move classical Database architecture to cloud Database architecture. Software as a Service (SaaS) allows specific application processing and functionalities via the internet. Other cloud DBSs are PNUTS, HBase, SimpleDB, and Google BigTable. The DBaaS framework is managing classical Databases potentially well. Nevertheless, varied natures of models followed by the different Databases degrade DBaaS performance, such as data consistency, confidentiality, integrity, availability, and lack of security. The outsourcing paradigm DBaaS could be commercially successful if it ensured data confidentiality and security. The study [157] proposed a virtualization advisor for private table Database clouds. Many application is incorporated into the service architecture via cloud computing. Cloud computing offers higher scalability and flexibility at a low cost.

[1] http://cloud-standards.org/.
[2] https://www.dmtf.org/standards/ovf

The article [134][135] investigated unified API frameworks called CDPort and SecloudDB for SQL and NoSQL Databases to ensure their higher data Interoperability, portability, and security during data porting. Their proposed API provided the users' sensitive data safety before being handed over to the third party (CSP). Through their proposed MCTool, the request is converted to the associated model and interacts with the particular Database based on the supported models. Storing the metadata in the various clouds can enhance security, and only legitimate users and DBA can access and alter the data via metadata encryption/decryption keys. Their proposed framework has the support of encryption/decryption.

## 4. Discussion

Several intimations are noticed for NoSQL Database technologies due to the lack of standard SQL in the implementation structure. Each type of NoSQL Database owns distinct query procedures and mechanisms. Software developers are commonly responsible for precisely constructing query accomplishment instead of depending on declarative query language that compiles queries and generates optimized query execution plans. The system's users must also check and analyze the data collection results. Furthermore, the developer's responsibility is increased significantly to tackle data consistencies, data replication and availability during concurrent updates in shared and replicated Databases. Extensive data systems and NoSQL Databases have numerous consequences for software architectures.

The Brewer's CAP method [7] specified foundational standard constraints for distributed Databases. Whenever occurring of network partitions (P: In the cluster, when the information is lost randomly between the nodes), consistency (C: display the identical data to all users) must be pursued by the system in opposition to availability (A: the acknowledge must be received by every client either success or failure). A framework essentially pursues the availability (A) in opposition to consistency (C) whenever partition (P) takes place; moreover, if there is no partition, the framework essentially pursues latency (L) in opposition to consistency stated by Abadi's PARCELS.

In the current era, the varied nature of astronomical data is demanded efficient data analytics and knowledge extraction rather than only depending on the structured data. SQL Database capabilities cannot process diverse kinds of big data efficiently. Big data can be processed efficiently via the capabilities of NoSQL Databases. The skills of NoSQL Databases are providing massive storage, dynamic schema, scalability, and data availability. The NoSQL Databases follow the BASE property. NoSQL Databases are more efficient in data reading and writing than focusing on data consistencies and securities. Data consistency is handled at the application level. Hence, it is the right choice for big data. Besides, NoSQL Databases do not use constraints at the data, column, and table levels as used in structured Databases.

The study investigated around 140 papers of SQL and NoSQL Database performance evaluations regarding reliability, data availability, and efficiency by considering big data[158][88]. According to our findings and analysis, NoSQL Databases are more scalable [21][37][158][68][159]. They can process massive, varied data efficiently, while SQL Database is more efficient for transaction systems and consumes more resources on data integrity and consistency. The same is not true for NoSQL Databases[95][22] because their main focus is on data availability. Based on our evaluations, NoSQL Databases are not the alternative to relational Databases. The particular Database can be used according to the organization's needs because both Databases have their pros and cons. For example, NoSQL Databases are more suitable for parallel computing[160][124] in a cluster environment and have the support MapReduce programming module.

NoSQL Databases follow dynamic and flexible schema [22][158][68][85][62][81][161], while relational Databases are heavily dependent on the schema (Tabular). For example, if you want to store student information, i.e., StdRegNo, StdName, StdAddress. So in relational Databases, you must first create the schema with appropriate domain and integrity constraints. After that, you can store the essential student information by following the required constraint. Consider it if you want to add two new columns to the current schema. You will bring changes in the current schema and migrate data from the old schema to the new one. In the case of large data, it will be time-consuming and make the application unsuitable. The big challenge in agile development because every time you will bring changes in the schema, whenever you incorporate the new changes, the same is not true for NoSQL Databases [23]. You can do manipulation in NoSQL Databases without any pre-defined schema. Table 7 (Appendix A) describes various used Databases in each selected study.

Relational Databases[22][24][37] also have some advantages over NoSQL Databases, such as data normalization to control anomalies, relational schema (attributes), domain constraints, check constraints, unique constraints, Not NULL constraints, Etc., to ensure the data integration. Relational Databases have established security and user authentication methods. Besides, considering durability performance, the SQL Database is better than a NoSQL Database. Relational Databases use one familiar standard interface, i.e., SQL, while the same is not valid for NoSQL Databases.

Various models are being followed by SQL and NoSQL Databases[68][60][126]. In addition, each NoSQL Database[72],[76] category follows different data model. Therefore, data

porting[134][138] is challenging across multiple CSPs. Another issue is the size limitation on the cloud for enterprise Databases as they are growing in size continuously. In contrast, AWS DBaaS[1] And Azure cloud Databases offer limited storage scalability. Table 9 describes each Database product category, architecture, Database type, and in-written language.

| **Database Name** | **Database Category** | **Database Architecture** | **Database Type** | **Witten In** |
|---|---|---|---|---|
| MongoDB | NoSQL-Document Store | Distributed Multi-Model | Open Source | C++, Go, JavaScript, Python |
| MySQL | SQL | | Open Source | C, C++ |
| Oracle | SQL | | Not | Assembly language, C, C++ |
| SQL Server | SQL | | Not | C, C++ |
| Neo4j | NoSQL-Graph Family | | Open Source | Java |
| Couchbase | NoSQL-Document Store | Distributed Multi-Model | Open Source | C++, Erlang, C, Go |
| CouchDB | NoSQL-Document Store | Distributed Multi-Model | Open Source | Erlang, JavaScript, C, C++ |
| Cassandra | NoSQL-Column Based | Distributed Multi-Model | Open Source | Java |
| Rethink | NoSQL-Document Store | Distributed Multi-Model | Open Source | C++, Python, Java, JavaScri pt, Bash |
| MariaDB | SQL | | Open Source | C, C++, Perl, Bash |
| RavenDB | NoSQL-Document Store | | Open Source | C# |
| BigTable | NoSQL-Column Based | | Not | C++ (core), Java, Python, Go , Ruby |
| DynamoDB | NoSQL-Key/Value Store | | Not | Java |
| SimpleDB | NoSQL-Key/Value Store | Distributed Database | Not | Erlang |
| PostgreSQL | SQL | | Open Source | C |
| PostGIS | SQL | | Open Source | C |
| Hbase | NoSQL-Column Based | Distributed Multi-Model | Open Source | Java |
| GraphDB | NoSQL-Graph Family | Distributed Database | | |
| Sybase | SQL | | Not | SQL |
| Redis | NoSQL-Key/Value Store | | Open Source | ANSI C |
| HyperTable | NoSQL-Column Based | | Open Source | C++ |
| JenusGraph | NoSQL-Graph Family | | Open Source | Java |
| VoltDB | in-memory DBMS | | Open Source | Java, C++ |
| VoldeMort | NoSQL-Key/Value Store | Distributed datastore | Open Source | Java |
| Infogrid | NoSQL-Graph Family | | Open Source | Java |
| H2 | SQL | | Open Source | Java |
| ArangoDB | NoSQLDB | | Open Source | C++, JavaScript |
| OrientDB | NoSQLDB | | Open Source | Java |
| Azure SQL | SQL | | Open Source | C, C++ |
| Azure DocumentDB | NoSQL-Document Store | | Open Source | SQL ( Core) API |

Table 9: Database product category, architecture & in-written languages.

## 5.1. Research Gap

Complex distribution systems are required to gain a significant level of availability and scalability. Besides, sharding and partitioning occur at the underline software architecture layer like the application tier, cached tier and back-end storage tier. However, achieving high scalability is challenging through the atomic abstraction image of a traditional framework using standard SQL non-procedural language. Moreover, the software must think intelligently to tackle data replicas and inconsistencies issues due to concurrent replicas update collisions. The tradeoffs inside the quality set of attributes are existed within each category of NoSQL Databases, particularly considering scalability, consistency, performance, and durability features. Furthermore, to satisfy the vendor requirements to choose the appropriate Database, the architect is essentially required to explore the nominee's Database characteristics.

Big data[162][88][158] term generally referred to as large sensor networks retrieved data or social networks' user-generated data. However, lack of knowledge, many of the technologies, approaches and disciplines around big data are new, so people lack the knowledge about how to work with data and accomplish a business result. While relational data stores cannot process such a large amount of data; hence new storage system capabilities are demanded to handle such a vast amount of data. NoSQL Databases are schemaless Databases and are distributed in nature. Due to the higher scalability, data availability, fault-tolerant, and efficient processing of massive unstructured data, characteristics of the NoSQL Databases are the right choice for big data. These tradeoffs between traditional and NoSQL data stores generated many research gaps.

According to the article's findings, NoSQL is not a replacement for relational Databases. Besides this, NoSQL is a suitable choice for big data of varied nature. There are many research gaps in both of them. Some vital research gaps are NoSQL Databases' design simplicity, scalability, and performance. The NoSQL Databases follow a flat-file or key/value data structure, while the relational Database follows a rigid schema data structure (tabular form). The abundant, varied nature of data can easily be scaled-up by the NoSQL Databases of its horizontal scalability nature. The same is not valid for SQL Databases because they are vertically scalable. Scalability and performance are the most critical factor of NoSQL Databases over relational Databases. Closing the functionalities [49] of SQL and NoSQL Databases is also an open research challenge. Besides, easy schema design [163] for NoSQL Databases is also the research gap.

Another research gap that has gained more attention in the last few years is the flexible data migration frameworks [65][76][54][62][50] from a relational Database into the particular NoSQL data store. Data migration from SQL to NoSQL datastore is significant. It plays a vital role because every organisation wants to analyse their acids like system resources critically, employee performance evaluations, Etc. for knowledge extraction and can make practical and correct decisions.

Furthermore, considering Cloud Platform, a distributed environment, NoSQL is the right choice for the cloud instead of relational Databases. The state-of-art demanded unified API [134][138][135] frameworks to reduce the cloud consumer effort during data movement across multiple cloud platforms as well as control the data Interoperability and portability issues.

## 5.2. Prediction Analysis of a Particular DBMS

Our data set-up is based on Table 7. First, we made an (x,y) pair of data. We used a combination formula to make a pair of (x,y) for a line with more than two Database names. In this way, we made 301 pairs. We used a label encoder to encode x and y to numbers from string for pre-processing. We used Gaussian Naïve Bayes to train on our encoded data. Once it has been prepared, we got a unique Database name from x. We encoded those unique names and then tested them on individual encoded data. Then we got the probability of all classes corresponding to individual encoded data. We represented the final result like a unique x encoded corresponding to the rest of all different Database probability. So table size is n by n, where n is a unique Database.

*"Naive Bayes[3] is a simple technique for constructing classifiers: models that assign class labels to problem instances, represented as vectors of feature values, where the class labels are drawn from some finite set."*

$$P(x_i|y) = \frac{1}{\sqrt{2\pi\sigma_y^2}} exp\left(-\frac{(x_i-\mu_y)^2}{2\sigma_y^2}\right)$$

**Analysis of collected data:** On the collected data, we train the Gaussian naive Bayes model. Nevertheless, the main issue is that it predicted the MongoDB for all Database names except itself. MongoDB in label set (y) are many as compared to others. The pie chart and table 11 of Appendix

[3] https://en.wikipedia.org/wiki/Naive_Bayes_classifier

A show that the MongoDB percentage is more than 22% of all data. So, the model is data biased. That is what data imbalance issue. To cater to this issue, we generated the data.

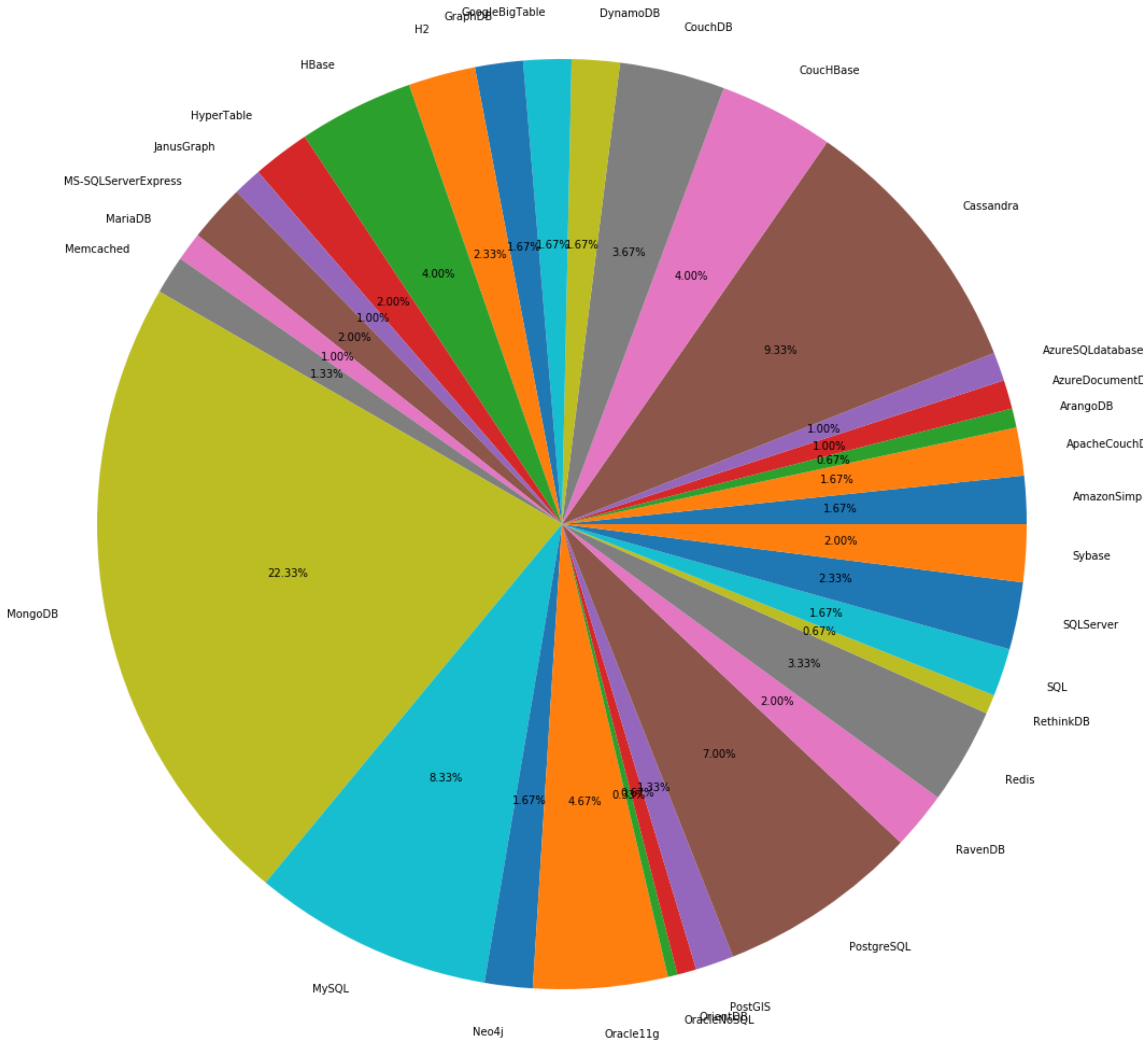

Pie Chart 1: Collected Data

**Analysis of collected plus generated data:** Analysis of the above pie chart, we concluded that these Databases “MongoDB, Cassandra, MySQL, HBase” significantly contribute to a collected dataset. So what we made the strategy, we ignore these Database names when creating a combination of other Database names. So that the percentage of these Database names can be decreased, another Database name can be increased to make the dataset balanced. As we can see from the below pie chart two and table 12 of Appendix A, it seems balanced, and the results differ from what we got on a collected dataset.

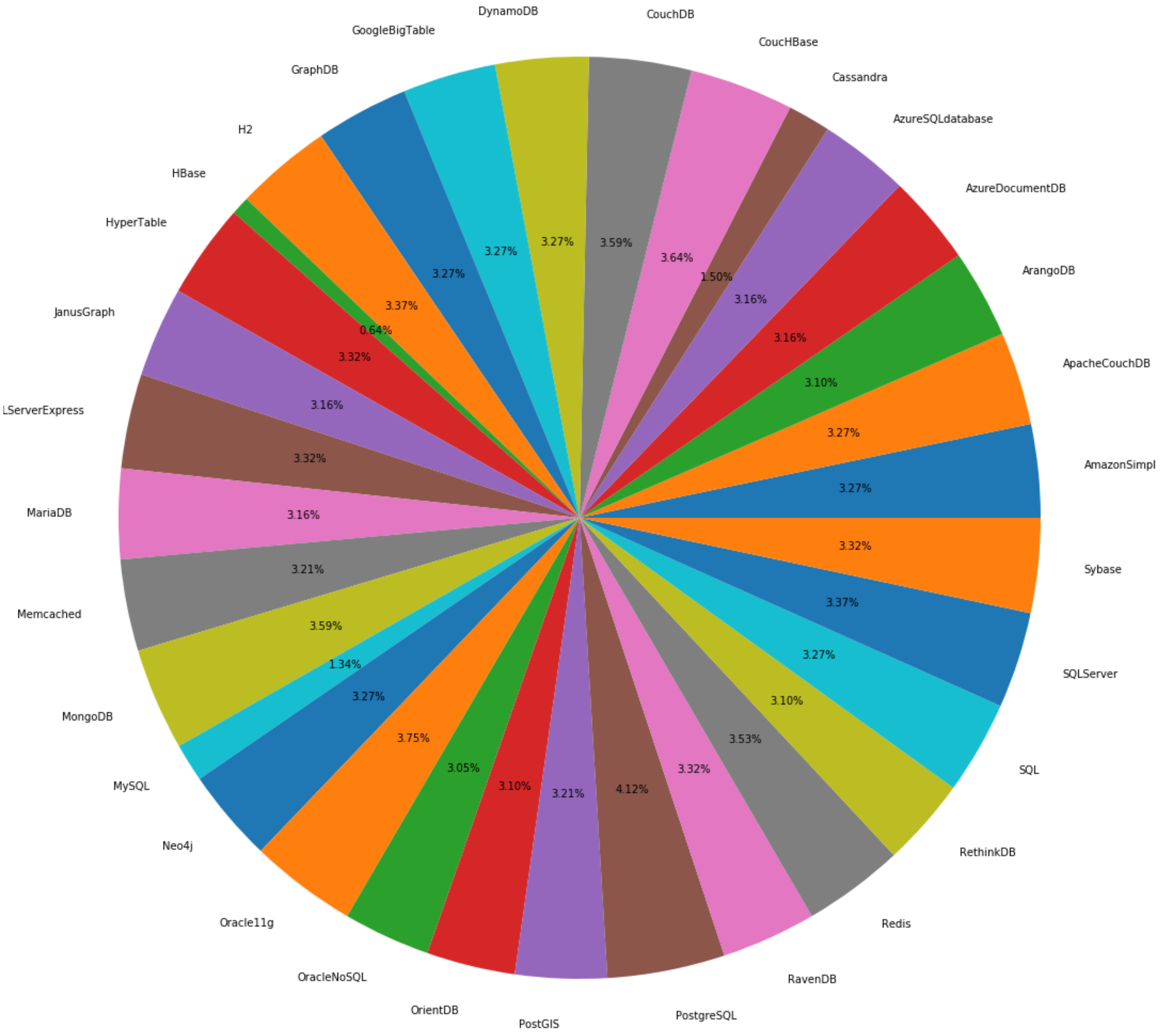

Pie Chart 2: Collected plus generated Data.

## 5.3. Prediction analysis via Frequent Pattern of a Particular DBMS

We used the Apriori Algorithm to find frequent patterns in our dataset of Database names. We used different matric evaluations to see how they affect rule generation. We use algorithms directly on the dataset, in which first X as rows and Y as columns, mapped from original data; if there is no association, we simple placed 0 and otherwise 1. Keep our dataset in mind; we set a support threshold (shown below) of 0.25 for all experiments since the dataset is too low. After frequent items finding, we have done the following operations using different evaluation. We did a total 9 experiments to see the effect of parameters.

1. **Using confidence with the minimu** $Support = \frac{frq(X,Y)}{N}$ **and 0.7,** we got the following results. The confidence formula is s

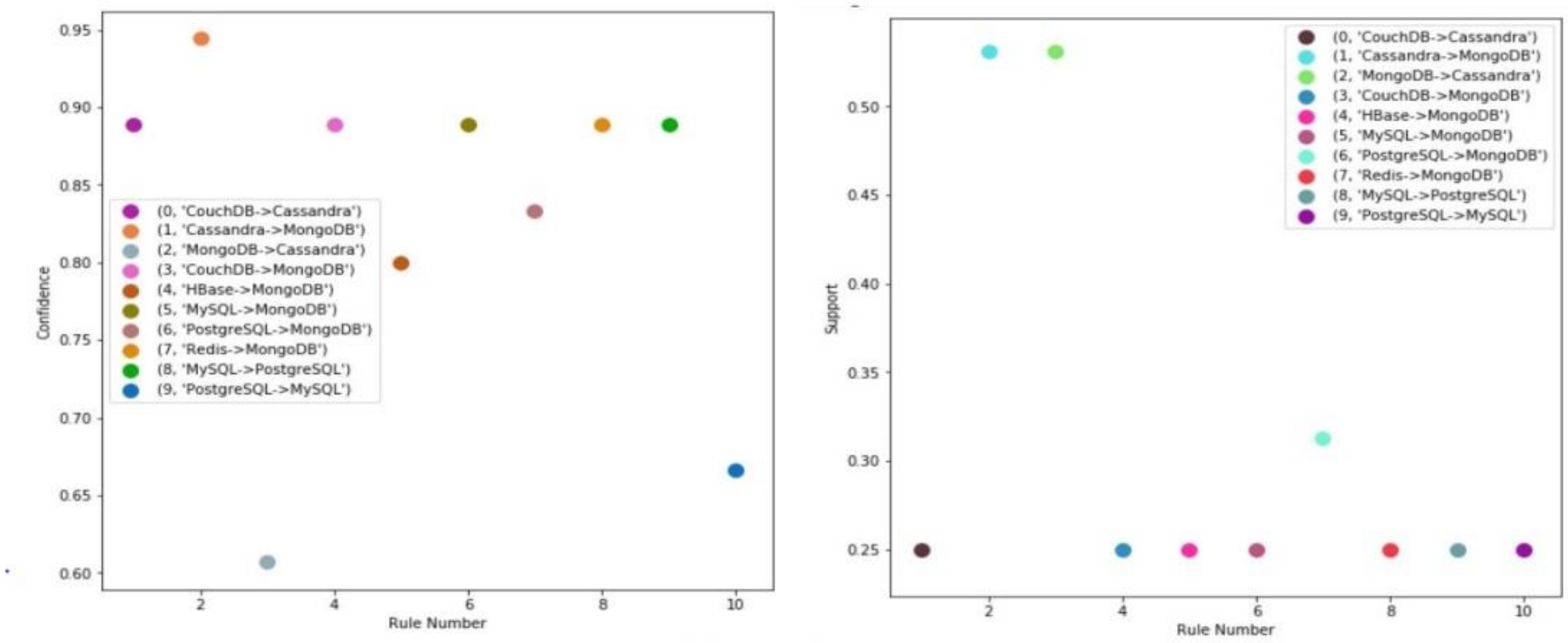

**Figure 1**

$$Confidence = \frac{frq(X,Y)}{frq(X)}$$

**Table 1 : Confidence=0.5**

| Rule | antecedents | consequents | antecedent support | consequent support | support | confidence |
|---|---|---|---|---|---|---|
| 0 | frozenset({'CouchDB'}) | frozenset({'Cassandra'}) | 0.28125 | 0.5625 | 0.25 | 0.888888889 |
| 1 | frozenset({'Cassandra'}) | frozenset({'MongoDB'}) | 0.5625 | 0.875 | 0.53125 | 0.944444444 |
| 2 | frozenset({'MongoDB'}) | frozenset({'Cassandra'}) | 0.875 | 0.5625 | 0.53125 | 0.607142857 |
| 3 | frozenset({'CouchDB'}) | frozenset({'MongoDB'}) | 0.28125 | 0.875 | 0.25 | 0.888888889 |
| 4 | frozenset({'HBase'}) | frozenset({'MongoDB'}) | 0.3125 | 0.875 | 0.25 | 0.8 |
| 5 | frozenset({'MySQL'}) | frozenset({'MongoDB'}) | 0.28125 | 0.875 | 0.25 | 0.888888889 |
| 6 | frozenset({'PostgreSQL'}) | frozenset({'MongoDB'}) | 0.375 | 0.875 | 0.3125 | 0.833333333 |
| 7 | frozenset({'Redis'}) | frozenset({'MongoDB'}) | 0.28125 | 0.875 | 0.25 | 0.888888889 |
| 8 | frozenset({'MySQL'}) | frozenset({'PostgreSQL'}) | 0.28125 | 0.375 | 0.25 | 0.888888889 |
| 9 | frozenset({'PostgreSQL'}) | frozenset({'MySQL'}) | 0.375 | 0.28125 | 0.25 | 0.666666667 |

**Table 2: Confidence=0.6**

| Rule | antecedents | consequents | antecedent support | consequent support | support | confidence |
|---|---|---|---|---|---|---|
| 0 | frozenset({'CouchDB'}) | frozenset({'Cassandra'}) | 0.28125 | 0.5625 | 0.25 | 0.888888889 |
| 1 | frozenset({'Cassandra'}) | frozenset({'MongoDB'}) | 0.5625 | 0.875 | 0.53125 | 0.944444444 |
| 2 | frozenset({'MongoDB'}) | frozenset({'Cassandra'}) | 0.875 | 0.5625 | 0.53125 | 0.607142857 |
| 3 | frozenset({'CouchDB'}) | frozenset({'MongoDB'}) | 0.28125 | 0.875 | 0.25 | 0.888888889 |
| 4 | frozenset({'HBase'}) | frozenset({'MongoDB'}) | 0.3125 | 0.875 | 0.25 | 0.8 |
| 5 | frozenset({'MySQL'}) | frozenset({'MongoDB'}) | 0.28125 | 0.875 | 0.25 | 0.888888889 |
| 6 | frozenset({'PostgreSQL'}) | frozenset({'MongoDB'}) | 0.375 | 0.875 | 0.3125 | 0.833333333 |
| 7 | frozenset({'Redis'}) | frozenset({'MongoDB'}) | 0.28125 | 0.875 | 0.25 | 0.888888889 |
| 8 | frozenset({'MySQL'}) | frozenset({'PostgreSQL'}) | 0.28125 | 0.375 | 0.25 | 0.888888889 |
| 9 | frozenset({'PostgreSQL'}) | frozenset({'MySQL'}) | 0.375 | 0.28125 | 0.25 | 0.666666667 |

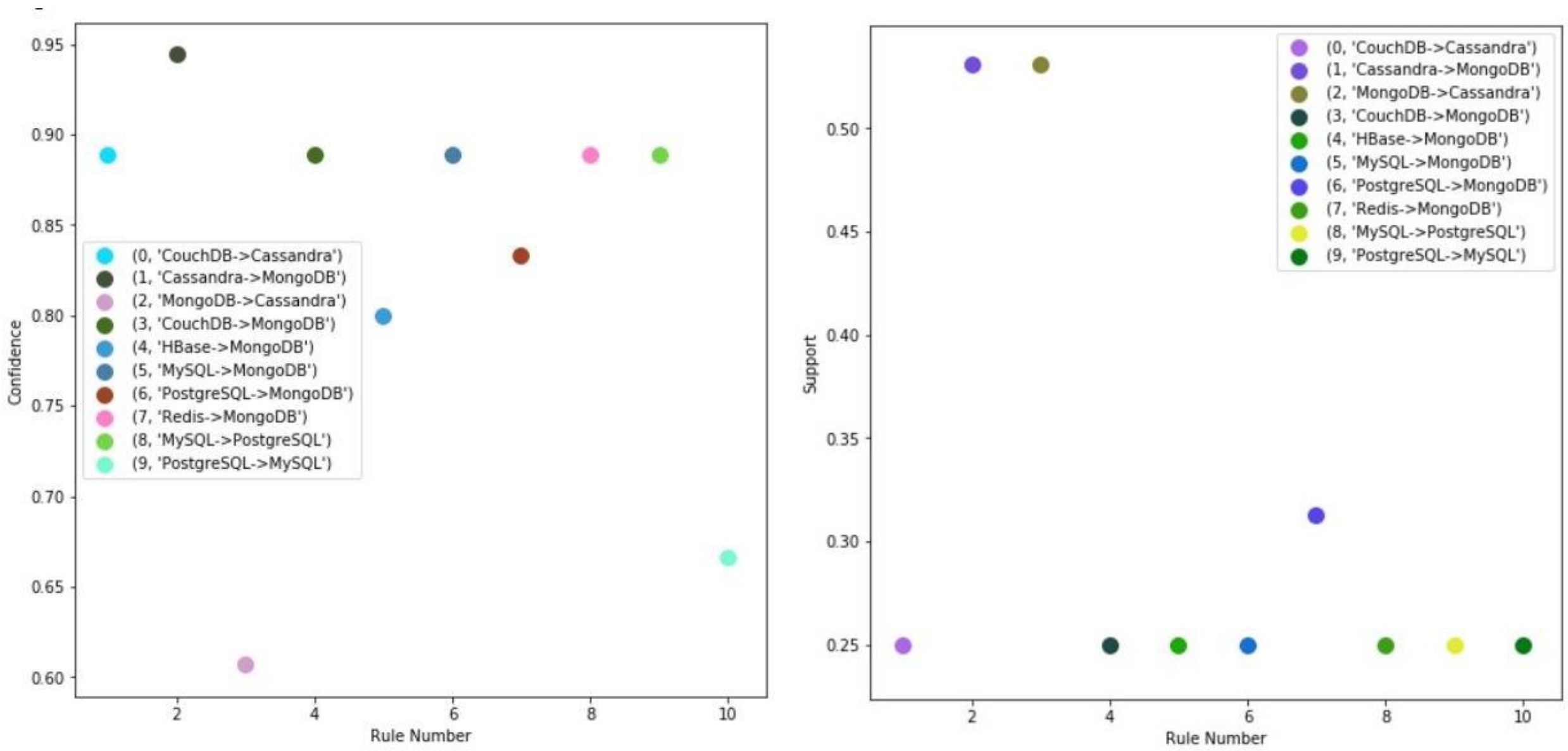

**Figure. 2**

**Table 3: Confidence=0.7**

| Rule | antecedents | consequents | antecedent support | consequent support | support | confidence |
|---|---|---|---|---|---|---|
| 0 | frozenset({'CouchDB'}) | frozenset({'Cassandra'}) | 0.28125 | 0.5625 | 0.25 | 0.88888889 |
| 1 | frozenset({'Cassandra'}) | frozenset({'MongoDB'}) | 0.5625 | 0.875 | 0.53125 | 0.94444444 |
| 2 | frozenset({'CouchDB'}) | frozenset({'MongoDB'}) | 0.28125 | 0.875 | 0.25 | 0.88888889 |
| 3 | frozenset({'HBase'}) | frozenset({'MongoDB'}) | 0.3125 | 0.875 | 0.25 | 0.8 |
| 4 | frozenset({'MySQL'}) | frozenset({'MongoDB'}) | 0.28125 | 0.875 | 0.25 | 0.88888889 |
| 5 | frozenset({'PostgreSQL'}) | frozenset({'MongoDB'}) | 0.375 | 0.875 | 0.3125 | 0.83333333 |
| 6 | frozenset({'Redis'}) | frozenset({'MongoDB'}) | 0.28125 | 0.875 | 0.25 | 0.88888889 |
| 7 | frozenset({'MySQL'}) | frozenset({'PostgreSQL'}) | 0.28125 | 0.375 | 0.25 | 0.88888889 |

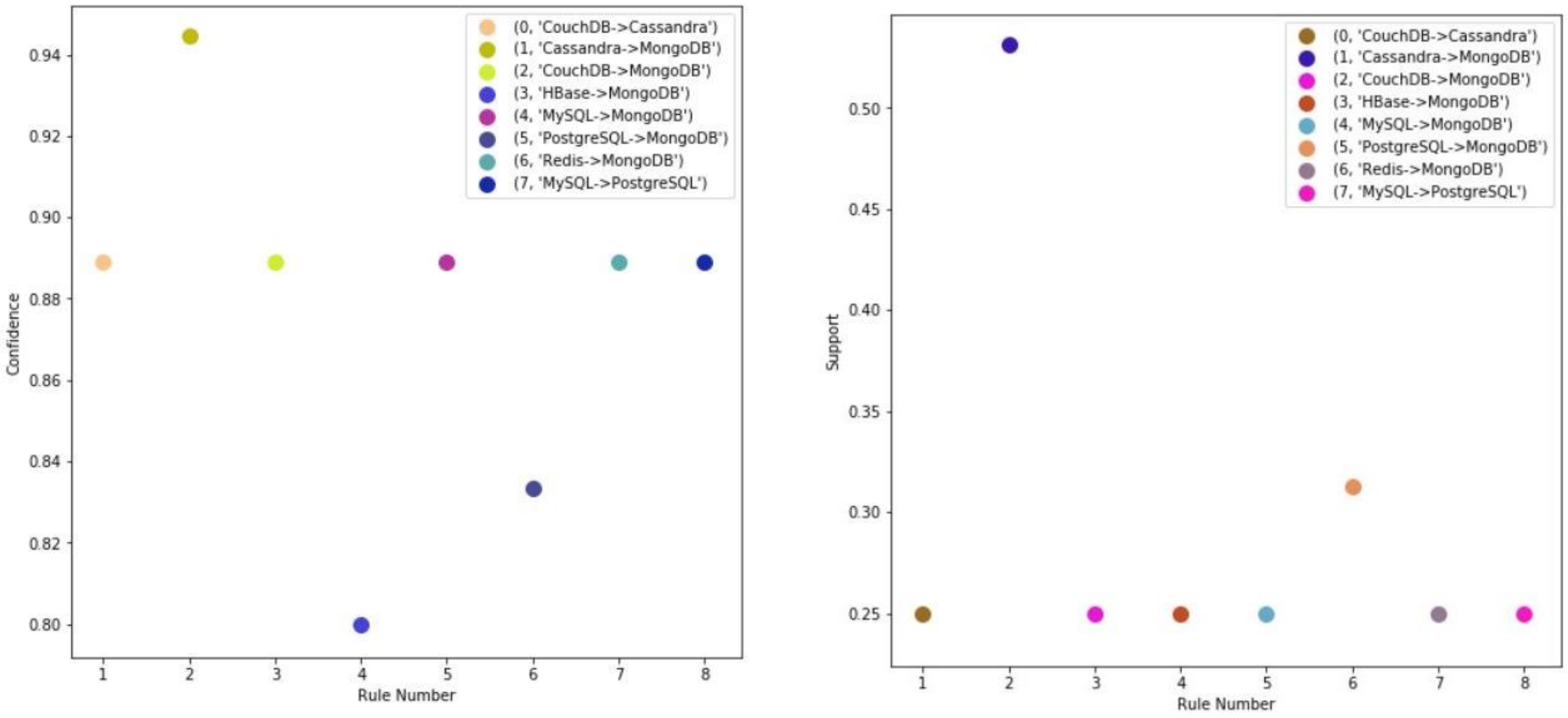

**Figure 3**

**Confidence results analysis:** We have used different thresholds for confidence. In each figure, we set the x-axis as the rule number, the y-axis as confidence on the left side, and the y-axis as support on the right side. From (tables 1, 2, and figures no. 1,2), the 1 and 2 rules have great support, and from table 3 (figure 3), rule 1 has the highest support, which means they are the most frequent in a dataset. In contrast, rule 1 has the highest confidence; it has high predictive power (meaning that whenever the antecedent occurs, the consequent tends to occur. These results look similar because the dataset is too small. For an algorithm, it is easy to find a pattern; even if we change confidence, almost the same pairs are selected under changing confidence value[4].

2. **Using Lift evaluation with a minimum threshold of 0.8, 1, and 1.1**, we got the following results. The lift formula is shown below.

$$Lift = \frac{Support}{Supp(X) \times Supp(Y)}$$

**Results:**

Table 4: Lift=0.8

| Rule | antecedents | consequents | antecedent support | consequent support | support | lift |
|---|---|---|---|---|---|---|

[4] https://support.bigml.com/hc/en-us/articles/207310215-How-do-I-know-which-rules-are-the-most-relevant-ones-Are-there-any-metrics-to-summarize-all-results-

| | | | | | | |
|---|---|---|---|---|---|---|
| 0 | frozenset({'Cassandra'}) | frozenset({'CouchDB'}) | 0.5625 | 0.28125 | 0.25 | 1.58024691 |
| 1 | frozenset({'CouchDB'}) | frozenset({'Cassandra'}) | 0.28125 | 0.5625 | 0.25 | 1.58024691 |
| 2 | frozenset({'Cassandra'}) | frozenset({'MongoDB'}) | 0.5625 | 0.875 | 0.53125 | 1.07936508 |
| 3 | frozenset({'MongoDB'}) | frozenset({'Cassandra'}) | 0.875 | 0.5625 | 0.53125 | 1.07936508 |
| 4 | frozenset({'CouchDB'}) | frozenset({'MongoDB'}) | 0.28125 | 0.875 | 0.25 | 1.01587302 |
| 5 | frozenset({'MongoDB'}) | frozenset({'CouchDB'}) | 0.875 | 0.28125 | 0.25 | 1.01587302 |
| 6 | frozenset({'MongoDB'}) | frozenset({'HBase'}) | 0.875 | 0.3125 | 0.25 | 0.91428571 |
| 7 | frozenset({'HBase'}) | frozenset({'MongoDB'}) | 0.3125 | 0.875 | 0.25 | 0.91428571 |
| 8 | frozenset({'MySQL'}) | frozenset({'MongoDB'}) | 0.28125 | 0.875 | 0.25 | 1.01587302 |
| 9 | frozenset({'MongoDB'}) | frozenset({'MySQL'}) | 0.875 | 0.28125 | 0.25 | 1.01587302 |
| 10 | frozenset({'PostgreSQL'}) | frozenset({'MongoDB'}) | 0.375 | 0.875 | 0.3125 | 0.95238095 |
| 11 | frozenset({'MongoDB'}) | frozenset({'PostgreSQL'}) | 0.875 | 0.375 | 0.3125 | 0.95238095 |
| 12 | frozenset({'Redis'}) | frozenset({'MongoDB'}) | 0.28125 | 0.875 | 0.25 | 1.01587302 |
| 13 | frozenset({'MongoDB'}) | frozenset({'Redis'}) | 0.875 | 0.28125 | 0.25 | 1.01587302 |
| 14 | frozenset({'MySQL'}) | frozenset({'PostgreSQL'}) | 0.28125 | 0.375 | 0.25 | 2.37037037 |
| 15 | frozenset({'PostgreSQL'}) | frozenset({'MySQL'}) | 0.375 | 0.28125 | 0.25 | 2.37037037 |

Table 5: Lift=1.0

| **Rule** | **antecedents** | **consequents** | **antecedent support** | **consequent support** | **support** | **lift** |
|---|---|---|---|---|---|---|
| 0 | frozenset({'Cassandra'}) | frozenset({'CouchDB'}) | 0.5625 | 0.28125 | 0.25 | 1.580247 |
| 1 | frozenset({'CouchDB'}) | frozenset({'Cassandra'}) | 0.28125 | 0.5625 | 0.25 | 1.580247 |
| 2 | frozenset({'Cassandra'}) | frozenset({'MongoDB'}) | 0.5625 | 0.875 | 0.53125 | 1.079365 |
| 3 | frozenset({'MongoDB'}) | frozenset({'Cassandra'}) | 0.875 | 0.5625 | 0.53125 | 1.079365 |
| 4 | frozenset({'CouchDB'}) | frozenset({'MongoDB'}) | 0.28125 | 0.875 | 0.25 | 1.015873 |
| 5 | frozenset({'MongoDB'}) | frozenset({'CouchDB'}) | 0.875 | 0.28125 | 0.25 | 1.015873 |
| 6 | frozenset({'MySQL'}) | frozenset({'MongoDB'}) | 0.28125 | 0.875 | 0.25 | 1.015873 |
| 7 | frozenset({'MongoDB'}) | frozenset({'MySQL'}) | 0.875 | 0.28125 | 0.25 | 1.015873 |
| 8 | frozenset({'Redis'}) | frozenset({'MongoDB'}) | 0.28125 | 0.875 | 0.25 | 1.015873 |
| 9 | frozenset({'MongoDB'}) | frozenset({'Redis'}) | 0.875 | 0.28125 | 0.25 | 1.015873 |
| 10 | frozenset({'MySQL'}) | frozenset({'PostgreSQL'}) | 0.28125 | 0.375 | 0.25 | 2.37037 |
| 11 | frozenset({'PostgreSQL'}) | frozenset({'MySQL'}) | 0.375 | 0.28125 | 0.25 | 2.37037 |

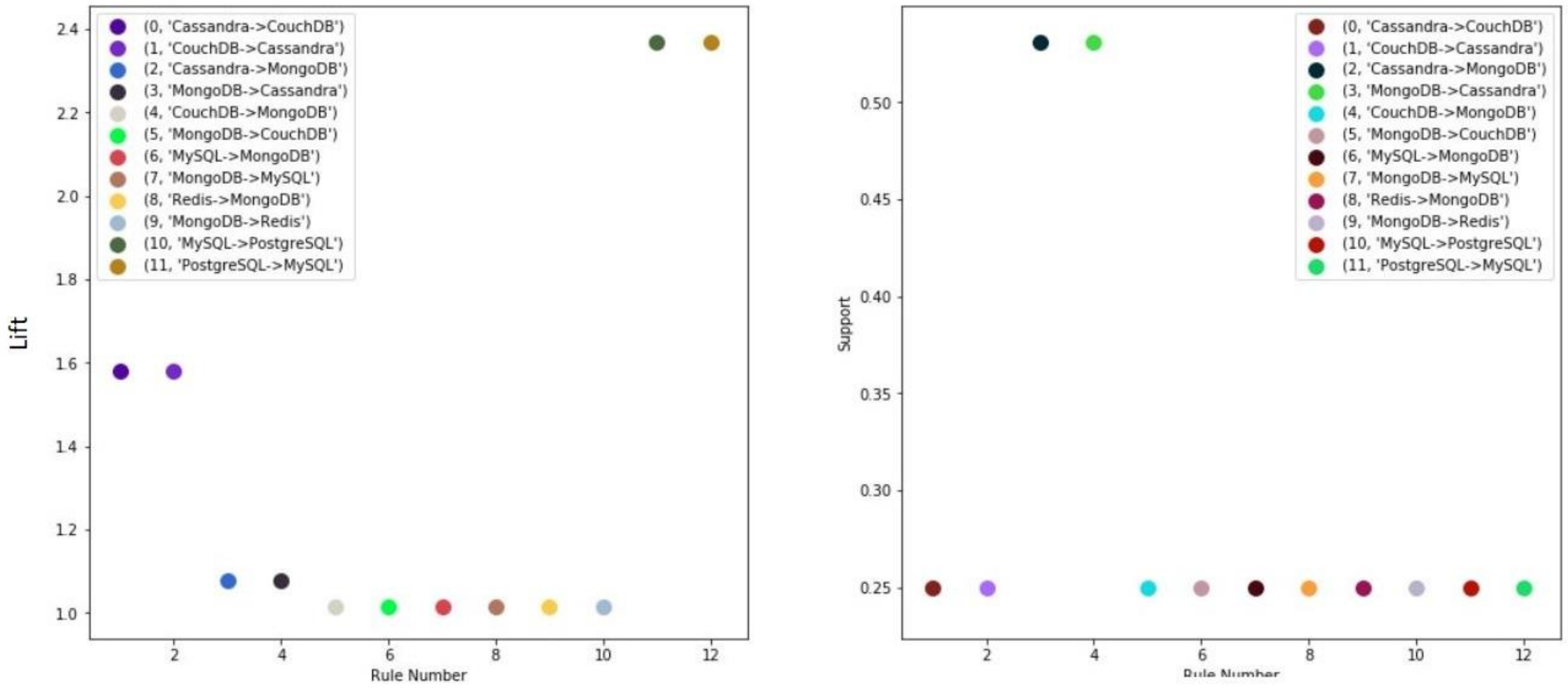

Figure 5

Table 6: Lift=1.1

| Rule | antecedents | consequents | antecedent support | consequent support | support | lift |
|---|---|---|---|---|---|---|
| 0 | frozenset({'Cassandra'}) | frozenset({'CouchDB'}) | 0.5625 | 0.28125 | 0.25 | 1.5802469 |
| 1 | frozenset({'CouchDB'}) | frozenset({'Cassandra'}) | 0.28125 | 0.5625 | 0.25 | 1.5802469 |
| 2 | frozenset({'MySQL'}) | frozenset({'PostgreSQL'}) | 0.28125 | 0.375 | 0.25 | 2.3703704 |
| 3 | frozenset({'PostgreSQL'}) | frozenset({'MySQL'}) | 0.375 | 0.28125 | 0.25 | 2.3703704 |

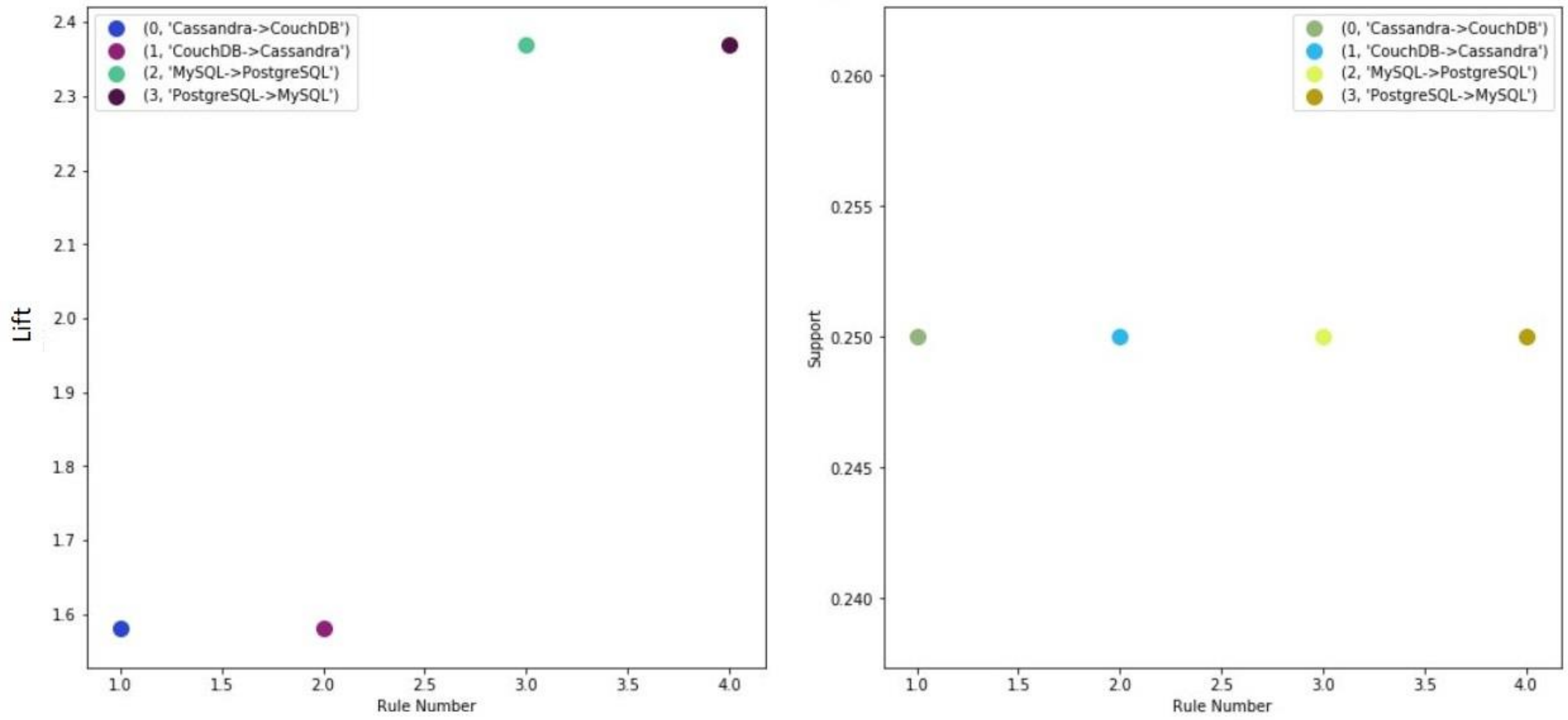

Figure 6

**Result Analysis:** We have used different values (0.8,1,1.1) for lift. In each figure, We set the x-axis as the rule number, the y-axis as the Lift on the left side, and the y-axis as support on the right side. From (table no. 4,5 and figure no. 4,5), 2 and 3 rules are the most frequent in the data. The Lift illustrates how much more accurate a rule is at foretelling the outcome than simply assuming it in the first place. Stronger associations are indicated by higher lift values[1]. According to Lift measurement, from table 6 (figure 6.) Rule 2 and 3 are the strong predictor due to their highest lift values, while tables 5, 10 and 11 are the weakest in prediction due to their lower lift value. Changing the lift value threshold is much more effective in selecting the rules; we should select this value based on how much stricter the rules we want. These results do not look similar because changing the lift threshold value effectively selects the rules.

**3. Using leverage measurement with a threshold of 0.05, 0.08, and 0.1 values**, we got the following results. The leverage formula is shown below.

$$\text{levarage}(A \to C) = \text{support}(A \to C) - \text{support}(A) \times \text{support}(C),$$

Results:

Table 7: Leverage=0.05

| Rule | antecedents | consequents | antecedent support | consequent support | support | leverage |
|---|---|---|---|---|---|---|
| 0 | frozenset({'Cassandra'}) | frozenset({'CouchDB'}) | 0.5625 | 0.28125 | 0.25 | 0.091797 |
| 1 | frozenset({'CouchDB'}) | frozenset({'Cassandra'}) | 0.28125 | 0.5625 | 0.25 | 0.091797 |
| 2 | frozenset({'MySQL'}) | frozenset({'PostgreSQL'}) | 0.28125 | 0.375 | 0.25 | 0.144531 |
| 3 | frozenset({'PostgreSQL'} | frozenset({'MySQL'}) | 0.375 | 0.28125 | 0.25 | 0.144531 |

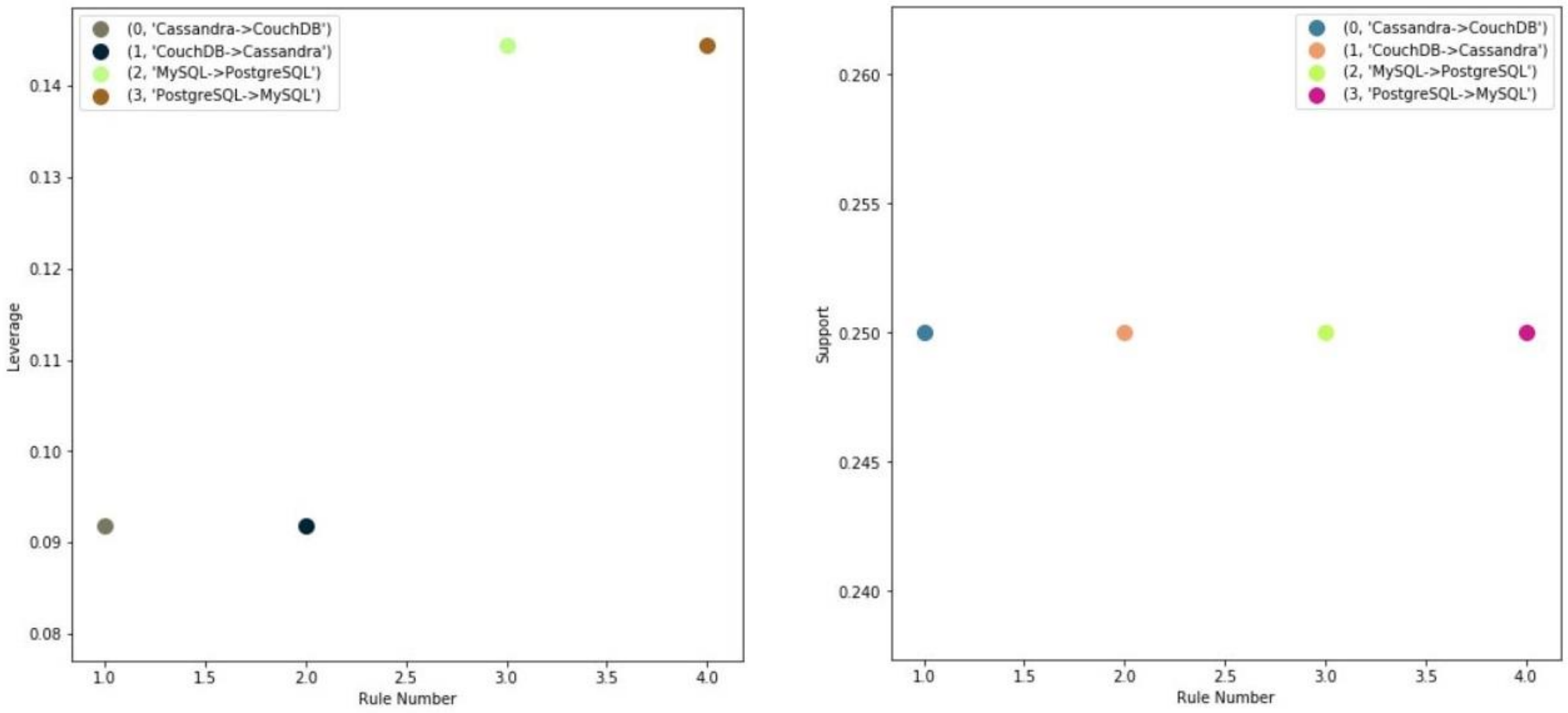

**Figure 7**

Table 8: Leverage=0.08

| Rule | antecedents | consequents | antecedent support | consequent support | support | leverage |
|---|---|---|---|---|---|---|
| 0 | frozenset({'Cassandra'}) | frozenset({'CouchDB'}) | 0.5625 | 0.28125 | 0.25 | 0.091797 |
| 1 | frozenset({'CouchDB'}) | frozenset({'Cassandra'}) | 0.28125 | 0.5625 | 0.25 | 0.091797 |
| 2 | frozenset({'MySQL'}) | frozenset({'PostgreSQL'}) | 0.28125 | 0.375 | 0.25 | 0.144531 |
| 3 | frozenset({'PostgreSQL'}) | frozenset({'MySQL'}) | 0.375 | 0.28125 | 0.25 | 0.144531 |

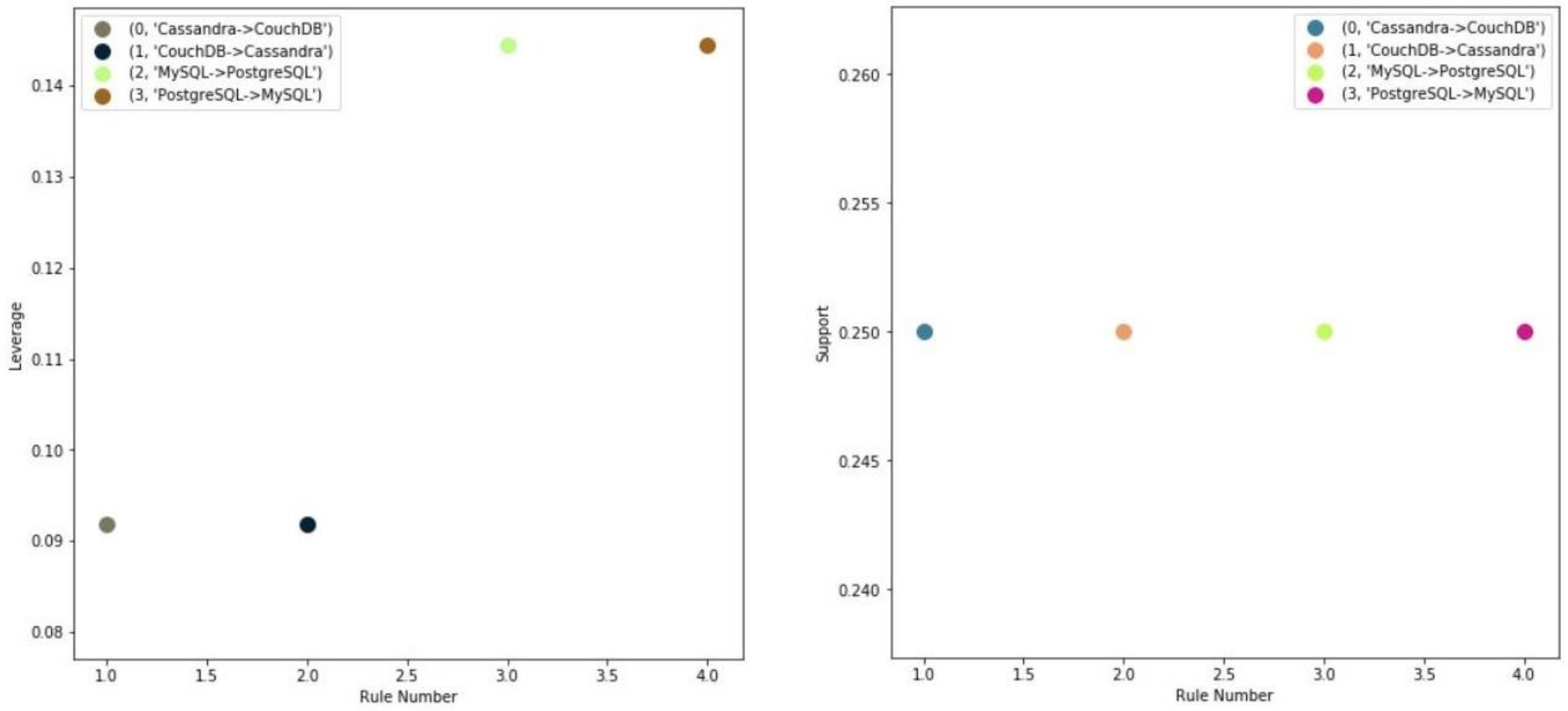

**Figure 8**

Table 9: Leverage=0.1

| Rule | antecedents | consequents | antecedent support | consequent support | support | leverage |
|---|---|---|---|---|---|---|
| 0 | frozenset({'MySQL'}) | frozenset({'PostgreSQL'}) | 0.28125 | 0.375 | 0.25 | 0.144531 |
| 1 | frozenset({'PostgreSQL'} | frozenset({'MySQL'}) | 0.375 | 0.28125 | 0.25 | 0.144531 |

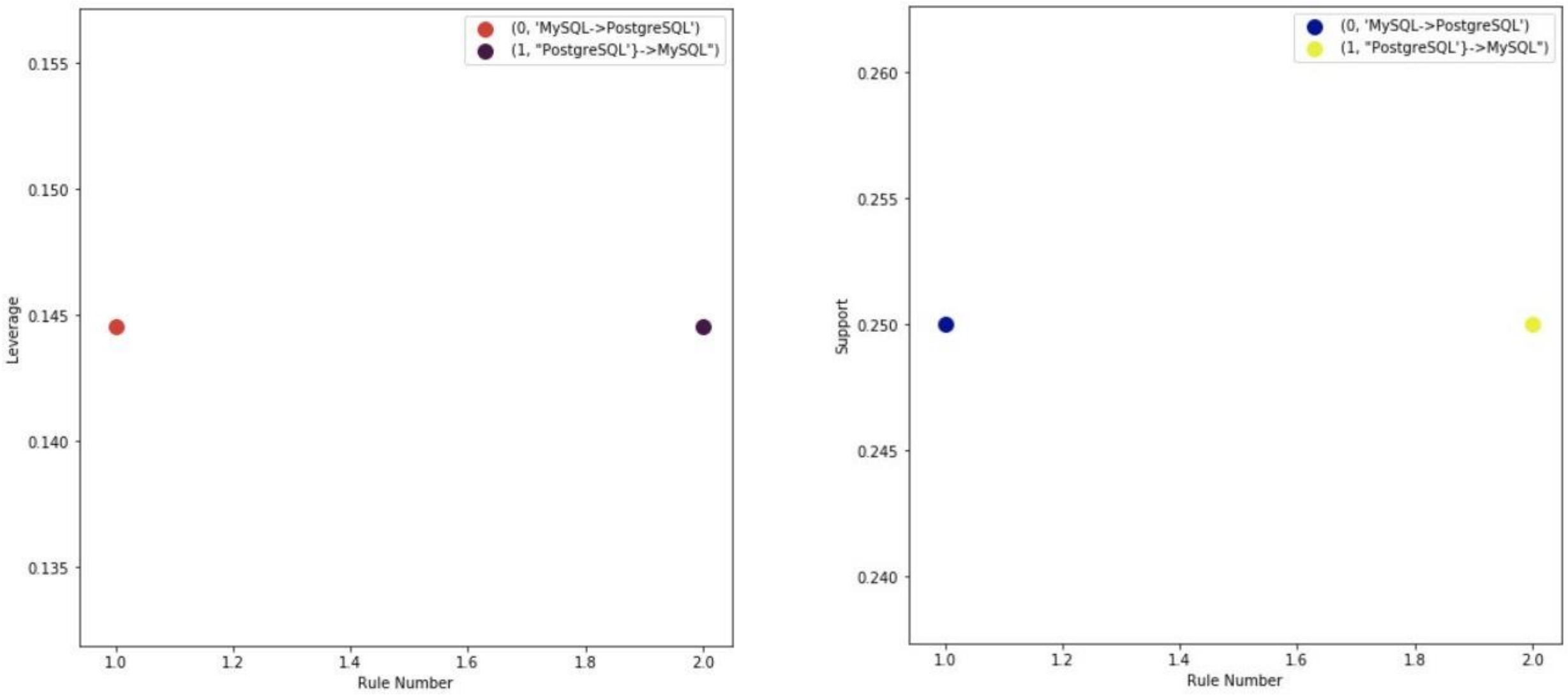

**Figure 8**

**Result analysis:** From table 7, 8, and 9 (Figures 7,8 and 9) have the same support but different leverage. In each figure, we set the x-axis as the rule number, the y-axis as leverage on the left side, and the y-axis as support on the right side. It indicates that it is not necessary to have frequent items selected on a leverage threshold value. From table 7, 8 (figure 7,8) rules 0,1. Furthermore, 2,3 are similar; in table 9 (figure 9), rules 0 and 1 are similar. All rules have equal support, while leverage computes the difference between the observed frequency of A and C appearing together and the frequency that would be expected if A and C were independent. A leverage value of 0 indicates independence[1]. So it means the higher the value, the better it is. From table 7 and figure 7 (rule 2 and 3), table 8 and figure 8 (rule 2 and 3) and table 9 and figure 9 (rule 0 and 1), rules have a strong prediction.

## 6. Appendix A

Table 4: Selected Primary studies: https://github.com/kmr2017/Database_SLR_selected_papers

Table 5: SQL and NoSQL Databases comparisons studies

| **Performance Comparisons** | **NoSQL Databases** |
|---|---|
| **Relational Databases** | [3], [5], [6]–[10], [12]–[14], [16]–[18],[47],[56],[24], [26], [28], [30], [37], [38], [108], |

Table 6: main strength of both Oracle relational Database and MongoDB NoSQL Database

| **Properties** | **Oracle RDBMS** | **MongoDB** |
|---|---|---|
| ACID | X | |
| BASE | | X |
| Large Data Scalability | | X |
| Data Shading | X | X |
| Partitioning | X | X |
| replication | X | X |
| Distributed | X | X |
| Vertical/Horizontal | Vertical | Horizontal |
| Schema | Rigid Schema | Schemaless / dynamic schema |
| Full SQL | X | |
| Indexing | X | X |
| Uni-Code characters | X | X |
| Built-in MapReduce | | X |
| Maximum Value Size | 4KB | 16MB |
| Sharing Support | | X |
| Open Source / Licensed | Licensed | Open Source |

Table 7: used Databases in each selected studies:
https://github.com/kmr2017/Database_SLR_selected_papers

Table 10: Each DBMS with its DBMSID

| DBMSID | DBMS-Name |
|---|---|
| 0 | AmazonSimpleDB |
| 1 | ApacheCouchDB |
| 2 | ArangoDB |
| 3 | AzureDocumentDB |
| 4 | AzureSQLdatabase |
| 5 | Cassandra |
| 6 | CoucHBase |
| 7 | CouchDB |
| 8 | DynamoDB |
| 9 | GoogleBigTable |
| 10 | GraphDB |
| 11 | H2 |
| 12 | HBase |
| 13 | HyperTable |
| 14 | JanusGraph |
| 15 | MS-SQLServerExpress |
| 16 | MariaDB |
| 17 | Memcached |
| 18 | MongoDB |
| 19 | MySQL |
| 20 | Neo4j |
| 21 | Oracle11g |
| 22 | OracleNoSQL |
| 23 | OrientDB |
| 24 | PostGIS |
| 25 | PostgreSQL |
| 26 | RavenDB |
| 27 | Redis |
| 28 | RethinkDB |
| 29 | SQL |
| 30 | SQLServer |
| 31 | Sybase |

Table 11: Occurrences of a particular DBMS against other DBMSs based on table 7 (Unbalanced data).

| DBMSID | 0 | 1 | 2 | 3 | 4 | 5 | 6 | 7 | 8 | 9 | 10 | 11 | 12 | 13 | 14 | 15 | 16 | 17 | 18 | 19 | 20 | 21 | 22 | 23 | 24 | 25 | 26 | 27 | 28 | 29 | 30 | 31 | Predicted Result |
|---|---|---|---|---|---|---|---|---|---|---|---|---|---|---|---|---|---|---|---|---|---|---|---|---|---|---|---|---|---|---|---|---|---|
| 0 | 0.05 | 0.05 | 0.01 | 0.01 | 0.02 | 0.10 | 0.05 | 0.05 | 0.03 | 0.08 | 0.03 | 0.01 | 0.07 | 0.04 | 0.00 | 0.04 | 0.00 | 0.02 | 0.21 | 0.01 | 0.00 | 0.00 | 0.00 | 0.03 | 0.03 | 0.01 | 0.02 | 0.01 | 0.01 | 0.02 | 0.00 | 0.00 | MongoDB |
| 1 | 0.05 | 0.05 | 0.01 | 0.01 | 0.01 | 0.10 | 0.05 | 0.05 | 0.03 | 0.07 | 0.03 | 0.01 | 0.07 | 0.03 | 0.00 | 0.04 | 0.00 | 0.02 | 0.21 | 0.01 | 0.00 | 0.00 | 0.00 | 0.05 | 0.03 | 0.01 | 0.03 | 0.01 | 0.01 | 0.03 | 0.00 | 0.00 | MongoDB |
| 2 | 0.05 | 0.05 | 0.01 | 0.01 | 0.01 | 0.10 | 0.05 | 0.05 | 0.03 | 0.06 | 0.03 | 0.01 | 0.06 | 0.03 | 0.01 | 0.03 | 0.00 | 0.02 | 0.20 | 0.01 | 0.00 | 0.00 | 0.00 | 0.07 | 0.03 | 0.01 | 0.03 | 0.01 | 0.01 | 0.03 | 0.00 | 0.00 | MongoDB |
| 3 | 0.05 | 0.05 | 0.01 | 0.01 | 0.01 | 0.10 | 0.05 | 0.05 | 0.02 | 0.06 | 0.03 | 0.01 | 0.06 | 0.03 | 0.01 | 0.03 | 0.00 | 0.02 | 0.20 | 0.01 | 0.00 | 0.00 | 0.00 | 0.08 | 0.03 | 0.01 | 0.03 | 0.01 | 0.01 | 0.04 | 0.00 | 0.00 | MongoDB |
| 4 | 0.05 | 0.05 | 0.01 | 0.01 | 0.01 | 0.10 | 0.05 | 0.05 | 0.02 | 0.05 | 0.02 | 0.01 | 0.06 | 0.03 | 0.01 | 0.03 | 0.00 | 0.01 | 0.20 | 0.01 | 0.00 | 0.00 | 0.00 | 0.08 | 0.03 | 0.01 | 0.04 | 0.01 | 0.01 | 0.04 | 0.00 | 0.00 | MongoDB |
| 5 | 0.04 | 0.04 | 0.01 | 0.01 | 0.01 | 0.11 | 0.05 | 0.05 | 0.02 | 0.05 | 0.02 | 0.02 | 0.06 | 0.03 | 0.01 | 0.03 | 0.00 | 0.02 | 0.20 | 0.02 | 0.00 | 0.00 | 0.00 | 0.06 | 0.03 | 0.02 | 0.04 | 0.01 | 0.01 | 0.04 | 0.00 | 0.00 | MongoDB |
| 6 | 0.04 | 0.04 | 0.01 | 0.01 | 0.01 | 0.11 | 0.05 | 0.05 | 0.02 | 0.04 | 0.02 | 0.02 | 0.06 | 0.03 | 0.01 | 0.03 | 0.00 | 0.02 | 0.21 | 0.02 | 0.00 | 0.00 | 0.00 | 0.04 | 0.03 | 0.02 | 0.05 | 0.02 | 0.01 | 0.05 | 0.00 | 0.00 | MongoDB |
| 7 | 0.04 | 0.04 | 0.01 | 0.01 | 0.01 | 0.11 | 0.05 | 0.05 | 0.02 | 0.04 | 0.02 | 0.02 | 0.06 | 0.03 | 0.01 | 0.03 | 0.00 | 0.02 | 0.21 | 0.03 | 0.00 | 0.00 | 0.00 | 0.02 | 0.03 | 0.03 | 0.05 | 0.02 | 0.01 | 0.05 | 0.00 | 0.00 | MongoDB |
| 8 | 0.04 | 0.04 | 0.01 | 0.01 | 0.01 | 0.12 | 0.05 | 0.05 | 0.02 | 0.03 | 0.02 | 0.02 | 0.05 | 0.03 | 0.01 | 0.03 | 0.00 | 0.02 | 0.22 | 0.04 | 0.00 | 0.00 | 0.00 | 0.01 | 0.02 | 0.04 | 0.05 | 0.02 | 0.01 | 0.05 | 0.00 | 0.00 | MongoDB |
| 9 | 0.04 | 0.03 | 0.01 | 0.01 | 0.01 | 0.12 | 0.05 | 0.05 | 0.02 | 0.03 | 0.02 | 0.02 | 0.05 | 0.03 | 0.01 | 0.03 | 0.01 | 0.02 | 0.22 | 0.04 | 0.00 | 0.00 | 0.00 | 0.00 | 0.02 | 0.04 | 0.05 | 0.02 | 0.01 | 0.05 | 0.00 | 0.00 | MongoDB |
| 10 | 0.03 | 0.03 | 0.01 | 0.01 | 0.01 | 0.12 | 0.05 | 0.05 | 0.02 | 0.03 | 0.02 | 0.02 | 0.05 | 0.03 | 0.01 | 0.03 | 0.01 | 0.02 | 0.22 | 0.05 | 0.00 | 0.00 | 0.00 | 0.00 | 0.02 | 0.05 | 0.05 | 0.02 | 0.01 | 0.04 | 0.00 | 0.00 | MongoDB |
| 11 | 0.03 | 0.03 | 0.01 | 0.01 | 0.01 | 0.12 | 0.05 | 0.05 | 0.02 | 0.02 | 0.02 | 0.03 | 0.05 | 0.03 | 0.01 | 0.03 | 0.01 | 0.02 | 0.22 | 0.06 | 0.00 | 0.01 | 0.00 | 0.00 | 0.02 | 0.06 | 0.04 | 0.02 | 0.01 | 0.04 | 0.00 | 0.00 | MongoDB |
| 12 | 0.03 | 0.02 | 0.01 | 0.01 | 0.01 | 0.12 | 0.05 | 0.04 | 0.02 | 0.02 | 0.02 | 0.03 | 0.05 | 0.03 | 0.01 | 0.03 | 0.01 | 0.02 | 0.23 | 0.07 | 0.00 | 0.01 | 0.00 | 0.00 | 0.02 | 0.07 | 0.04 | 0.03 | 0.01 | 0.03 | 0.00 | 0.00 | MongoDB |
| 13 | 0.02 | 0.02 | 0.01 | 0.01 | 0.01 | 0.12 | 0.04 | 0.04 | 0.02 | 0.02 | 0.02 | 0.03 | 0.05 | 0.02 | 0.01 | 0.03 | 0.01 | 0.02 | 0.23 | 0.08 | 0.00 | 0.02 | 0.00 | 0.00 | 0.02 | 0.07 | 0.03 | 0.03 | 0.01 | 0.03 | 0.01 | 0.00 | MongoDB |
| 14 | 0.02 | 0.02 | 0.01 | 0.01 | 0.01 | 0.11 | 0.04 | 0.04 | 0.02 | 0.01 | 0.02 | 0.03 | 0.04 | 0.02 | 0.01 | 0.02 | 0.01 | 0.01 | 0.23 | 0.09 | 0.00 | 0.02 | 0.00 | 0.00 | 0.02 | 0.08 | 0.03 | 0.03 | 0.01 | 0.02 | 0.01 | 0.00 | MongoDB |
| 15 | 0.02 | 0.01 | 0.01 | 0.01 | 0.01 | 0.11 | 0.04 | 0.04 | 0.02 | 0.01 | 0.02 | 0.03 | 0.04 | 0.02 | 0.01 | 0.02 | 0.01 | 0.01 | 0.23 | 0.10 | 0.00 | 0.03 | 0.00 | 0.00 | 0.01 | 0.09 | 0.02 | 0.03 | 0.01 | 0.02 | 0.01 | 0.01 | MongoDB |
| 16 | 0.01 | 0.01 | 0.01 | 0.01 | 0.01 | 0.11 | 0.04 | 0.04 | 0.02 | 0.01 | 0.02 | 0.03 | 0.04 | 0.02 | 0.01 | 0.02 | 0.01 | 0.01 | 0.22 | 0.11 | 0.01 | 0.04 | 0.00 | 0.00 | 0.01 | 0.09 | 0.02 | 0.03 | 0.01 | 0.01 | 0.02 | 0.01 | MongoDB |
| 17 | 0.01 | 0.01 | 0.01 | 0.01 | 0.01 | 0.10 | 0.04 | 0.04 | 0.02 | 0.01 | 0.01 | 0.03 | 0.04 | 0.02 | 0.01 | 0.02 | 0.02 | 0.01 | 0.22 | 0.11 | 0.02 | 0.05 | 0.00 | 0.00 | 0.01 | 0.10 | 0.02 | 0.04 | 0.01 | 0.01 | 0.02 | 0.01 | MongoDB |
| 18 | 0.00 | 0.00 | 0.00 | 0.00 | 0.00 | 0.00 | 0.00 | 0.00 | 0.00 | 0.00 | 0.00 | 0.00 | 0.00 | 0.00 | 0.00 | 0.00 | 0.00 | 0.00 | 0.00 | 0.00 | 0.00 | 0.00 | 0.99 | 0.00 | 0.00 | 0.00 | 0.00 | 0.00 | 0.00 | 0.00 | 0.00 | 0.00 | OracleNoSQL |
| 19 | 0.01 | 0.01 | 0.01 | 0.01 | 0.01 | 0.09 | 0.04 | 0.03 | 0.01 | 0.01 | 0.01 | 0.03 | 0.03 | 0.02 | 0.01 | 0.02 | 0.02 | 0.01 | 0.21 | 0.12 | 0.04 | 0.07 | 0.00 | 0.00 | 0.01 | 0.10 | 0.01 | 0.04 | 0.01 | 0.00 | 0.03 | 0.02 | MongoDB |
| 20 | 0.00 | 0.00 | 0.01 | 0.01 | 0.01 | 0.08 | 0.03 | 0.03 | 0.01 | 0.00 | 0.01 | 0.03 | 0.03 | 0.01 | 0.01 | 0.01 | 0.02 | 0.01 | 0.21 | 0.12 | 0.06 | 0.08 | 0.00 | 0.00 | 0.01 | 0.10 | 0.01 | 0.04 | 0.00 | 0.00 | 0.04 | 0.03 | MongoDB |
| 21 | 0.00 | 0.00 | 0.01 | 0.01 | 0.01 | 0.08 | 0.03 | 0.03 | 0.01 | 0.00 | 0.01 | 0.03 | 0.03 | 0.01 | 0.01 | 0.01 | 0.02 | 0.01 | 0.20 | 0.12 | 0.06 | 0.09 | 0.00 | 0.00 | 0.01 | 0.10 | 0.00 | 0.04 | 0.00 | 0.00 | 0.04 | 0.04 | MongoDB |
| 22 | 0.00 | 0.00 | 0.01 | 0.01 | 0.01 | 0.07 | 0.03 | 0.03 | 0.01 | 0.00 | 0.01 | 0.02 | 0.03 | 0.01 | 0.01 | 0.01 | 0.01 | 0.01 | 0.21 | 0.12 | 0.06 | 0.10 | 0.00 | 0.00 | 0.01 | 0.09 | 0.00 | 0.04 | 0.00 | 0.00 | 0.05 | 0.04 | MongoDB |
| 23 | 0.00 | 0.00 | 0.01 | 0.01 | 0.01 | 0.07 | 0.03 | 0.03 | 0.01 | 0.00 | 0.01 | 0.02 | 0.03 | 0.01 | 0.01 | 0.01 | 0.01 | 0.01 | 0.21 | 0.12 | 0.05 | 0.10 | 0.00 | 0.00 | 0.01 | 0.09 | 0.00 | 0.04 | 0.00 | 0.00 | 0.05 | 0.05 | MongoDB |
| 24 | 0.00 | 0.00 | 0.01 | 0.01 | 0.01 | 0.07 | 0.03 | 0.03 | 0.01 | 0.00 | 0.01 | 0.02 | 0.03 | 0.01 | 0.01 | 0.01 | 0.01 | 0.01 | 0.22 | 0.12 | 0.03 | 0.11 | 0.00 | 0.00 | 0.00 | 0.09 | 0.00 | 0.04 | 0.00 | 0.00 | 0.06 | 0.06 | MongoDB |
| 25 | 0.00 | 0.00 | 0.01 | 0.01 | 0.01 | 0.07 | 0.03 | 0.02 | 0.01 | 0.00 | 0.01 | 0.02 | 0.03 | 0.01 | 0.01 | 0.01 | 0.01 | 0.01 | 0.22 | 0.11 | 0.02 | 0.11 | 0.00 | 0.00 | 0.00 | 0.09 | 0.00 | 0.05 | 0.00 | 0.00 | 0.06 | 0.06 | MongoDB |
| 26 | 0.00 | 0.00 | 0.01 | 0.01 | 0.01 | 0.06 | 0.03 | 0.02 | 0.01 | 0.00 | 0.01 | 0.02 | 0.02 | 0.01 | 0.01 | 0.01 | 0.01 | 0.01 | 0.23 | 0.11 | 0.01 | 0.11 | 0.00 | 0.00 | 0.00 | 0.08 | 0.00 | 0.05 | 0.00 | 0.00 | 0.07 | 0.06 | MongoDB |
| 27 | 0.00 | 0.00 | 0.01 | 0.01 | 0.01 | 0.06 | 0.03 | 0.02 | 0.01 | 0.00 | 0.01 | 0.02 | 0.02 | 0.01 | 0.01 | 0.01 | 0.01 | 0.01 | 0.25 | 0.11 | 0.01 | 0.11 | 0.00 | 0.00 | 0.00 | 0.08 | 0.00 | 0.05 | 0.00 | 0.00 | 0.07 | 0.07 | MongoDB |
| 28 | 0.00 | 0.00 | 0.01 | 0.01 | 0.01 | 0.06 | 0.03 | 0.02 | 0.01 | 0.00 | 0.01 | 0.02 | 0.02 | 0.01 | 0.01 | 0.01 | 0.01 | 0.01 | 0.26 | 0.11 | 0.00 | 0.10 | 0.00 | 0.00 | 0.00 | 0.08 | 0.00 | 0.06 | 0.00 | 0.00 | 0.07 | 0.06 | MongoDB |
| 29 | 0.00 | 0.00 | 0.01 | 0.01 | 0.01 | 0.06 | 0.04 | 0.02 | 0.01 | 0.00 | 0.01 | 0.02 | 0.02 | 0.01 | 0.01 | 0.01 | 0.01 | 0.01 | 0.28 | 0.10 | 0.00 | 0.10 | 0.00 | 0.00 | 0.00 | 0.07 | 0.00 | 0.06 | 0.00 | 0.00 | 0.06 | 0.06 | MongoDB |
| 30 | 0.00 | 0.00 | 0.01 | 0.01 | 0.01 | 0.06 | 0.04 | 0.02 | 0.01 | 0.00 | 0.01 | 0.02 | 0.02 | 0.01 | 0.01 | 0.01 | 0.01 | 0.01 | 0.30 | 0.10 | 0.00 | 0.09 | 0.00 | 0.00 | 0.00 | 0.07 | 0.00 | 0.07 | 0.00 | 0.00 | 0.06 | 0.06 | MongoDB |
| 31 | 0.00 | 0.00 | 0.01 | 0.01 | 0.01 | 0.06 | 0.04 | 0.02 | 0.02 | 0.00 | 0.02 | 0.02 | 0.03 | 0.01 | 0.01 | 0.01 | 0.01 | 0.01 | 0.32 | 0.09 | 0.00 | 0.08 | 0.00 | 0.00 | 0.00 | 0.06 | 0.00 | 0.07 | 0.00 | 0.00 | 0.06 | 0.05 | MongoDB |

Table 12: Occurences of a particular against other DBMSs based on table 7 data and generated data (Balanced data)

| DBMSID | 0 | 1 | 2 | 3 | 4 | 5 | 6 | 7 | 8 | 9 | 10 | 11 | 12 | 13 | 14 | 15 | 16 | 17 | 18 | 19 | 20 | 21 | 22 | 23 | 24 | 25 | 26 | 27 | 28 | 29 | 30 | 31 | Predicted Result |
|---|---|---|---|---|---|---|---|---|---|---|---|---|---|---|---|---|---|---|---|---|---|---|---|---|---|---|---|---|---|---|---|---|---|
| 0 | 0.04 | 0.04 | 0.03 | 0.03 | 0.03 | 0.01 | 0.04 | 0.04 | 0.04 | 0.04 | 0.04 | 0.03 | 0.01 | 0.04 | 0.03 | 0.04 | 0.03 | 0.03 | 0.02 | 0.00 | 0.03 | 0.03 | 0.03 | 0.04 | 0.04 | 0.03 | 0.04 | 0.03 | 0.03 | 0.04 | 0.03 | 0.03 | GoogleBigTable |
| 1 | 0.04 | 0.04 | 0.03 | 0.03 | 0.03 | 0.01 | 0.04 | 0.04 | 0.04 | 0.04 | 0.04 | 0.03 | 0.01 | 0.04 | 0.03 | 0.04 | 0.03 | 0.03 | 0.03 | 0.00 | 0.03 | 0.03 | 0.03 | 0.04 | 0.04 | 0.03 | 0.04 | 0.03 | 0.03 | 0.04 | 0.03 | 0.03 | GoogleBigTable |
| 2 | 0.04 | 0.04 | 0.03 | 0.03 | 0.03 | 0.01 | 0.04 | 0.04 | 0.04 | 0.04 | 0.04 | 0.03 | 0.01 | 0.04 | 0.03 | 0.04 | 0.03 | 0.03 | 0.03 | 0.00 | 0.03 | 0.03 | 0.03 | 0.04 | 0.04 | 0.03 | 0.04 | 0.03 | 0.03 | 0.04 | 0.03 | 0.03 | CouchDB |
| 3 | 0.04 | 0.04 | 0.03 | 0.03 | 0.03 | 0.01 | 0.04 | 0.04 | 0.03 | 0.04 | 0.04 | 0.03 | 0.01 | 0.04 | 0.03 | 0.04 | 0.03 | 0.03 | 0.03 | 0.00 | 0.03 | 0.03 | 0.03 | 0.03 | 0.04 | 0.04 | 0.04 | 0.03 | 0.03 | 0.04 | 0.03 | 0.03 | CouchDB |
| 4 | 0.04 | 0.04 | 0.03 | 0.03 | 0.03 | 0.02 | 0.04 | 0.04 | 0.03 | 0.04 | 0.03 | 0.03 | 0.01 | 0.04 | 0.03 | 0.04 | 0.03 | 0.03 | 0.03 | 0.00 | 0.03 | 0.03 | 0.03 | 0.03 | 0.03 | 0.04 | 0.04 | 0.03 | 0.03 | 0.04 | 0.03 | 0.03 | CouchDB |
| 5 | 0.04 | 0.04 | 0.03 | 0.03 | 0.03 | 0.02 | 0.04 | 0.04 | 0.03 | 0.04 | 0.03 | 0.03 | 0.01 | 0.04 | 0.03 | 0.04 | 0.03 | 0.03 | 0.03 | 0.00 | 0.03 | 0.03 | 0.03 | 0.03 | 0.03 | 0.04 | 0.04 | 0.03 | 0.03 | 0.04 | 0.03 | 0.03 | CouchDB |
| 6 | 0.04 | 0.04 | 0.03 | 0.03 | 0.03 | 0.02 | 0.04 | 0.04 | 0.03 | 0.04 | 0.03 | 0.03 | 0.01 | 0.04 | 0.03 | 0.04 | 0.03 | 0.03 | 0.03 | 0.00 | 0.03 | 0.03 | 0.03 | 0.03 | 0.03 | 0.04 | 0.04 | 0.03 | 0.03 | 0.04 | 0.03 | 0.03 | PostgreSQL |
| 7 | 0.04 | 0.04 | 0.03 | 0.03 | 0.03 | 0.02 | 0.04 | 0.04 | 0.03 | 0.04 | 0.03 | 0.03 | 0.01 | 0.03 | 0.03 | 0.04 | 0.03 | 0.03 | 0.03 | 0.00 | 0.03 | 0.03 | 0.03 | 0.03 | 0.03 | 0.04 | 0.04 | 0.03 | 0.03 | 0.04 | 0.03 | 0.03 | PostgreSQL |
| 8 | 0.03 | 0.03 | 0.03 | 0.03 | 0.03 | 0.02 | 0.04 | 0.04 | 0.03 | 0.03 | 0.03 | 0.03 | 0.01 | 0.03 | 0.03 | 0.03 | 0.03 | 0.03 | 0.03 | 0.01 | 0.03 | 0.03 | 0.03 | 0.03 | 0.03 | 0.04 | 0.04 | 0.03 | 0.03 | 0.03 | 0.03 | 0.03 | PostgreSQL |
| 9 | 0.03 | 0.03 | 0.03 | 0.03 | 0.03 | 0.02 | 0.04 | 0.04 | 0.03 | 0.03 | 0.03 | 0.03 | 0.01 | 0.03 | 0.03 | 0.03 | 0.03 | 0.03 | 0.04 | 0.01 | 0.03 | 0.03 | 0.03 | 0.03 | 0.03 | 0.04 | 0.04 | 0.03 | 0.03 | 0.03 | 0.03 | 0.03 | PostgreSQL |
| 10 | 0.03 | 0.03 | 0.03 | 0.03 | 0.03 | 0.02 | 0.04 | 0.04 | 0.03 | 0.03 | 0.03 | 0.03 | 0.01 | 0.03 | 0.03 | 0.03 | 0.03 | 0.03 | 0.04 | 0.01 | 0.03 | 0.04 | 0.03 | 0.03 | 0.03 | 0.04 | 0.03 | 0.03 | 0.03 | 0.03 | 0.03 | 0.03 | PostgreSQL |
| 11 | 0.03 | 0.03 | 0.03 | 0.03 | 0.03 | 0.02 | 0.04 | 0.04 | 0.03 | 0.03 | 0.03 | 0.03 | 0.01 | 0.03 | 0.03 | 0.03 | 0.03 | 0.03 | 0.04 | 0.01 | 0.03 | 0.04 | 0.03 | 0.03 | 0.03 | 0.04 | 0.03 | 0.03 | 0.03 | 0.03 | 0.03 | 0.03 | PostgreSQL |
| 12 | 0.03 | 0.03 | 0.03 | 0.03 | 0.03 | 0.02 | 0.04 | 0.04 | 0.03 | 0.03 | 0.03 | 0.03 | 0.01 | 0.03 | 0.03 | 0.03 | 0.03 | 0.03 | 0.04 | 0.01 | 0.03 | 0.04 | 0.03 | 0.03 | 0.03 | 0.04 | 0.03 | 0.03 | 0.03 | 0.03 | 0.03 | 0.03 | PostgreSQL |
| 13 | 0.03 | 0.03 | 0.03 | 0.03 | 0.03 | 0.02 | 0.04 | 0.04 | 0.03 | 0.03 | 0.03 | 0.03 | 0.01 | 0.03 | 0.03 | 0.03 | 0.03 | 0.03 | 0.04 | 0.01 | 0.03 | 0.04 | 0.03 | 0.03 | 0.03 | 0.04 | 0.03 | 0.03 | 0.03 | 0.03 | 0.03 | 0.03 | PostgreSQL |
| 14 | 0.03 | 0.03 | 0.03 | 0.03 | 0.03 | 0.02 | 0.04 | 0.04 | 0.03 | 0.03 | 0.03 | 0.03 | 0.01 | 0.03 | 0.03 | 0.03 | 0.03 | 0.03 | 0.04 | 0.02 | 0.03 | 0.04 | 0.03 | 0.03 | 0.03 | 0.04 | 0.03 | 0.04 | 0.03 | 0.03 | 0.03 | 0.03 | PostgreSQL |
| 15 | 0.03 | 0.03 | 0.03 | 0.03 | 0.03 | 0.02 | 0.04 | 0.04 | 0.03 | 0.03 | 0.03 | 0.03 | 0.01 | 0.03 | 0.03 | 0.03 | 0.03 | 0.03 | 0.04 | 0.02 | 0.03 | 0.04 | 0.03 | 0.03 | 0.03 | 0.04 | 0.03 | 0.04 | 0.03 | 0.03 | 0.03 | 0.03 | PostgreSQL |
| 16 | 0.03 | 0.03 | 0.03 | 0.03 | 0.03 | 0.02 | 0.04 | 0.04 | 0.03 | 0.03 | 0.03 | 0.03 | 0.01 | 0.03 | 0.03 | 0.03 | 0.03 | 0.03 | 0.04 | 0.02 | 0.03 | 0.04 | 0.03 | 0.03 | 0.03 | 0.04 | 0.03 | 0.04 | 0.03 | 0.03 | 0.03 | 0.03 | PostgreSQL |
| 17 | 0.03 | 0.03 | 0.03 | 0.03 | 0.03 | 0.02 | 0.04 | 0.04 | 0.03 | 0.03 | 0.03 | 0.03 | 0.01 | 0.03 | 0.03 | 0.03 | 0.03 | 0.03 | 0.04 | 0.02 | 0.03 | 0.04 | 0.03 | 0.03 | 0.03 | 0.04 | 0.03 | 0.04 | 0.03 | 0.03 | 0.03 | 0.03 | PostgreSQL |
| 18 | 0.03 | 0.03 | 0.03 | 0.03 | 0.03 | 0.02 | 0.04 | 0.04 | 0.03 | 0.03 | 0.03 | 0.03 | 0.01 | 0.03 | 0.03 | 0.03 | 0.03 | 0.03 | 0.04 | 0.02 | 0.03 | 0.04 | 0.03 | 0.03 | 0.03 | 0.04 | 0.03 | 0.04 | 0.03 | 0.03 | 0.03 | 0.03 | PostgreSQL |
| 19 | 0.03 | 0.03 | 0.03 | 0.03 | 0.03 | 0.02 | 0.04 | 0.04 | 0.03 | 0.03 | 0.03 | 0.03 | 0.01 | 0.03 | 0.03 | 0.03 | 0.03 | 0.03 | 0.04 | 0.02 | 0.03 | 0.04 | 0.03 | 0.03 | 0.03 | 0.04 | 0.03 | 0.04 | 0.03 | 0.03 | 0.03 | 0.03 | PostgreSQL |
| 20 | 0.03 | 0.03 | 0.03 | 0.03 | 0.03 | 0.02 | 0.04 | 0.04 | 0.03 | 0.03 | 0.03 | 0.03 | 0.01 | 0.03 | 0.03 | 0.03 | 0.03 | 0.03 | 0.04 | 0.02 | 0.03 | 0.04 | 0.03 | 0.03 | 0.03 | 0.04 | 0.03 | 0.04 | 0.03 | 0.03 | 0.03 | 0.03 | PostgreSQL |
| 21 | 0.03 | 0.03 | 0.03 | 0.03 | 0.03 | 0.02 | 0.04 | 0.04 | 0.03 | 0.03 | 0.03 | 0.03 | 0.01 | 0.03 | 0.03 | 0.03 | 0.03 | 0.03 | 0.04 | 0.02 | 0.03 | 0.04 | 0.03 | 0.03 | 0.03 | 0.04 | 0.03 | 0.04 | 0.03 | 0.03 | 0.04 | 0.03 | PostgreSQL |
| 22 | 0.03 | 0.03 | 0.03 | 0.03 | 0.03 | 0.01 | 0.04 | 0.04 | 0.03 | 0.03 | 0.03 | 0.03 | 0.01 | 0.03 | 0.03 | 0.03 | 0.03 | 0.03 | 0.04 | 0.02 | 0.03 | 0.04 | 0.03 | 0.03 | 0.03 | 0.04 | 0.03 | 0.04 | 0.03 | 0.03 | 0.04 | 0.03 | PostgreSQL |
| 23 | 0.03 | 0.03 | 0.03 | 0.03 | 0.03 | 0.01 | 0.04 | 0.04 | 0.03 | 0.03 | 0.03 | 0.03 | 0.01 | 0.03 | 0.03 | 0.03 | 0.03 | 0.03 | 0.04 | 0.02 | 0.03 | 0.04 | 0.03 | 0.03 | 0.03 | 0.04 | 0.03 | 0.04 | 0.03 | 0.03 | 0.04 | 0.04 | PostgreSQL |
| 24 | 0.03 | 0.03 | 0.03 | 0.03 | 0.03 | 0.01 | 0.04 | 0.04 | 0.03 | 0.03 | 0.03 | 0.03 | 0.01 | 0.03 | 0.03 | 0.03 | 0.03 | 0.03 | 0.04 | 0.02 | 0.03 | 0.04 | 0.03 | 0.03 | 0.03 | 0.04 | 0.03 | 0.04 | 0.03 | 0.03 | 0.04 | 0.04 | PostgreSQL |
| 25 | 0.03 | 0.03 | 0.03 | 0.03 | 0.03 | 0.01 | 0.04 | 0.04 | 0.03 | 0.03 | 0.03 | 0.03 | 0.00 | 0.03 | 0.03 | 0.03 | 0.03 | 0.03 | 0.04 | 0.02 | 0.03 | 0.04 | 0.03 | 0.03 | 0.03 | 0.04 | 0.03 | 0.04 | 0.03 | 0.03 | 0.04 | 0.04 | PostgreSQL |
| 26 | 0.03 | 0.03 | 0.03 | 0.03 | 0.03 | 0.01 | 0.04 | 0.04 | 0.03 | 0.03 | 0.03 | 0.03 | 0.00 | 0.03 | 0.03 | 0.03 | 0.03 | 0.03 | 0.04 | 0.02 | 0.03 | 0.04 | 0.03 | 0.03 | 0.03 | 0.04 | 0.03 | 0.04 | 0.03 | 0.03 | 0.04 | 0.04 | PostgreSQL |
| 27 | 0.03 | 0.03 | 0.03 | 0.03 | 0.03 | 0.01 | 0.04 | 0.03 | 0.03 | 0.03 | 0.03 | 0.03 | 0.00 | 0.03 | 0.03 | 0.03 | 0.03 | 0.03 | 0.04 | 0.02 | 0.03 | 0.04 | 0.03 | 0.03 | 0.03 | 0.04 | 0.03 | 0.04 | 0.03 | 0.03 | 0.04 | 0.04 | PostgreSQL |
| 28 | 0.03 | 0.03 | 0.03 | 0.03 | 0.03 | 0.01 | 0.04 | 0.03 | 0.03 | 0.03 | 0.03 | 0.03 | 0.00 | 0.03 | 0.03 | 0.03 | 0.03 | 0.03 | 0.04 | 0.02 | 0.04 | 0.04 | 0.03 | 0.03 | 0.03 | 0.04 | 0.03 | 0.04 | 0.03 | 0.03 | 0.04 | 0.04 | PostgreSQL |
| 29 | 0.03 | 0.03 | 0.03 | 0.03 | 0.03 | 0.01 | 0.04 | 0.03 | 0.03 | 0.03 | 0.03 | 0.03 | 0.00 | 0.03 | 0.03 | 0.03 | 0.03 | 0.03 | 0.04 | 0.01 | 0.04 | 0.04 | 0.03 | 0.03 | 0.03 | 0.04 | 0.03 | 0.04 | 0.03 | 0.03 | 0.04 | 0.04 | Oracle11g |
| 30 | 0.03 | 0.03 | 0.03 | 0.03 | 0.03 | 0.01 | 0.04 | 0.03 | 0.03 | 0.03 | 0.03 | 0.03 | 0.00 | 0.03 | 0.03 | 0.03 | 0.03 | 0.03 | 0.04 | 0.01 | 0.04 | 0.04 | 0.03 | 0.03 | 0.03 | 0.04 | 0.03 | 0.04 | 0.03 | 0.03 | 0.04 | 0.04 | Oracle11g |
| 31 | 0.03 | 0.03 | 0.03 | 0.03 | 0.03 | 0.01 | 0.04 | 0.03 | 0.03 | 0.03 | 0.03 | 0.04 | 0.00 | 0.03 | 0.03 | 0.03 | 0.03 | 0.03 | 0.03 | 0.01 | 0.04 | 0.05 | 0.03 | 0.03 | 0.03 | 0.04 | 0.03 | 0.04 | 0.03 | 0.03 | 0.04 | 0.04 | Oracle11g |

## 7. Conclusion

The study concludes that NoSQL Databases are not the replacement for relational Databases. Both Databases have their advantages and disadvantages. According to the organizational need, one can choose the appropriate DBMS. If data consistency and normalization are the high priority of a particular organization, then the SQL Database is the right choice. In the case of extensive unstructured data and data availability is the high precedence of a company, NoSQL is the appropriate selection. In the aggregation scenario, the relational Database aggregation may outperform the NoSQL Database aggregation for the small datasets and vice versa in the case of big data analytics. MapReduce is mainly designed for distributed computing in clusters. Therefore, MapReduce is slow in aggregation, but it is developed for handling big data, particularly massive unstructured data, and is more efficient in parallel computing. NoSQL Databases are the appropriate choice where applications generate massive and varied nature data because of their distributed nature and high scalability. In contrast, relational Database scalability is better than NoSQL Databases considering geospatial data. At the same time, NoSQL Databases outperformed relational Databases in data response time, particularly for extensive geospatial data.

Furthermore, due to the dynamic/flexible nature of the schema of NoSQL Databases, rapid development becomes easy. NoSQL Databases are flexible models and architectures, while relational Databases are based on a monolithic structure.

NoSQL Databases follow various models, making it difficult for a developer to move from one storage model to another. CSPs are not compatible with each other, following various standards and interfaces. Therefore, data movement becomes difficult among various CSPs because each CSP designed different APIs for their intended services. NoSQL Databases demand one common cloud solution to handle data portability and Interoperability issues efficiently.

The future work directions are while considering structured data: If we follow the Denormalized approach for SQL RDBMS and then evaluate the results of insertion, updating, and data retrieval against MongoDB in a particular scenario. Alternatively, if we monitor the normalized approach for MongoDB NoSQL Database, then assess the results of insertion, updating, and data retrieval in contradiction to SQL RDBMS in a precise situation. NoSQL Database's parallel geospatial approaches require more attention to handle many user requests efficiently.

**Acknowledgment**

This work is partially supported by the National High Technology Research and Development Program of China (863 programs) under Grant 2014AA012204, the NSFC under Grant 61671018 and the Chinese Government Scholarship (CSC) for International Scholars.

## References

[1] A. Siddiqa *et al.*, "A survey of big data management: Taxonomy and state-of-the-art," *J. Netw. Comput. Appl.*, vol. 71, pp. 151–166, 2016.

[2] X. Kong, Y. Shi, S. Yu, J. Liu, and F. Xia, "Academic social networks: Modeling, analysis, mining and applications," *J. Netw. Comput. Appl.*, vol. 132, pp. 86–103, 2019.

[3] C. Ordonez, "Optimization of linear recursive queries in SQL," *IEEE Trans. Knowl. Data Eng.*, vol. 22, no. 2, pp. 264–277, 2009.

[4] D. Obasanjo, "Building scalable Databases: Denormalization, the NoSQL movement and Digg," 2009.

[5] C. Strozzi, "NoSQL–A relational Database management system. 2007–2010," *WWW page Artic. http//www. strozzi. it/cgi-bin/CSA/tw7/I/en_US/nosql/Home% 20Page. Accessed*, vol. 10, 2013.

[6] S. George, "NoSQL–NOT ONLY SQL," *Int. J. Enterp. Comput. Bus. Syst.*, vol. 2, no. 2, 2013.

[7] E. A. Brewer, "Towards robust distributed systems," in *PODC*, 2000, vol. 7.

[8] M. Díaz, C. Martín, and B. Rubio, "State-of-the-art, challenges, and open issues in the integration of Internet of things and cloud computing," *J. Netw. Comput. Appl.*, vol. 67, pp. 99–117, 2016.

[9] B. T. Rao, "A study on data storage security issues in cloud computing," *Procedia Comput. Sci.*, vol. 92, pp. 128–135, 2016.

[10] Y. Mansouri, V. Prokhorenko, and M. A. Babar, "An automated implementation of hybrid cloud for performance evaluation of distributed Databases," *J. Netw. Comput. Appl.*, vol. 167, 2020, doi: 10.1016/j.jnca.2020.102740.

[11] K. Ravi, Y. Khandelwal, B. S. Krishna, and V. Ravi, "Analytics in/for cloud-an interdependence: A review," *J. Netw. Comput. Appl.*, vol. 102, pp. 17–37, 2018.

[12] L. Wiese, T. Waage, and M. Brenner, "CloudDBGuard: A framework for encrypted data storage in NoSQL wide column stores," *Data Knowl. Eng.*, p. 101732, 2019.

[13] M. Ribas *et al.*, "A Petri net-based decision-making framework for assessing cloud services adoption: The use of spot instances for cost reduction," *J. Netw. Comput. Appl.*, vol. 57, pp. 102–118, 2015.

[14] A. Kumari, S. Tanwar, S. Tyagi, N. Kumar, R. M. Parizi, and K.-K. R. Choo, "Fog data analytics: A taxonomy and process model," *J. Netw. Comput. Appl.*, vol. 128, pp. 90–104, 2019.

[15] B. Kitchenham and S. Charters, "Guidelines for performing systematic literature reviews in software engineering," 2007.

[16] T. Dyba, B. A. Kitchenham, and M. Jorgensen, “Evidence-based software engineering for practitioners,” *IEEE Softw.*, vol. 22, no. 1, pp. 58–65, 2005.

[17] S. Hosseinzadeh *et al.*, “Diversification and obfuscation techniques for software security: A systematic literature review,” *Inf. Softw. Technol.*, vol. 104, pp. 72–93, 2018.

[18] H. Alsolai and M. Roper, “A systematic literature review of machine learning techniques for software maintainability prediction,” *Inf. Softw. Technol.*, vol. 119, p. 106214, 2020.

[19] A. A. Imam, S. Basri, R. Ahmad, and M. T. González-Aparicio, “Literature Review on Database Design Testing Techniques,” in *Computer Science On-line Conference*, 2019, pp. 1–13.

[20] J. Han, E. Haihong, G. Le, and J. Du, “Survey on NoSQL Database,” in *2011 6th international conference on pervasive computing and applications*, 2011, pp. 363–366.

[21] R. P. Padhy, M. R. Patra, and S. C. Satapathy, “RDBMS to NoSQL: reviewing some next-generation non-relational Database’s,” *Int. J. Adv. Eng. Sci. Technol.*, vol. 11, no. 1, pp. 15–30, 2011.

[22] M. Stonebraker, “SQL Databases v. NoSQL Databases,” *Commun. ACM*, vol. 53, no. 4, pp. 10–11, 2010.

[23] C. Băzăr and C. S. Iosif, “The transition from rdbms to nosql. a comparative analysis of three popular non-relational solutions: Cassandra, mongodb and couchbase,” *Database Syst. J.*, vol. 5, no. 2, pp. 49–59, 2014.

[24] S. Mukherjee, “The battle between NoSQL Databases and RDBMS,” *Available SSRN 3393986*, 2019.

[25] R. Chopade and V. K. Pachghare, “Ten years of critical review on Database forensics research,” *Digit. Investig.*, 2019.

[26] K. Petersen, S. Vakkalanka, and L. Kuzniarz, “Guidelines for conducting systematic mapping studies in software engineering: An update,” *Inf. Softw. Technol.*, vol. 64, pp. 1–18, 2015.

[27] B. Kitchenham and P. Brereton, “A systematic review of systematic review process research in software engineering,” *Inf. Softw. Technol.*, vol. 55, no. 12, pp. 2049–2075, 2013.

[28] D. Badampudi, C. Wohlin, and K. Petersen, “Experiences from using snowballing and Database searches in systematic literature studies,” in *Proceedings of the 19th International Conference on Evaluation and Assessment in Software Engineering*, 2015, p. 17.

[29] K. Petersen and C. Gencel, “Worldviews, research methods, and their relationship to validity in empirical software engineering research,” in *2013 Joint Conference of the 23rd International Workshop on Software Measurement and the 8th International Conference on Software Process and Product Measurement*, 2013, pp. 81–89.

[30] J. Maxwell, “Understanding and validity in qualitative research,” *Harv. Educ. Rev.*, vol. 62, no. 3, pp. 279–301, 1992.

[31] M. Rodrigues, M. Y. Santos, and J. Bernardino, “Big data processing tools: an experimental performance evaluation,” *Wiley Interdiscip. Rev. Data Min. Knowl. Discov.*, vol. 9, no. 2, p. e1297, 2019.

[32] B. Hou, K. Qian, L. Li, Y. Shi, L. Tao, and J. Liu, “MongoDB NoSQL injection analysis and detection,” in *2016 IEEE 3rd International Conference on Cyber Security and Cloud Computing (CSCloud)*, 2016, pp. 75–78.

[33] C. Győrödi, R. Győrödi, G. Pecherle, and A. Olah, “A comparative study: MongoDB vs. MySQL,” in *2015 13th International Conference on Engineering of Modern Electric Systems (EMES)*, 2015, pp. 1–6.

[34] Z. Parker, S. Poe, and S. V Vrbsky, “Comparing nosql mongodb to an sql db,” in *Proceedings of the 51st ACM Southeast Conference*, 2013, p. 5.

[35] Z. Wei-Ping, L. I. Ming-Xin, and C. Huan, “Using MongoDB to implement textbook management system instead of MySQL,” in *2011 IEEE 3rd International Conference on Communication Software and Networks*, 2011, pp. 303–305.

[36] A. Boicea, F. Radulescu, and L. I. Agapin, “MongoDB vs Oracle--Database comparison,” in *2012 third international conference on emerging intelligent data and web technologies*, 2012, pp. 330–335.

[37] Y. Li and S. Manoharan, “A performance comparison of SQL and NoSQL Databases,” in *2013 IEEE Pacific Rim Conference on Communications, Computers and Signal Processing (PACRIM)*, 2013, pp. 15–19.

[38] W. Khan, W. Ahmad, B. Luo, and E. Ahmed, “SQL Database with physical Database tuning technique and NoSQL graph Database comparisons,” 2019. doi: 10.1109/ITNEC.2019.8729264.

[39] W. Khan and W. Shahzad, “Predictive Performance Comparison Analysis of Relational & NoSQL Graph Databases,” *Int. J. Adv. Comput. Sci. Appl*, vol. 8, pp. 523–530, 2017.

[40] A. Faraj, B. Rashid, and T. Shareef, “Comparative study of relational and non-relations Database performances using Oracle and MongoDB systems,” *J. Impact Factor*, vol. 5, no. 11, pp. 11–22, 2014.

[41] C. Vicknair, M. Macias, Z. Zhao, X. Nan, Y. Chen, and D. Wilkins, “A comparison of a graph Database and a relational Database: a data provenance perspective,” in *Proceedings of the 48th annual Southeast regional conference*, 2010, p. 42.

[42] S. Khan and V. Mane, “SQL support over MongoDB using metadata,” *Int. J. Sci. Res. Publ.*, vol. 3, no. 10, pp. 1–5, 2013.

[43] L. Kumar, S. Rajawat, and K. Joshi, “Comparative analysis of nosql (mongodb) with mysql Database,” *Int. J. Mod. Trends Eng. Res.*, vol. 2, no. 5, pp. 120–127, 2015.

[44] R. Aghi, S. Mehta, R. Chauhan, S. Chaudhary, and N. Bohra, "A comprehensive comparison of SQL and MongoDB Databases," *Int. J. Sci. Res. Publ.*, vol. 5, no. 2, 2015.

[45] M. B. Ayub and N. Ali, "Performance comparison of in-memory and disk-based Databases using transaction processing performance council (TPC) benchmarking," *J. Internet Inf. Syst.*, vol. 8, no. 1, pp. 1–8, 2018.

[46] R. Deari, X. Zenuni, J. Ajdari, F. Ismaili, and B. Raufi, "ANALYSIS AND COMPARISON OF DOCUMENT-BASED DATABASES WITH SQL RELATIONAL DATABASES: MONGODB VS MYSQL," 2018.

[47] M. Sharma, V. D. Sharma, and M. M. Bundele, "Performance Analysis of RDBMS and No SQL Databases: PostgreSQL, MongoDB and Neo4j," in *2018 3rd International Conference and Workshops on Recent Advances and Innovations in Engineering (ICRAIE)*, 2018, pp. 1–5.

[48] R. Čerešňák and M. Kvet, "Comparison of query performance in relational a non-relation Databases," *Transp. Res. Procedia*, vol. 40, pp. 170–177, 2019.

[49] Z. H. Liu, B. Hammerschmidt, D. McMahon, Y. Liu, and H. J. Chang, "Closing the functional and performance gap between SQL and NoSQL," in *Proceedings of the 2016 International Conference on Management of Data*, 2016, pp. 227–238.

[50] H.-J. Kim, E.-J. Ko, Y.-H. Jeon, and K.-H. Lee, "Migration from rdbms to column-oriented nosql: Lessons learned and open problems," in *Proceedings of the 7th International Conference on Emerging Databases*, 2018, pp. 25–33.

[51] R. C. McColl, D. Ediger, J. Poovey, D. Campbell, and D. A. Bader, "A performance evaluation of open source graph Databases," in *Proceedings of the first workshop on Parallel programming for analytics applications*, 2014, pp. 11–18.

[52] D. Anikin, O. Borisenko, and Y. Nedumov, "Labeled Property Graphs: SQL or NoSQL?," in *2019 Ivannikov Memorial Workshop (IVMEM)*, 2019, pp. 7–13.

[53] Z. A. El Mouden, A. Jakimi, M. Hajar, and M. Boutahar, "Graph Schema Storage in SQL Object-Relational Database and NoSQL Document-Oriented Database: A Comparative Study," in *International Conference Europe Middle East & North Africa Information Systems and Technologies to Support Learning*, 2019, pp. 176–183.

[54] V. Rathika, "Graph-Based Denormalization for Migrating Big Data from SQL Database to NoSQL Database," in *Intelligent Communication Technologies and Virtual Mobile Networks*, 2019, pp. 546–556.

[55] Y. Zhu, E. Yan, and I. Song, "The use of a graph-based system to improve bibliographic information retrieval: System design, implementation, and evaluation," *J. Assoc. Inf. Sci. Technol.*, vol. 68, no. 2, pp. 480–490, 2017.

[56] M.-G. Jung, S.-A. Youn, J. Bae, and Y.-L. Choi, "A study on data input and output performance comparison of MongoDB and PostgreSQL in the big data environment," in *2015 8th International Conference on Database Theory and Application (DTA)*, 2015, pp.

14–17.

[57] H. Fatima and K. Wasnik, “Comparison of SQL, NoSQL and NewSQL Databases for internet of things,” in *2016 IEEE Bombay Section Symposium (IBSS)*, 2016, pp. 1–6.

[58] P. P. Ray, D. Dash, and D. De, “Edge computing for Internet of Things: A survey, e-healthcare case study and future direction,” *J. Netw. Comput. Appl.*, vol. 140, pp. 1–22, 2019.

[59] A. Singh, “Data Migration from Relational Database to MongoDB,” *Glob. J. Comput. Sci. Technol.*, 2019.

[60] G. Zhao, W. Huang, S. Liang, and Y. Tang, “Modeling MongoDB with relational model,” in *2013 Fourth International Conference on Emerging Intelligent Data and Web Technologies*, 2013, pp. 115–121.

[61] L. Stanescu, M. Brezovan, and D. D. Burdescu, “Automatic mapping of MySQL Databases to NoSQL MongoDB,” in *2016 Federated Conference on Computer Science and Information Systems (FedCSIS)*, 2016, pp. 837–840.

[62] F. Yassine and M. A. Awad, “Migrating from SQL to NOSQL Database: Practices and Analysis,” in *2018 International Conference on Innovations in Information Technology (IIT)*, 2018, pp. 58–62.

[63] Y.-T. Liao *et al.*, “Data adapter for querying and transformation between SQL and NoSQL Database,” *Futur. Gener. Comput. Syst.*, vol. 65, pp. 111–121, 2016.

[64] D. Tomar, J. P. Bhati, P. Tomar, and G. Kaur, “Migration of healthcare relational Database to NoSQL cloud Database for healthcare analytics and management,” in *Healthcare Data Analytics and Management*, Elsevier, 2019, pp. 59–87.

[65] L. Rocha, F. Vale, E. Cirilo, D. Barbosa, and F. Mourão, “A framework for migrating relational datasets to NoSQL,” *Procedia Comput. Sci.*, vol. 51, pp. 2593–2602, 2015.

[66] S. Ghule and R. Vadali, “Transformation of SQL system to NoSQL system and performing data analytics using SVM,” in *2017 International Conference on Trends in Electronics and Informatics (ICEI)*, 2017, pp. 883–887.

[67] J. C. Hsu, C. H. Hsu, S. C. Chen, and Y. C. Chung, “Correlation Aware Technique for SQL to NoSQL Transformation,” in *2014 7th International Conference on Ubi-Media Computing and Workshops*, 2014, pp. 43–46.

[68] G. B. Solanke and K. Rajeswari, “SQL to NoSQL transformation system using data adapter and analytics,” in *2017 IEEE International Conference on Technological Innovations in Communication, Control and Automation (TICCA)*, 2017, pp. 59–63.

[69] R. Lawrence, “Integration and virtualization of relational SQL and NoSQL systems including MySQL and MongoDB,” in *2014 International Conference on Computational Science and Computational Intelligence*, 2014, vol. 1, pp. 285–290.

[70] D. A. Pereira, W. Ourique de Morais, and E. Pignaton de Freitas, “NoSQL real-time Database performance comparison,” *Int. J. Parallel, Emergent Distrib. Syst.*, vol. 33, no. 2, pp. 144–156, 2018.

[71] V. Anand and C. M. Rao, “MongoDB and Oracle NoSQL: A technical critique for design decisions,” in *2016 International Conference on Emerging Trends in Engineering, Technology and Science (ICETETS)*, 2016, pp. 1–4.

[72] A. B. M. Moniruzzaman and S. A. Hossain, “Nosql Database: New era of Databases for big data analytics-classification, characteristics and comparison,” *arXiv Prepr. arXiv1307.0191*, 2013.

[73] H. T. A. Simanjuntak, L. Simanjuntak, G. Situmorang, and A. Saragih, “Query Response Time Comparison NOSQLDB MONGODB with SQLDB Oracle,” *JUTI J. Ilm. Teknol. Inf.*, vol. 13, no. 1, pp. 95–105, 2015.

[74] A. L. Almeida, V. J. Schettino, T. J. R. Barbosa, P. F. Freitas, P. G. S. Guimarães, and W. A. ARBEX, “Relative scalability of NoSQL Databases for genotype data manipulation.,” *Embrapa Gado Leite-Artigo em periódico indexado*, 2018.

[75] R. Cattell, “Scalable SQL and NoSQL data stores,” *Acm Sigmod Rec.*, vol. 39, no. 4, pp. 12–27, 2011.

[76] C.-H. Lee and Y.-L. Zheng, “SQL-to-NoSQL schema denormalization and migration: a study on content management systems,” in *2015 IEEE International Conference on Systems, Man, and Cybernetics*, 2015, pp. 2022–2026.

[77] A. Meier and M. Kaufmann, “SQL & NoSQL Databases”.

[78] J. Pokorný, “Integration of Relational and NoSQL Databases,” in *Asian Conference on Intelligent Information and Database Systems*, 2018, pp. 35–45.

[79] A. V Miranskyy, Z. Al-zanbouri, D. Godwin, and A. B. Bener, “Database engines: Evolution of greenness,” *J. Softw. Evol. Process*, vol. 30, no. 4, p. e1915, 2018.

[80] M. Chapple, “The acid model,” *About. com,[online][retrieved Jun. 4, 2013] http//Databases. about. com/od/specificproducts/a/acid. htm*, p. 1, 2007.

[81] D. G. Chandra, “BASE analysis of NoSQL Database,” *Futur. Gener. Comput. Syst.*, vol. 52, pp. 13–21, 2015.

[82] M. T. Gonzalez-Aparicio, M. Younas, J. Tuya, and R. Casado, “Evaluation of ACE properties of traditional SQL and NoSQL big data systems,” in *Proceedings of the 34th ACM/SIGAPP Symposium on Applied Computing*, 2019, pp. 1988–1995.

[83] H. Sun, B. Xiao, X. Wang, and X. Liu, “Adaptive trade-off between consistency and performance in data replication,” *Softw. Pract. Exp.*, vol. 47, no. 6, pp. 891–906, 2017.

[84] C.-H. Lee and Y.-L. Zheng, “Automatic SQL-to-NoSQL schema transformation over the MySQL and HBase Databases,” in *2015 IEEE International Conference on Consumer*

*Electronics-Taiwan*, 2015, pp. 426–427.

[85] A. Kuzochkina, M. Shirokopetleva, and Z. Dudar, “Analyzing and Comparison of NoSQL DBMS,” in *2018 International Scientific-Practical Conference Problems of Infocommunications. Science and Technology (PIC S&T)*, 2018, pp. 560–564.

[86] A. T. Kabakus and R. Kara, “A performance evaluation of in-memory Databases,” *J. King Saud Univ. Inf. Sci.*, vol. 29, no. 4, pp. 520–525, 2017.

[87] J. Dean and S. Ghemawat, “MapReduce: simplified data processing on large clusters,” *Commun. ACM*, vol. 51, no. 1, pp. 107–113, 2008.

[88] J. Pokorný, “Database technologies in the world of big data,” in *Proceedings of the 16th International Conference on Computer Systems and Technologies*, 2015, pp. 1–12.

[89] H. Yang, A. Dasdan, R.-L. Hsiao, and D. S. Parker, “Map-reduce-merge: simplified relational data processing on large clusters,” in *Proceedings of the 2007 ACM SIGMOD international conference on Management of data*, 2007, pp. 1029–1040.

[90] T. Nykiel, M. Potamias, C. Mishra, G. Kollios, and N. Koudas, “MRShare: sharing across multiple queries in MapReduce,” *Proc. VLDB Endow.*, vol. 3, no. 1–2, pp. 494–505, 2010.

[91] C.-H. Lee and Z.-W. Shih, “A Comparison of NoSQL and SQL Databases over the Hadoop and Spark Cloud Platforms using Machine Learning Algorithms,” in *2018 IEEE International Conference on Consumer Electronics-Taiwan (ICCE-TW)*, 2018, pp. 1–2.

[92] S. Son, M.-S. Gil, Y.-S. Moon, and H.-S. Won, “Performance Analysis of Hadoop-Based SQL and NoSQL for Processing Log Data,” in *International Conference on Database Systems for Advanced Applications*, 2015, pp. 293–299.

[93] T. Ivanov and M. Pergolesi, “The impact of columnar file formats on SQL-on-hadoop engine performance: A study on ORC and Parquet,” *Concurr. Comput. Pract. Exp.*, p. e5523, 2019.

[94] F. Diaz and R. Freato, “Working with NoSQL Alternatives,” in *Cloud Data Design, Orchestration, and Management Using Microsoft Azure*, Springer, 2018, pp. 169–262.

[95] N. Zeng, G.-Q. Zhang, X. Li, and L. Cui, “Evaluation of relational and NoSQL approaches for patient cohort identification from heterogeneous data sources,” in *2017 IEEE International Conference on Bioinformatics and Biomedicine (BIBM)*, 2017, pp. 1135–1140.

[96] J.-G. Lee and M. Kang, “Geospatial big data: challenges and opportunities,” *Big Data Res.*, vol. 2, no. 2, pp. 74–81, 2015.

[97] Z. Liu, H. Guo, and C. Wang, “Considerations on geospatial big data,” in *IOP Conf. Ser.: Earth Environ. Sci*, 2016, vol. 46, no. 1, p. 12058.

[98] A. Albert, J. Kaur, and M. C. Gonzalez, “Using convolutional networks and satellite imagery to identify patterns in urban environments at a large scale,” in *Proceedings of the*

*23rd ACM SIGKDD international conference on knowledge discovery and data mining*, 2017, pp. 1357–1366.

[99] D. Ahlers and E. Wilde, "Report on the Seventh International Workshop on Location and the Web (LocWeb 2017)," in *ACM SIGIR Forum*, 2017, vol. 51, no. 1, pp. 52–57.

[100] N. Bari, G. Mani, and S. Berkovich, "Internet of things as a methodological concept," in *2013 Fourth International Conference on Computing for Geospatial Research and Application*, 2013, pp. 48–55.

[101] M. Mendoza, B. Poblete, and C. Castillo, "Twitter under crisis: Can we trust what we RT?," in *Proceedings of the first workshop on social media analytics*, 2010, pp. 71–79.

[102] C. Aubrecht, P. Meier, and H. Taubenböck, "Speeding up the clock in remote sensing: identifying the 'black spots' in exposure dynamics by capitalizing on the full spectrum of joint high spatial and temporal resolution," *Nat. Hazards*, vol. 86, no. 1, pp. 177–182, 2017.

[103] M. D. McCoy, "Geospatial Big Data and archaeology: Prospects and problems too great to ignore," *J. Archaeol. Sci.*, vol. 84, pp. 74–94, 2017.

[104] M. Burzańska and P. Wiśniewski, "How Poor Is the 'Poor Man's Search Engine'?," in *International Conference: Beyond Databases, Architectures and Structures*, 2018, pp. 294–305.

[105] K. Harezlak and R. Skowron, "Performance aspects of migrating a web application from a relational to a NoSQL Database," in *International Conference: Beyond Databases, Architectures and Structures*, 2015, pp. 107–115.

[106] S. Rautmare and D. M. Bhalerao, "MySQL and NoSQL Database comparison for IoT application," in *2016 IEEE International Conference on Advances in Computer Applications (ICACA)*, 2016, pp. 235–238.

[107] A.-S. Aya, H. Qattous, and M. Hijjawi, "A proposed performance evaluation of NoSQL Databases in the field of IoT," in *2018 8th International Conference on Computer Science and Information Technology (CSIT)*, 2018, pp. 32–37.

[108] D. Bartoszewski, A. Piorkowski, and M. Lupa, "The Comparison of Processing Efficiency of Spatial Data for PostGIS and MongoDB Databases," in *International Conference: Beyond Databases, Architectures and Structures*, 2019, pp. 291–302.

[109] A. Tear, "SQL or NoSQL? Contrasting approaches to the storage, manipulation and analysis of spatio-temporal online social network data," in *International Conference on Computational Science and Its Applications*, 2014, pp. 221–236.

[110] K. Fraczek and M. Plechawska-Wojcik, "Comparative analysis of relational and non-relational Databases in the context of performance in web applications," in *International Conference: Beyond Databases, Architectures and Structures*, 2017, pp. 153–164.

[111] R. Hricov, A. Šenk, P. Kroha, and M. Valenta, "Evaluation of XPath queries over XML

documents using SparkSQL framework," in *International Conference: Beyond Databases, Architectures and Structures*, 2017, pp. 28–41.

[112] E. Płuciennik and K. Zgorzałek, "The multi-model Databases–a review," in *International Conference: Beyond Databases, Architectures and Structures*, 2017, pp. 141–152.

[113] P. Yue and Z. Tan, "1.06 GIS Databases and NoSQL Databases," *Compr. Geogr. Inf. Syst.*, p. 50, 2017.

[114] P. A. Longley, M. F. Goodchild, D. J. Maguire, and D. W. Rhind, *Geographic information systems and science*. John Wiley & Sons, 2005.

[115] P. Bajerski and S. Kozielski, "Computational model for efficient processing of geofield queries," in *Man-Machine Interactions*, Springer, 2009, pp. 573–583.

[116] M. Chromiak and K. Stencel, "A data model for heterogeneous data integration architecture," in *International Conference: Beyond Databases, Architectures and Structures*, 2014, pp. 547–556.

[117] P. K. Akulakrishna, J. Lakshmi, and S. K. Nandy, "Efficient storage of big-data for real-time gps applications," in *2014 IEEE Fourth International Conference on Big Data and Cloud Computing*, 2014, pp. 1–8.

[118] M. Lupa, K. Kozioł, and A. Leśniak, "An attempt to automate the simplification of building objects in multiresolution Databases," in *International Conference: Beyond Databases, Architectures and Structures*, 2015, pp. 448–459.

[119] K. Kozioł, M. Lupa, and A. Krawczyk, "The extended structure of multi-resolution Database," in *International Conference: Beyond Databases, Architectures and Structures*, 2014, pp. 435–443.

[120] M. Wyszomirski, "Przegląd możliwości zastosowania wybranych baz danych NoSQL do zarządzania danymi przestrzennymi," *Rocz. Geomatyki-Annals Geomatics*, vol. 16, no. 1 (80), pp. 55–69, 2018.

[121] A. Czerepicki, "Perspektywy zastosowania baz danych NoSQL w inteligentnych systemach transportowych," *Pr. Nauk. Politech. Warsz. Transp.*, vol. 92, pp. 29–38, 2013.

[122] P. Martins, J. Cecílio, M. Abbasi, and P. Furtado, "GISB: a benchmark for geographic map information extraction," in *Beyond Databases, Architectures and Structures. Advanced Technologies for Data Mining and Knowledge Discovery*, Springer, 2015, pp. 600–609.

[123] A. Inglot and K. Koziol, "The importance of contextual topology in the process of harmonization of the spatial Databases on example BDOT500," in *2016 Baltic Geodetic Congress (BGC Geomatics)*, 2016, pp. 251–256.

[124] M. Chuchro, A. Franczyk, M. Dwornik, and A. Leśniak, "A Big Data processing strategy for hybrid interpretation of flood embankment multisensor data," *Geol. Geophys. Environ.*, vol. 42, no. 3, pp. 269–277, 2016.

[125] W. L. Schulz, B. G. Nelson, D. K. Felker, T. J. S. Durant, and R. Torres, "Evaluation of relational and NoSQL Database architectures to manage genomic annotations," *J. Biomed. Inform.*, vol. 64, pp. 288–295, 2016.

[126] J. Lian, S. Miao, M. McGuire, and Z. Tang, "SQL or NoSQL? Which Is the Best Choice for Storing Big Spatio-Temporal Climate Data?," in *International Conference on Conceptual Modeling*, 2018, pp. 275–284.

[127] A. Piórkowski, "Mysql spatial and postgis–implementations of spatial data standards," *EJPAU*, vol. 14, no. 1, p. 3, 2011.

[128] R. Kothuri, A. Godfrind, and E. Beinat, *Pro oracle spatial for oracle Database 11g*. Dreamtech Press, 2008.

[129] E. Baralis, A. Dalla Valle, P. Garza, C. Rossi, and F. Scullino, "SQL versus NoSQL Databases for geospatial applications," in *2017 IEEE International Conference on Big Data (Big Data)*, 2017, pp. 3388–3397.

[130] N. Roy-Hubara and A. Sturm, "Exploring the Design Needs for the New Database Era," in *Enterprise, Business-Process and Information Systems Modeling*, Springer, 2018, pp. 276–290.

[131] J. Yoon, D. Jeong, C. Kang, and S. Lee, "Forensic investigation framework for the document store NoSQL DBMS: MongoDB as a case study," *Digit. Investig.*, vol. 17, pp. 53–65, 2016.

[132] E. Mehmood and T. Anees, "Performance Analysis of Not Only SQL Semi-Stream Join Using MongoDB for Real-Time Data Warehousing," *IEEE Access*, vol. 7, pp. 134215–134225, 2019.

[133] L. Okman, N. Gal-Oz, Y. Gonen, E. Gudes, and J. Abramov, "Security issues in nosql Databases," in *2011IEEE 10th International Conference on Trust, Security and Privacy in Computing and Communications*, 2011, pp. 541–547.

[134] E. Alomari, A. Barnawi, and S. Sakr, "Cdport: A portability framework for nosql datastores," *Arab. J. Sci. Eng.*, vol. 40, no. 9, pp. 2531–2553, 2015.

[135] E. Alomari and A. Noaman, "SeCloudDB: A Unified API for Secure SQL and NoSQL Cloud Databases," in *Proceedings of the 2019 3rd International Conference on Cloud and Big Data Computing*, 2019, pp. 38–42.

[136] K. Stravoskoufos, A. Preventis, S. Sotiriadis, and E. G. M. Petrakis, "A Survey on Approaches for Interoperability and Portability of Cloud Computing Services.," in *CLOSER*, 2014, pp. 112–117.

[137] M. N. Shirazi, H. C. Kuan, and H. Dolatabadi, "Design patterns to enable data portability between clouds' Databases," in *2012 12th International Conference on Computational Science and Its Applications*, 2012, pp. 117–120.

[138] E. Alomari, A. Barnawi, and S. Sakr, "Cdport: A framework of data portability in cloud

platforms," in *Proceedings of the 16th International Conference on Information Integration and Web-based Applications & Services*, 2014, pp. 126–133.

[139] I. Indu, R. A. PM, and V. Bhaskar, "Encrypted token based authentication with adapted SAML technology for cloud web services," *J. Netw. Comput. Appl.*, vol. 99, pp. 131–145, 2017.

[140] B. Vanelli *et al.*, "Internet of things data storage infrastructure in the cloud using NoSQL Databases," *IEEE Lat. Am. Trans.*, vol. 15, no. 4, pp. 737–743, 2017.

[141] K. Grolinger, W. A. Higashino, A. Tiwari, and M. A. M. Capretz, "Data management in cloud environments: NoSQL and NewSQL data stores," *J. Cloud Comput. Adv. Syst. Appl.*, vol. 2, no. 1, p. 22, 2013.

[142] S. Marston, Z. Li, S. Bandyopadhyay, J. Zhang, and A. Ghalsasi, "Cloud computing—The business perspective," *Decis. Support Syst.*, vol. 51, no. 1, pp. 176–189, 2011.

[143] B. Kolev, C. Bondiombouy, P. Valduriez, R. Jiménez-Peris, R. Pau, and J. Pereira, "The cloudmdsql multistore system," in *Proceedings of the 2016 International Conference on Management of Data*, 2016, pp. 2113–2116.

[144] N. Loutas, E. Kamateri, and K. Tarabanis, "A semantic Interoperability framework for cloud platform as a service," in *2011 IEEE Third International Conference on Cloud Computing Technology and Science*, 2011, pp. 280–287.

[145] L. Zhou, A. Fu, S. Yu, M. Su, and B. Kuang, "Data integrity verification of the outsourced big data in the cloud environment: A survey," *J. Netw. Comput. Appl.*, vol. 122, pp. 1–15, 2018.

[146] M. Kostoska, M. Gusev, S. Ristov, and K. Kiroski, "Cloud Computing Interoperability Approaches-Possibilities and Challenges.," in *BCI (Local)*, 2012, pp. 30–34.

[147] N. Loutas, E. Kamateri, F. Bosi, and K. Tarabanis, "Cloud computing Interoperability: the state of play," in *2011 IEEE Third International Conference on Cloud Computing Technology and Science*, 2011, pp. 752–757.

[148] M. F. P. Escalera and M. A. L. Chávez, "UML model of a standard API for cloud computing application development," in *2012 9th International Conference on Electrical Engineering, Computing Science and Automatic Control (CCE)*, 2012, pp. 1–8.

[149] D. Petcu *et al.*, "Experiences in building a mOSAIC of clouds," *J. Cloud Comput. Adv. Syst. Appl.*, vol. 2, no. 1, p. 12, 2013.

[150] E. Di Nitto *et al.*, "Supporting the development and operation of multi-cloud applications: The modaclouds approach," in *2013 15th International Symposium on Symbolic and Numeric Algorithms for Scientific Computing*, 2013, pp. 417–423.

[151] D. Petcu, "Portability and Interoperability between clouds: challenges and case study," in *European Conference on a Service-Based Internet*, 2011, pp. 62–74.

[152] C.-S. Liao, J.-M. Shih, and R.-S. Chang, "Simplifying MapReduce data processing," *Int. J. Comput. Sci. Eng.*, vol. 8, no. 3, pp. 219–226, 2013.

[153] L. A. B. Silva, C. Costa, and J. L. Oliveira, "A common API for delivering services over multi-vendor cloud resources," *J. Syst. Softw.*, vol. 86, no. 9, pp. 2309–2317, 2013.

[154] S. Sakr, A. Liu, D. M. Batista, and M. Alomari, "A survey of large scale data management approaches in cloud environments," *IEEE Commun. Surv. Tutorials*, vol. 13, no. 3, pp. 311–336, 2011.

[155] C. Curino *et al.*, "Relational cloud: A Database-as-a-service for the cloud," 2011.

[156] W. Lehner and K.-U. Sattler, "Database as a service (DBaaS)," in *2010 IEEE 26th International Conference on Data Engineering (ICDE 2010)*, 2010, pp. 1216–1217.

[157] T. Kiefer and W. Lehner, "Private table Database virtualization for dbaas," in *2011 Fourth IEEE International Conference on Utility and Cloud Computing*, 2011, pp. 328–329.

[158] R. Zafar, E. Yafi, M. F. Zuhairi, and H. Dao, "Big data: the NoSQL and RDBMS review," in *2016 International Conference on Information and Communication Technology (ICICTM)*, 2016, pp. 120–126.

[159] K. Sahatqija, J. Ajdari, X. Zenuni, B. Raufi, and F. Ismaili, "Comparison between relational and NOSQL Databases," in *2018 41st International Convention on Information and Communication Technology, Electronics and Microelectronics (MIPRO)*, 2018, pp. 216–221.

[160] N. Alias *et al.*, "Parallel computing of numerical schemes and big data analytic for solving real life applications," *J. Teknol.*, vol. 78, no. 8–2, 2016.

[161] M.-L. E. Chang and H. N. Chua, "SQL and NoSQL Database Comparison," in *Future of Information and Communication Conference*, 2018, pp. 294–310.

[162] J. Frizzo-Barker, P. A. Chow-White, M. Mozafari, and D. Ha, "An empirical study of the rise of big data in business scholarship," *Int. J. Inf. Manage.*, vol. 36, no. 3, pp. 403–413, 2016.

[163] S. Chickerur, A. Goudar, and A. Kinnerkar, "Comparison of relational Database with document-oriented Database (mongodb) for big data applications," in *2015 8th International Conference on Advanced Software Engineering & Its Applications (ASEA)*, 2015, pp. 41–47.

[164] M. S. Kumar, "Comparison of NoSQL Database and Traditional Database-An emphatic analysis," *JOIV Int. J. Informatics Vis.*, vol. 2, no. 2, pp. 51–55, 2018.

[165] H. Ansari, "Performance Comparison of Two Database Management Systems MySQL vs MongoDB." 2018.